\documentclass[11pt, a4paper]{article}
\usepackage[a4paper]{geometry} 
\usepackage{amsmath}
\usepackage{algorithm}
\usepackage{algpseudocode}
\usepackage{mathtools}
\usepackage{amssymb}
\usepackage{bm}
\usepackage{physics}
\usepackage{booktabs}
\usepackage{tabularx}
\usepackage{enumitem}
\usepackage{pifont}
\usepackage{subcaption}
\usepackage{float}
\usepackage{graphicx} 
\usepackage[affil-it]{authblk} 
\newcommand{\pderiv}[3]{\left(\frac{\partial #1}{\partial #2}\right)_{#3}}

\title{An Oscillation-Free Real Fluid Quasi-Conservative Finite Volume Method for Transcritical and Phase-Change Flows}
\author[1,2]{Haotong Bai}
\author[2]{Wenjia Xie\thanks{Corresponding author: xiewenjia@nudt.edu.cn}}
\author[1,2]{Yixin Yang\thanks{Corresponding author: yangyixin@nudt.edu.cn}}
\author[3]{Ping Yi}
\author[1,2]{Mingbo Sun\thanks{Corresponding author: sunmingbo@nudt.edu.cn}}

\affil[1]{Hypersonic Technology Laboratory, National University of Defense Technology, Kaifu District, Changsha, Hunan 41073, China}
\affil[2]{College of Aerospace Science and Engineering, National University of Defense Technology, Kaifu District, Changsha, Hunan 41073, China}
\affil[3]{Institute of Power Plants and Automation, Shanghai Jiao Tong University, Shanghai, 200240, China}

\date{}
\begin{document}

\begingroup
\renewcommand\thefootnote{}
\footnotetext{
\textcopyright\ 2026. This manuscript version is made available under the
CC-BY-NC-ND 4.0 license. 

This is the accepted manuscript version for publication in \textit{Journal of Computational Physics}.
}
\endgroup
\maketitle
\begin{abstract}
\noindent \textbf{Abstract~~~}
A new Real Fluid Quasi-Conservative (RFQC) finite volume method is developed to address the numerical simulation of real fluids involving shock waves in transcritical and phase-change flows. To eliminate the spurious pressure oscillations inherent in fully conservative schemes, we extend the classic quasi-conservative method, originally designed for two-phase flows, to real fluids governed by arbitrary equations of state (EoS). The RFQC method locally linearizes the real fluid EoS at each grid point and time step, constructing and evolving the frozen Grüneisen coefficient $\Gamma$ and the linearization remainder $E_0$ via two advection equations. At the end of each time step, the evolved $\Gamma$ and $E_0$ are utilized to reconstruct the oscillation-free pressure field, followed by a thermodynamic re-projection applied to the conserved variables. Theoretical analysis demonstrates that, in smooth regions, the energy conservation error introduced by the RFQC method is a second-order small term dominated by the time-step. In discontinuous regions, this error is determined by the entropy increase rate, thereby maintaining consistency with the inherent truncation error of shock-capturing methods. A series of numerical tests verifies that the method can robustly simulate complex flow processes with only minor energy conservation errors, including transcritical flows, phase transitions, and shock-interface interactions. The RFQC method is proven to be both accurate and robust in capturing shock waves and phase transitions.
\\
\textbf{Keywords~~~}Transcritical flow, Phase transition, Two-phase flow, Pressure oscillation, Finite volume method
\end{abstract}

\section{Introduction}
\begin{figure}[h]
\centering 
\includegraphics[width=0.85\textwidth]{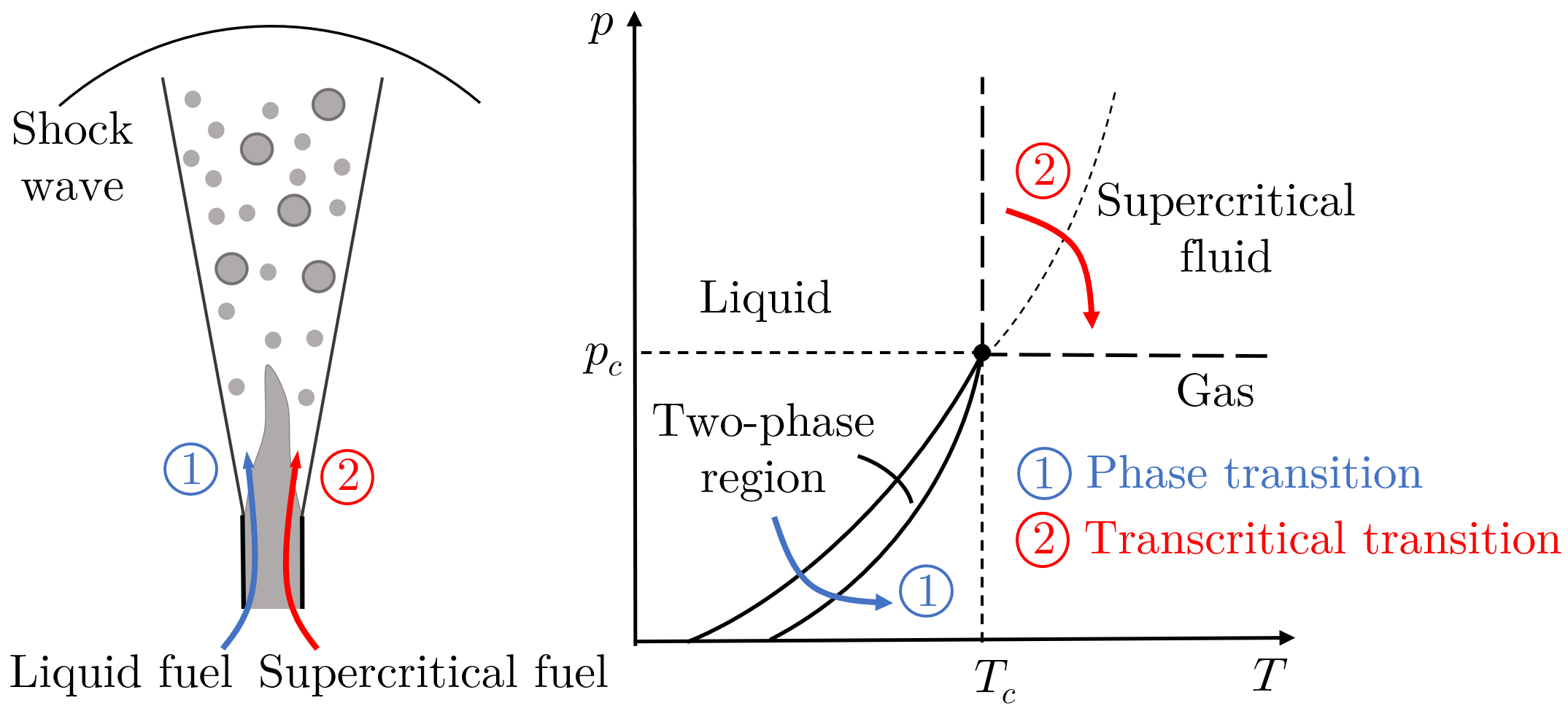} 
\caption{Fuel spray and phase diagram.}
\label{fig1}
\end{figure}

Complex real-fluid flows involving transcritical and phase-change phenomena are prevalent in advanced propulsion systems, such as transverse injection in scramjet combustors \cite{Tang2023,Li2023a,Li2023b}, spray jets in high-pressure internal combustion engines \cite{Yi2019,Yi2025,Cong2026}, and cavitating flows in rocket engine turbopumps \cite{Popp2004,Caze2024}. In the specific context of scramjets, fuel is typically injected into the supersonic combustor crossflow in a liquid or supercritical state. During this process, the fuel undergoes transcritical transitions or phase changes, as shown in Fig. \ref{fig1}. Meanwhile, the interaction between the supersonic airflow and the spray jet induces complex shock structures. To elucidate the mechanisms of such complex flows and provide reliable predictions, it is essential to develop accurate and robust shock-capturing schemes for real fluids.

It is well established that traditional conservative schemes often induce severe non-physical pressure oscillations when simulating transcritical real-fluid flows. This issue arises from the high nonlinearity of the equation of state (EoS), leading to computational errors or even numerical instability \cite{terashima2012,Ma2017}. This phenomenon is analogous to the spurious oscillations observed at phase interfaces in two-phase flows simulated by conservative schemes \cite{Abgrall1996,Abgrall2001}. In fact, the pressure oscillation problem in real fluids can be interpreted as a manifestation of the classic interface problem in two-phase flows. Over the past few decades, extensive research has been conducted to address this issue, primarily including pressure-based, enthalpy-based, double-flux, and hybrid conservative/non-conservative methods. Among these, pressure-based (PB) methods \cite{Karni1992,Karni1994,terashima2012,Kitamura2018} were the first to be proposed. These methods evolve a pressure equation in place of the total energy conservation equation. While they have shown promise in simulating transcritical turbulent flows \cite{Kawai2015}, PB methods inherently violate conservation laws by discarding the energy equation, leading to calculation errors in shock capturing. To mitigate this, Lacaze et al. \cite{Lacaze2019} proposed an enthalpy-based equation to replace the pressure equation, thereby improving energy conservation compared to pressure-based approaches.
The Double Flux (DF) method, proposed by Ma et al. \cite{Ma2017}, is currently one of the most concise and widely used methods for simulating transcritical and phase-change flows \cite{rodriguez2018,yatsuyanagi2022}. This method freezes the isentropic index and the linearization remainder of the real fluid within each time step, extending the ideal-gas double-flux scheme to real fluids to maintain formal conservation. However, this approach introduces flux inconsistencies at grid interfaces, limiting the spatial accuracy of energy conservation to first order. Consequently, significant errors persist when simulating flows with severe property variations, such as vapor-liquid first-order phase transitions.
Hybrid conservative/non-conservative methods \cite{Karni1996,Xu2022,Xu2026} apply non-conservative evolution in regions with transcritical contact discontinuities and conservative evolution near shock waves. While this approach improves energy conservation, it relies on ad-hoc shock sensors  and is computationally expensive.
Recently, researchers have also focused on fully conservative schemes for real fluids. Ching et al. \cite{ching2023,ching2025} proposed a fully conservative discontinuous Galerkin method for supercritical fluids, utilizing primitive variable projection to mitigate pressure oscillations. Fujiwara et al. \cite{fujiwara2023} derived pressure equilibrium conditions from the Euler equations and constructed a conservative, oscillation-free numerical flux, which was subsequently extended to real fluids by Terashima et al. \cite{terashima2025}. However, generalizing these methods to phase-change flows involving discontinuities requires further investigation.

In summary, the numerical simulation of real fluids involving shocks, transcritical states, and phase changes remains challenging. It requires an algorithm that balances pressure stability, energy conservation, algorithmic complexity, and computational cost. Thus, we propose an oscillation-free Real Fluid Quasi-Conservative (RFQC) finite volume method. Inspired by the thermodynamic coefficient freezing strategy in the DF method of Ma\cite{Ma2017}, the RFQC method freezes two thermodynamic coefficients—the Grüneisen coefficient $\Gamma$ and the linearization remainder $E_0$—to achieve a local linearization of the pressure-internal energy relation, thereby reconstructing an oscillation-free pressure. Building upon the quasi-conservative framework of Abgrall and Shyue\cite{Abgrall1996,Shyue1998,Johnsen2006}, we evolve these frozen thermodynamic coefficients via advection equations within each time step. This strategy eliminates the spatial flux inconsistency inherent in the DF method, achieving higher conservation accuracy. Furthermore, the proposed method is not limited to a specific form of the EoS and remains algorithmically concise. Compared to Shyue's original scheme, it requires only one additional thermodynamic projection step at the end of each time step, incurring only minimal computational overhead.

The structure of this paper is organized as follows: Section 2 introduces the governing equations and thermodynamic relations for fluids under the Homogeneous Equilibrium and Vapor-Liquid Equilibrium assumptions. Section 3 analyzes the mathematical origin of pressure oscillations, details the construction and implementation of the proposed algorithm, and provides a theoretical analysis of the evolution and convergence properties of conservation errors. In Section 4, within the Peng-Robinson EoS \cite{Peng1976}, we present numerical results for transcritical and flash evaporation Riemann problems \cite{Bai2026} of n-dodecane. These results are validated against pressure-based methods, double-flux methods, and exact solutions. Finally, a one-dimensional shock-interface interaction, a two-dimensional advection and shock-droplet interaction, as well as two-dimensional transcritical and phase-change jet flows are simulated to demonstrate the superior robustness of the proposed method.

\section{Governing Equations and Basic Thermodynamic Quantities}
\label{sec:euler equ}
We consider the one-dimensional Euler equations for a single-component fluid under the assumptions of the Homogeneous Equilibrium Model (HEM) \cite{Brennen2005} and Vapor-Liquid Equilibrium (VLE) \cite{Yi2019}. The conservation laws for mass, momentum, and total energy are given by:

\begin{equation}
\begin{dcases}
\frac{\partial \rho}{\partial t} + \frac{\partial}{\partial x}(\rho u) = 0, \\
\frac{\partial (\rho u)}{\partial t} +\frac{\partial}{\partial x}(\rho u^2 + p) = 0, \\
\frac{\partial (\rho e_t)}{\partial t} + \frac{\partial}{\partial x}((\rho e_t + p)u) = 0.
\label{equ:euler}
\end{dcases}
\end{equation}
where $\rho$ denotes density, $u$ velocity, and $p$ pressure. The specific total energy $e_t$ is related to the specific internal energy $e$ by:
\begin{equation}
    e_t = e+\frac{u^2}{2}.
\end{equation}\par

The Euler system is closed by two EoS. We provide only the general forms here. Using $\rho$ and $T$ as independent variables, the thermal EoS is:
\begin{equation}
    p = p(\rho,T)
\end{equation}\par
and the caloric EoS is:
\begin{equation}
    e = e(\rho,T)
\end{equation}\par
Additionally, the speed of sound $c$ is also a function of $\rho$ and $T$:
\begin{equation}
    c = c(\rho,T)
\end{equation}\par
Detailed definitions of internal energy, entropy, and speed of sound for single-phase and two-phase fluids under HEM and VLE assumptions are provided in \ref{app:thermo}. These governing equations and thermodynamic relations extend naturally to multi-component cases. As the proposed method imposes no restrictions on the number of components, further elaboration is omitted for brevity.

\section{Numerical Method}
\subsection{Spurious Pressure Oscillations}
Spurious pressure oscillation at interface or contact discontinuity is a classic problem in two-phase, transcritical, and phase-change flows. The root cause lies in the inconsistency between the averaging operation in the projection step of finite volume methods (or the Reconstruction-Evolution-Average/Projection methods\cite{Toro2009}) and the nonlinearity of the internal energy-pressure relationship. Specifically, the internal energy is averaged  within a cell in the finite volume method. However, due to the nonlinearity of the EoS, the pressure computed from the averaged internal energy differs from the average of the pressures, violating the pressure equilibrium condition across the contact discontinuity. To elucidate this mechanism and motivate the proposed solution, we analyze the pressure-internal energy relationship below, extending the seminal work of Abgrall \cite{Abgrall1996} and Shyue \cite{Shyue1998}. For brevity, we focus on single-component fluids, though the analysis applies naturally to multi-component fluids.

Consider the energy conservation equation of the 1D Euler system. Across a contact discontinuity with constant velocity $u=u_0$ and pressure $p=p_0$, the kinetic energy and pressure work terms vanish. Consequently, the energy equation reduces to a pure advection equation for the volumetric internal energy $\rho e$:
\begin{equation}
    \frac{\partial (\rho e)}{\partial t} + u_0 \frac{\partial (\rho e)}{\partial x} = 0.
    \label{equ:rho_e_eq}
\end{equation}
Consider a first-order finite volume update from time step $n$ to $n+1$. The conserved variables (density $\rho$ and volumetric internal energy $\rho e$) in a grid cell are updated as weighted averages of the left ($L$) and right ($R$) states. Let $\lambda$ denote the weighting coefficient determined by the CFL condition. The discrete update is given by:
\begin{equation}
    \begin{aligned}
       \rho_i^{n+1} &= (1-\lambda)\rho_L + \lambda \rho_R = \bar{\rho}, \\
        (\rho e)_i^{n+1} &= (1-\lambda)(\rho e)_L + \lambda (\rho e)_R = \overline{\rho e}.
    \end{aligned}
\end{equation}
At time $n+1$, we must solve for pressure $p^{n+1}$ via the EoS such that the equilibrium condition is satisfied:
 \begin{equation}
     p^{n+1} = p\left( \rho^{n+1}, e^{n+1} \right) = p\left( \bar{\rho}, \frac{\overline{\rho e}}{\bar{\rho}} \right) = p_0.
 \end{equation}
Define $E(\rho) = \rho e(\rho, p_0)$ as the volumetric internal energy at constant pressure $p_0$. To maintain the updated pressure at $p^{n+1} = p_0$, the updated conservative variables must also lie on this isobar, which yields $(\rho e)_i^{n+1} = E(\rho_i^{n+1})$. Thus, the oscillation-free condition is equivalent to the following:
\begin{equation}
    E( (1-\lambda)\rho_L + \lambda \rho_R) = (1-\lambda)E(\rho_L) + \lambda E(\rho_R).
\end{equation}
This implies that pressure equilibrium is maintained only if $E(\rho)$ is linear with respect to $\rho$. However, for stiff fluids in different states (or components) or for general real fluids, $E(\rho)$ is typically highly nonlinear. Therefore, standard conservative schemes inevitably generate spurious pressure oscillations when resolving contact discontinuities.

\subsection{Shyue's Quasi-Conservative Model and Local Freezing Strategy}

To address pressure oscillations at interfaces in conservative schemes, Shyue proposed a quasi-conservative five-equation model for two-phase flows governed by the stiffened gas EoS \cite{Shyue1998}, satisfying the Abgrall consistency condition \cite{Abgrall1996}. The stiffened gas EoS is expressed as:
\begin{equation}
    \rho e = \frac{p + \gamma p_{\infty}}{\gamma - 1}
    \label{equ:stiff_eos}
\end{equation}
Substituting Eq. (\ref{equ:stiff_eos}) into the pressure equilibrium condition (Eq. (\ref{equ:rho_e_eq})) yields:
\begin{equation}
    p_0 \left( \frac{\partial}{\partial t}\left(\frac{1}{\gamma-1} \right)+ u_0 \frac{\partial}{\partial x}\left(\frac{1}{\gamma-1} \right) \right) + \left( \frac{\partial}{\partial t}\left(\frac{\gamma p_{\infty} }{\gamma-1} \right) + u_0 \frac{\partial}{\partial x}\left(\frac{\gamma p_{\infty} }{\gamma-1} \right) \right) = 0
    \label{equ:gamma_eq}
\end{equation}

For Eq. (\ref{equ:gamma_eq}) to hold for arbitrary pressure $p$ and velocity $u$, the terms within the parentheses must vanish independently. This leads to the advection equations for $1/(\gamma-1)$ and $\gamma p_{\infty}/(\gamma-1)$:
\begin{equation}
\begin{dcases}
     \frac{\partial}{\partial t}\left(\frac{1}{\gamma-1} \right)+ u \frac{\partial}{\partial x}\left(\frac{1}{\gamma-1} \right) = 0\\
     \frac{\partial}{\partial t}\left(\frac{\gamma p_{\infty} }{\gamma-1} \right) + u \frac{\partial}{\partial x}\left(\frac{\gamma p_{\infty} }{\gamma-1} \right) = 0
     \label{equ:gamma_eq2}
\end{dcases}
\end{equation}

These advection equations serve as the two additional equations in Shyue's model, independent of the Euler equations. After solving these advection equations, the oscillation-free pressure can be recovered via the pressure-internal energy relation: 

\begin{equation}
    p = \frac{\rho e - \frac{\gamma p_{\infty}}{\gamma - 1}}{\gamma - 1}
    \label{equ:stiff_eos2}
\end{equation}

Shyue's model was originally developed for the stiffened gas EoS. Theoretically, this method can be extend to other EoS, such as the Mie-Grüneisen EoS \cite{Shyue2001}, by deriving advection equations for specific auxiliary variables. However, for complex real fluid EoS—such as Cubic EoS\cite{Peng1976,RedlichKwong1949}, tabulated EoS\cite{Zhang2024}, or neural network EoS \cite{Srinivasan2024}—deriving exact advection equations for auxiliary variables is often analytically intractable or impractical. Therefore, given the remarkable success of the quasi-conservative scheme in two-phase flows, we extend this approach to arbitrary real fluid EoS by locally freezing thermodynamic coefficients.

By analogy with the stiffened gas EoS (Eq. (\ref{equ:stiff_eos})), we decompose the internal energy-pressure relation of a real fluid into a linear term and a remainder term:
\begin{equation}
    \rho e = p\xi + E_0
    \label{equ:liner1}
\end{equation}
Here, corresponding to $1/(\gamma-1)$ in the stiffened gas EoS, $\xi$ denotes the reciprocal of the Grüneisen coefficient, serving as the linearization coefficient for the internal energy-pressure relation:
\begin{equation}
    \xi = \frac{1}{\Gamma}= \frac{h}{c^2} = \frac{e+p/\rho}{c^2} 
    \label{equ:xi}
\end{equation}
Corresponding to $\gamma p_{\infty}/(\gamma-1)$, $E_0$ is the linearization remainder:
\begin{equation}
    E_0 = \rho e - \frac{p(e+p/\rho) }{c^2}
    \label{equ:E0}
\end{equation}

Consistent with Shyue's framework, we evolve $\xi$ and $E_0$ via advection equations within each time step:
\begin{equation}
\begin{dcases}
 &\frac{\partial \xi}{\partial t}+ u \frac{\partial \xi}{\partial x} = 0\\
 & \frac{\partial E_0}{\partial t} + u \frac{\partial E_0}{\partial x} = 0
     \label{equ:xi_eq}
\end{dcases}
\end{equation}
Subsequently, the oscillation-free pressure can be reconstructed via the evolved $\xi$ and $E_0$:
\begin{equation}
    p = \frac{\rho e - E_0}{\xi}
    \label{equ:liner2}
\end{equation}

\begin{figure}[htbp]
\centering
\begin{subfigure}{0.49\textwidth}
    \includegraphics[width=\linewidth]{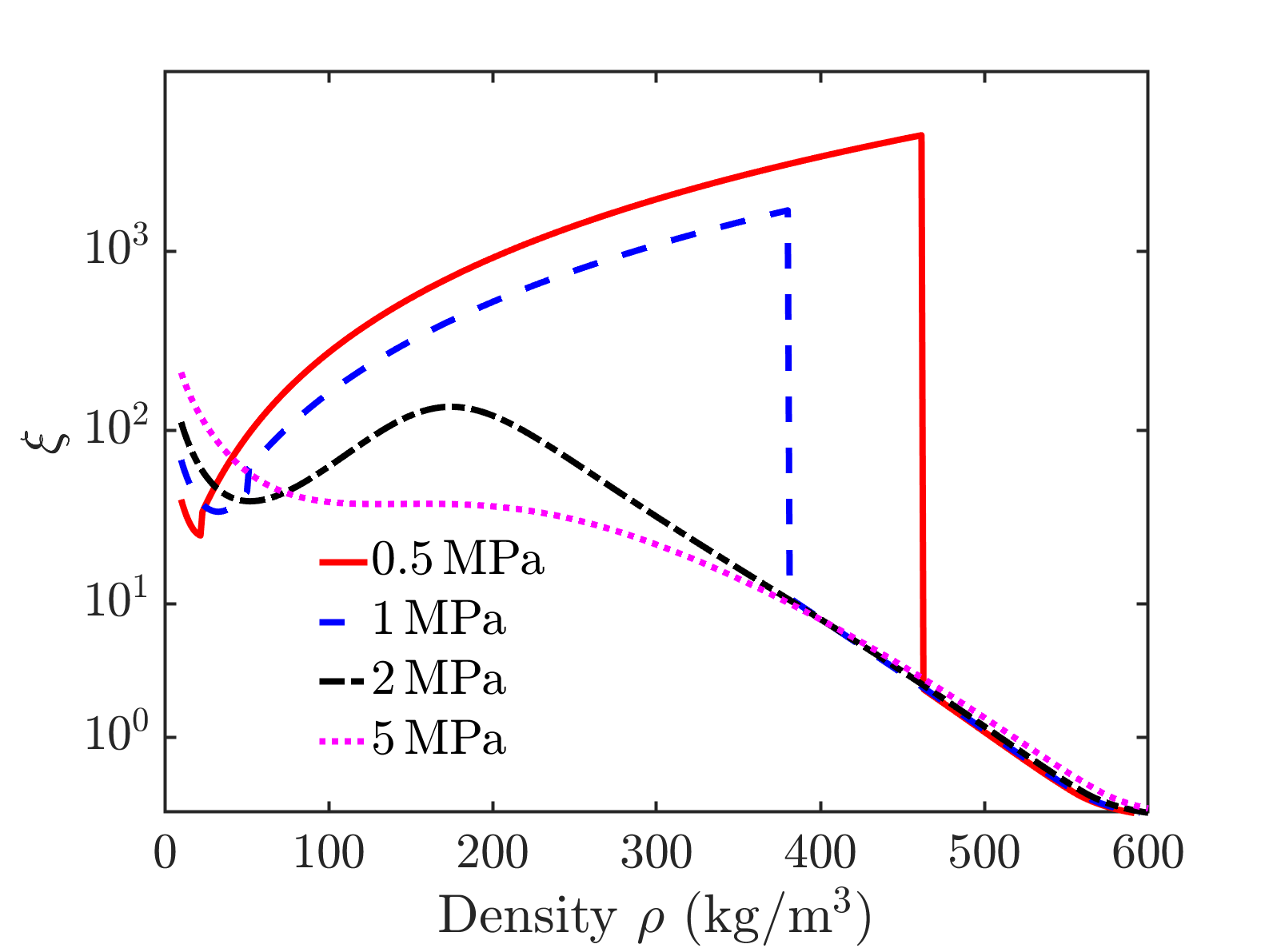}
    \caption{$\xi$ in the RFQC method}
    \label{fig2a}
\end{subfigure}
%\hfill
\begin{subfigure}{0.49\textwidth}
    \includegraphics[width=\linewidth]{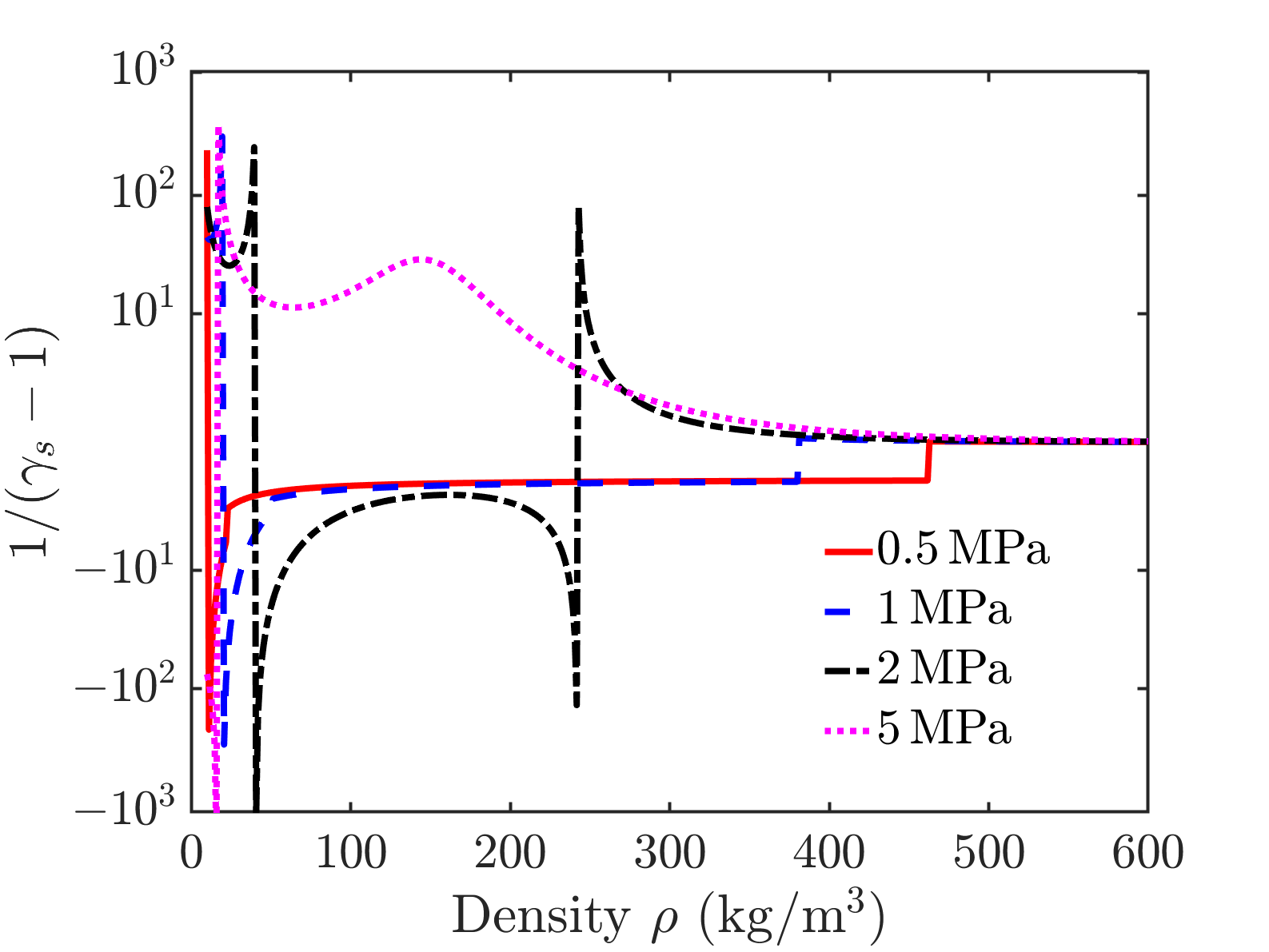}
    \caption{$ 1/(\gamma_s-1)$ in the DF method}
    \label{fig2b}
\end{subfigure}
\caption{Comparison of frozen thermodynamic coefficients between the two methods.}
\label{fig2}
\end{figure}

We briefly analyze the properties of the frozen linearization coefficient $\xi$. Taking n-Dodecane as a representative fluid, Fig. \ref{fig2a} illustrates the $\xi-\rho$ distribution curves at various pressures. Due to phase transitions, $\xi$ exhibits discontinuities below the critical pressure ($p_c = 1.82$ MPa), indicating strong nonlinearity. Conversely, the variation of $\xi$ remains relatively smooth under supercritical pressures. For comparison, Fig. \ref{fig2b} shows the frozen coefficient $1/(\gamma_s-1)$ utilized in the DF method \cite{Ma2017}, where $\gamma_s = \rho c^2 / p$ denotes the isentropic index. Compared to the coefficient in the DF method, $\xi$ demonstrates more favorable mathematical properties, characterized by smoother variations. It is important to note that $\xi = 1/\Gamma = h/c^2$ corresponds to the Grüneisen coefficient definition for stiffened gases. However, for general real fluids, the thermodynamic Grüneisen parameter $\Gamma = (\partial p/\partial e)_{\rho}/\rho$ is not strictly equal to $c^2/h$. Consequently, the value of $\xi$ depends on the internal energy reference state. In practical computations, an appropriate reference state should be selected to ensure favorable mathematical properties for $\xi$.

\subsection{Real Fluid Quasi-Conservative Scheme}
In summary, the proposed Real Fluid Quasi-Conservative scheme comprises the Euler equations and two advection equations for thermodynamic coefficients:
\begin{equation}
\begin{dcases}
\frac{\partial \rho}{\partial t} + \frac{\partial}{\partial x}(\rho u) = 0 \\
\frac{\partial (\rho u)}{\partial t} +\frac{\partial}{\partial x}(\rho u^2 + p) = 0 \\
\frac{\partial (\rho e_t)}{\partial t} + \frac{\partial}{\partial x}((\rho e_t + p)u) = 0 \\
\frac{\partial \xi}{\partial t} + u \frac{\partial \xi}{\partial x} = 0 \\
\frac{\partial E_0}{\partial t} + u \frac{\partial E_0}{\partial x} = 0
\end{dcases}
\end{equation}
Defining the conservative variables $\mathbf{U} = (\rho \quad \rho u \quad \rho e_t)^T$, fluxes $\mathbf{F} = (\rho u \quad \rho u^2 + p \quad (\rho e_t + p)u)^T$, primitive variables $\mathbf{W} = (\rho \quad  u \quad p)^T$, and thermodynamic coefficient variables $\boldsymbol{\Phi} = (\xi \quad E_0)^T$, the system can be written as:
\begin{equation}
\begin{dcases}
\frac{\partial \mathbf{U}}{\partial t} + \frac{\partial \mathbf{F}}{\partial x} =0 \\
\frac{\partial \boldsymbol{\Phi}}{\partial t} + u \frac{\partial \boldsymbol{\Phi}}{\partial x} = 0
\end{dcases}
\end{equation}

The computational procedure advances as follows:
\begin{itemize}
    \item \textbf{Initialization:} At time $t^0$, given primitive variables $\mathbf{W}^0$, compute internal energy and sound speed ($c^0, e^0$) via the EoS. Then, calculate conservative variables $\mathbf{U}^0$ and thermodynamic coefficients $\boldsymbol{\Phi}^0$ according to Eqs. (\ref{equ:xi}-\ref{equ:E0}).
    
    \item \textbf{Reconstruction and Evolution:} At time $t^n$, reconstruct (first-order or high-order) primitive variables $\mathbf{W}^n$. Combined with thermodynamic coefficients $\boldsymbol{\Phi}^n$, compute fluxes $\mathbf{F}^n$. Evolve via a first-order Godunov scheme to obtain the temporary conservative variables $\mathbf{U}^{n+1}_{temp}$. Concurrently, evolve the temporary thermodynamic coefficients $\boldsymbol{\Phi}^{n+1}_{temp}$ according to the advection equations.
    
    \item \textbf{Primitive Variable Recovery:} Based on the evolved conservative variables $\mathbf{U}^{n+1}_{temp}$ and thermodynamic coefficients $\boldsymbol{\Phi}^{n+1}_{temp}$, recover the primitive variables $\mathbf{W}^{n+1}$ at time $t^{n+1}$ according to Eq. (\ref{equ:liner2}).
    
   \item  \textbf{Thermodynamic Re-projection:} From the primitive variables $\mathbf{W}^{n+1}$, calculate the internal energy and sound speed ($c^{n+1}, e^{n+1}$) strictly via the EoS. Recalculate the conservative variables $\mathbf{U}^{n+1}$ and update the thermodynamic coefficients $\boldsymbol{\Phi}^{n+1}$ for the next time step according to Eqs. (\ref{equ:xi}-\ref{equ:E0}).
   
   \item \textbf{Time Stepping:} After projection, proceed to reconstruction and evolution for time $t^{n+1}$ until the target calculation time is reached.
\end{itemize}

It is particularly important to note that, distinct from classic quasi-conservative methods, the conservative variables $\mathbf{U}^{n+1}$ calculated strictly from $\mathbf{W}^{n+1}$ via the thermodynamic EoS are the actual quantities required at time $t^{n+1}$.  This step, termed \textbf{Thermodynamic Re-projection}, is the core of our algorithm and the origin of the energy conservation error. This is because at time $t^n$, the real fluid is locally frozen as a stiffened gas. At time $t^{n+1}$, we must update the thermodynamic coefficients $\boldsymbol{\Phi}^{n+1}$ based on the EoS. To obtain the oscillation-free pressure $p^{n+2}$ for the subsequent step, similar to the previous step,  strict thermodynamic consistency must be maintained among $\mathbf{W}^{n+1}$, $\mathbf{U}^{n+1}$, and $\boldsymbol{\Phi}^{n+1}$ at time $t^{n+1}$. Therefore, $\mathbf{U}^{n+1}$ must be recalculate via thermodynamic projection from $\mathbf{W}^{n+1}$, rather than naturally evolved via the Godunov method. The above algorithm is summarized in Algorithm \ref{alg:qcr_scheme}.

\begin{algorithm}[t]
\small 
\caption{Real-Fluid Quasi-Conservative Scheme}
\label{alg:qcr_scheme}
\begin{algorithmic}[1]
\Require Initial $\mathbf{W}^0$, $T_{end}$, CFL.
\textbf{Ensure} $\mathbf{W}$ at $T_{end}$.
\State \textbf{Init:} Compute $\mathbf{U}^0, \boldsymbol{\Phi}^0$ from $\mathbf{W}^0$ using Eqs. (\ref{equ:xi}-\ref{equ:E0}) and Real fluid EoS.
\While{$t < T_{end}$}
    \State Calculate $\Delta t$ based on CFL condition.

    \State \textbf{1. Reconstruction and Evolution}: \Comment{Solve Euler equations with frozen $\boldsymbol{\Phi}^n$}
    \State Compute fluxes $\mathbf{F}_{i+1/2}$ using Riemann solver with $(\mathbf{W}^n, \boldsymbol{\Phi}^n)$.
\State Update: $\mathbf{U}^{n+1}_{temp} \gets \mathbf{U}^n - \frac{\Delta t}{\Delta x}\Delta \mathbf{F}$, \quad $\boldsymbol{\Phi}^{n+1}_{temp} \gets \boldsymbol{\Phi}^n - \frac{\Delta t}{\Delta x}\Delta \mathbf{F}_{\Phi}$.
\State \textbf{2. Primitive Variable Recovery}:
    \State $\mathbf{W}^{n+1} \gets \text{Decode}(\mathbf{U}^{n+1}_{temp}, \boldsymbol{\Phi}^{n+1}_{temp})$ \Comment{Obtain oscillation-free $p^{n+1}$}

    \State \textbf{3. Thermodynamic Re-projection}: \Comment{Enforce real-fluid consistency}
    \State $(e, c)_{real} \gets \text{Real fluid EoS}(\rho^{n+1}, p^{n+1})$
    \State $(\rho e_t)^{n+1} \gets \rho^{n+1} e_{real} + \frac{1}{2}\rho^{n+1} (u^{n+1})^2$ \Comment{Energy re-projection}
    \State $\mathbf{U}^{n+1} \gets (\rho, \rho u, \rho e_t)^{n+1}$, \quad $\boldsymbol{\Phi}^{n+1} \gets \text{Update}(\mathbf{W}^{n+1}, c_{real}, e_{real})$

    \State $t \gets t + \Delta t$
\EndWhile
\end{algorithmic}
\end{algorithm}

\subsection{Re-Projection Error Analysis}
In the thermodynamic re-projection step, only the total energy is modified. We define the resulting energy conservation error, $\epsilon_p = (\rho e_t)^{n+1} - (\rho e_t)^{n+1}_{temp}$, as the re-projection error. We now analyze the convergence properties of this error.

The energy $(\rho e_t)^{n+1}_{temp}$ obtained from the natural evolution of the conservation laws satisfies:
\begin{equation}
   (\rho e_t)^{n+1}_{temp} = (\rho e_t)^n - \frac{\Delta t}{\Delta x} \left( \mathbf{F}_{S, j+1/2} - \mathbf{F}_{S, j-1/2} \right) 
\end{equation}
where $\mathbf{F}_{S}$ denotes the stiffened gas flux computed via the frozen parameters $(\xi, E_{0})$. Expressing this energy in terms of the recovered oscillation-free pressure $p^{n+1}$ and the stiffened gas EoS relation yields:
\begin{equation}
    (\rho e_t)^{n+1}_{temp} = \frac{1}{2}\rho^{n+1} (u^{n+1})^2 + \left( p^{n+1}(\xi)^{n+1}_{temp} + (E_0)^{n+1}_{temp} \right)
\end{equation}

In contrast, the energy $(\rho e_t)^{n+1}$ recalculated strictly from thermodynamic relations satisfies:
\begin{equation}
    (\rho e_t)^{n+1} = \frac{1}{2}\rho^{n+1} (u^{n+1})^2 + \rho e_{real}(\rho^{n+1}, p^{n+1}) 
\end{equation}
The error between the two is:
\begin{equation}
    \epsilon_p = \rho^{n+1} e_{real}(\rho^{n+1}, p^{n+1}) - \left( {p^{n+1} (\xi)^{n+1}_{temp} + (E_0)^{n+1}_{temp}} \right).
    \label{equ:error1}
\end{equation}

To provide a discrete analysis of the error, let $\psi(p,s)$ represent the internal energy $(\rho e)_{real}$ as a function of $(p,s)$:
\begin{equation} 
\psi(p,s)= (\rho e)_{real}(p,s).
\end{equation}

A Taylor expansion of $\psi$ at the state $(p^n, s^n)$ yields:
\begin{equation}
 \psi^{n+1} = \psi^n + \psi_p^n \Delta p + \psi_s^n \Delta s + \frac{1}{2}\psi_{pp}^n(\Delta p)^2 + \psi_{ps}^n \Delta p\,\Delta s + \frac{1}{2}\psi_{ss}^n(\Delta s)^2 + O(\|\Delta\|^3),
    \label{equ:psi_taylor}
\end{equation}
where the subscripts $p$ and $s$ denote partial derivatives, with $\Delta p=p^{n+1}-p^n$ and $\Delta s=s^{n+1}-s^n$.

As previously noted, $\xi$ in the real fluid thermodynamic relation is not strictly equal to the Grüneisen parameter. From the isentropic sound speed $c^2 = (\partial p /\partial \rho)_s$ and the thermodynamic relation $(\partial (\rho e) /\partial \rho)_s=h$, we have:
\begin{equation}
    \left(\frac{\partial(\rho e)}{\partial p}\right)_s = \frac{(\partial (\rho e) /\partial \rho)_s}{(\partial p /\partial \rho)_s} = \frac{h}{c^2} = \xi = \psi_p.
    \label{equ:xi_slope}
\end{equation}
This indicates that the linearized internal energy-pressure relation $\rho e= p\xi+E_0$ matches both the function value and the first-order pressure derivative along the isentrope, ensuring the correct wave speeds and structures in the Riemann solver.

Considering the advection relation and neglecting discretization higher-order errors, the parameters along a streamline satisfy:
\begin{equation}
    (\xi)^{n+1}_{temp} = \xi^{n},\quad (E_0)^{n+1}_{temp} = E_0^{n}.
\end{equation}
Consequently, the second term in Eq. \ref{equ:error1} becomes:
\begin{equation}
\begin{aligned}
    p^{n+1} (\xi)^{n+1}_{temp} + (E_0)^{n+1}_{temp} &= p^{n+1}\xi^n + E_0^n \\
    &= (p^n\xi^n+E_0^n) + \xi^n \Delta p \\
    &=\psi^n+\psi_p^n \Delta p.
    \label{equ:psi_frozen}
\end{aligned}
\end{equation}

Substituting Eqs. \ref{equ:psi_taylor} and \ref{equ:psi_frozen} into the error definition (Eq. \ref{equ:error1}) yields the local error expansion:
\begin{equation}
    \epsilon_p = \left( \psi_s^n \Delta s + \frac{1}{2}\psi_{pp}^n(\Delta p)^2 + \psi_{ps}^n \Delta p\,\Delta s + \frac{1}{2}\psi_{ss}^n(\Delta s)^2 \right) + O(\|\Delta\|^3).
    \label{equ:error2}
\end{equation}

Eq. \ref{equ:error2} demonstrates that the re-projection error stems primarily from the first-order entropy term $\psi_s^n\Delta s$ due to non-isentropic deviations and the second-order pressure term $\psi_{pp}^n(\Delta p)^2$ caused by the nonlinear curvature of the EoS. The error characteristics in different regions are summarized as follows:

\begin{itemize}
    \item \textbf{Isentropic regions}: Analytically, $\Delta s=0$, and the error is dominated by the EoS curvature:
    \begin{equation}
        \epsilon_p = \frac{1}{2}\psi_{pp}^n(\Delta p)^2 = \frac{1}{2}\left(\frac{1}{\rho c^2}-\frac{\xi}{c^2}\left(\frac{\partial c^2}{\partial p}\right)_s\right)^n(\Delta p)^2.
    \end{equation}
    In smooth flows, $\Delta p=O(\Delta t)$, resulting in a local error of $O(\Delta t^2) \sim O(\Delta x^2)$. The accumulated global error satisfies $\sum\epsilon_p=O(\Delta t)$, indicating first-order global conservation behavior.

    \item \textbf{Non-isentropic regions}: In the presence of shocks or numerical dissipation, the error is dominated by the first-order entropy change:
   \begin{equation}
        \epsilon_p = \psi_s^n \Delta s = \left(\rho T-\xi \left(\frac{\partial p}{\partial s}\right)_{\rho} \right)^n \Delta s.
    \end{equation}
    Here, the thermodynamic relations $\mathrm{d}(\rho e)=h\,\mathrm{d}\rho+\rho T\,\mathrm{d}s$ and $\mathrm{d}p=c^2\mathrm{d}\rho+(\partial p/\partial s)_\rho \mathrm{d}s$ are utilized. In shock-capturing schemes, the entropy change is controlled by numerical viscosity; thus, the integrated error remains of the same order as the inherent truncation error of the shock-capturing scheme.
\end{itemize}

In summary, the thermodynamic conservation error of the proposed RFQC scheme is locally second-order and globally first-order dominated by the time step in smooth isentropic regions. In discontinuous regions, the error growth rate is proportional to the entropy change $\Delta s$, which is consistent with the numerical dissipation required for shock capturing. This analysis indicates that, unlike the DF methods, the present approach avoids spatial energy conservation errors arising from flux inconsistency. All thermodynamic inconsistencies are isolated at the end of each time step within the re-projection step. 
Furthermore, the RFQC method strictly follows conservation laws to evolve conserved variables within each time step, which fundamentally distinguishes it from non-conservative methods that directly evolve pressure. The superior conservation properties of the proposed method will be further demonstrated in the numerical examples in the next section.

\section{Numerical Verification}
In this section, we present a comprehensive numerical verification of the proposed RFQC scheme with n-dodecane as the test fluid. First, we quantitatively assess the energy conservation error and convergence properties through a smooth density wave advection case where the fluid state crosses the saturation line, involving vapor-liquid mixtures. Subsequently, we examine Riemann problems for n-dodecane—including transcritical, flash evaporation, and subcritical impinging cases—to validate the accuracy of the proposed method against the Pressure-Based (PB) and Double Flux (DF) methods . The exact solutions for these Riemann problems are calculated via the algorithms from our previous works \cite{Bai2025,Bai2026}. 

The Peng-Robinson (P-R) EoS is adopted throughout this section. Internal energy is computed via analytical integration of the P-R EoS, as detailed in \ref{app:PR_EOS}. Thermodynamic parameters for n-dodecane are provided in \ref{app:dodecane}. Fugacity equilibrium and saturation properties are calculated following the procedures in \cite{Bai2026}. For the numerical implementation, all methods employ the HLLC Riemann solver \cite{Toro2009} with first-order discretization in both time and space. The detailed discretization method is provided in \ref{app:Discrete}

\subsection{Convergence Test of Energy Conservation Error}
\label{sec:convergencetest}
To evaluate the convergence characteristics of the energy conservation error (re-projection error) in the RFQC scheme, we simulate a density wave advection case near the saturation liquid line of n-dodecane. The flow is initialized with a uniform pressure $p_0 = 1.0 \times 10^6 \text{ Pa}$ and a uniform velocity $u = 100 \text{ m/s}$. At this pressure, the saturation liquid density of n-dodecane is $\rho_{\text{sat},l} \approx 389 \text{ kg/m}^3$. The initial density profile is given by a large-amplitude sinusoidal perturbation:
\begin{equation}
\rho(x) = 400 + 100 \sin\left(\frac{2\pi x}{L}\right)
\end{equation}
where the domain length $L=0.5 \text{ m}$, and the simulation time $t=0.001 \text{ s}$.

A distinct feature of this case is that the fluid state crosses the saturation line, transitioning between the liquid phase and the two-phase mixture region. Specifically, the fluid at the density peak remains in the liquid state, whereas at the trough, the density drops significantly below the saturation value, corresponding to a two-phase mixture state. Despite the smooth flow field, the sound speed and other thermodynamic derivatives exhibit strong nonlinear spatial variations due to the first-order phase transition. Under such conditions, standard conservative methods typically lead to spurious pressure oscillations.

\begin{figure}[ht]
\centering
\begin{subfigure}[b]{0.4\textwidth}
    \includegraphics[width=\textwidth]{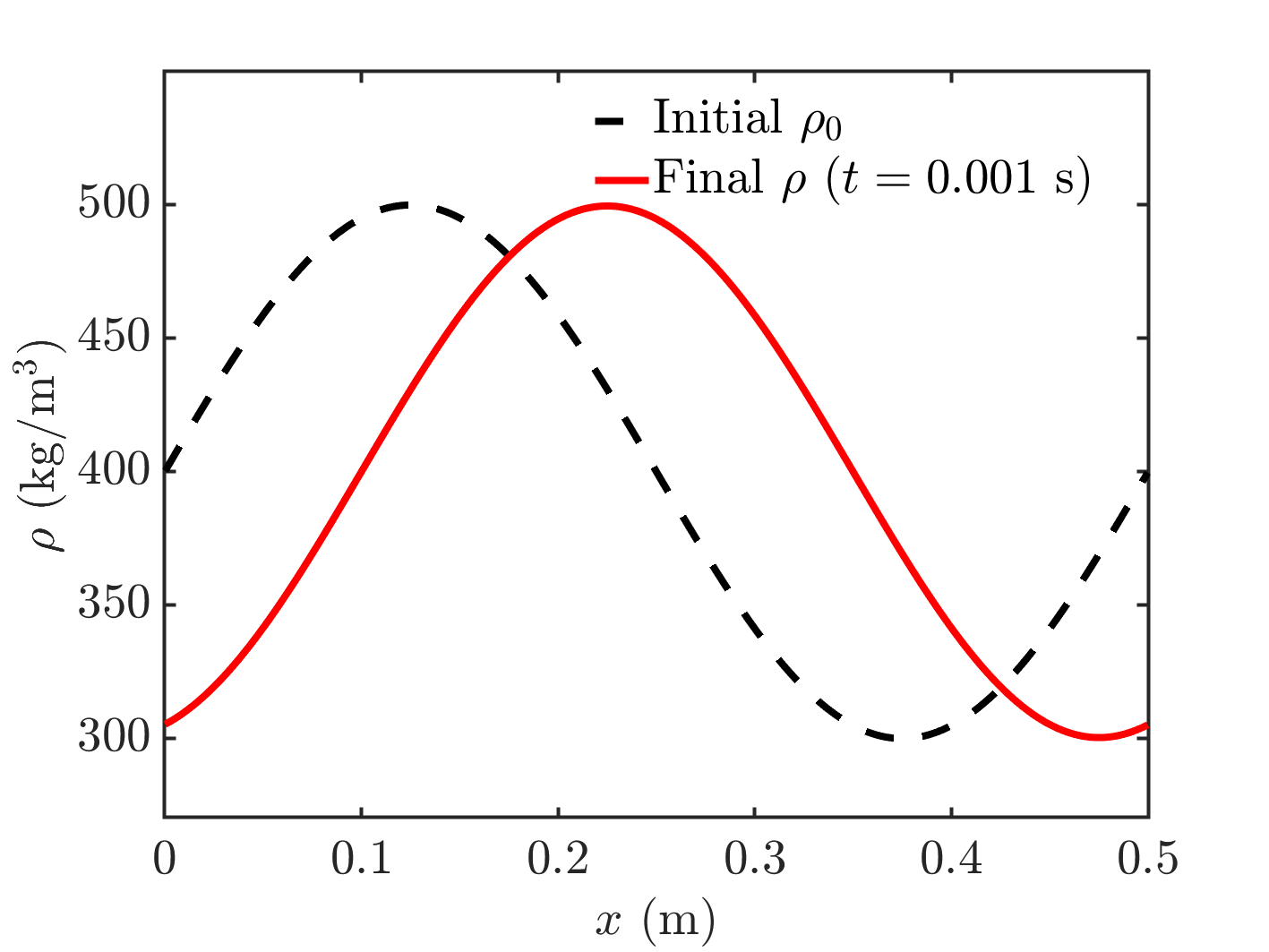}
    \caption{Density ($N=1024$)}
    \label{fig3a}
\end{subfigure}
%\hfill
\hspace{0.05\textwidth}
\begin{subfigure}[b]{0.4\textwidth}
    \includegraphics[width=\textwidth]{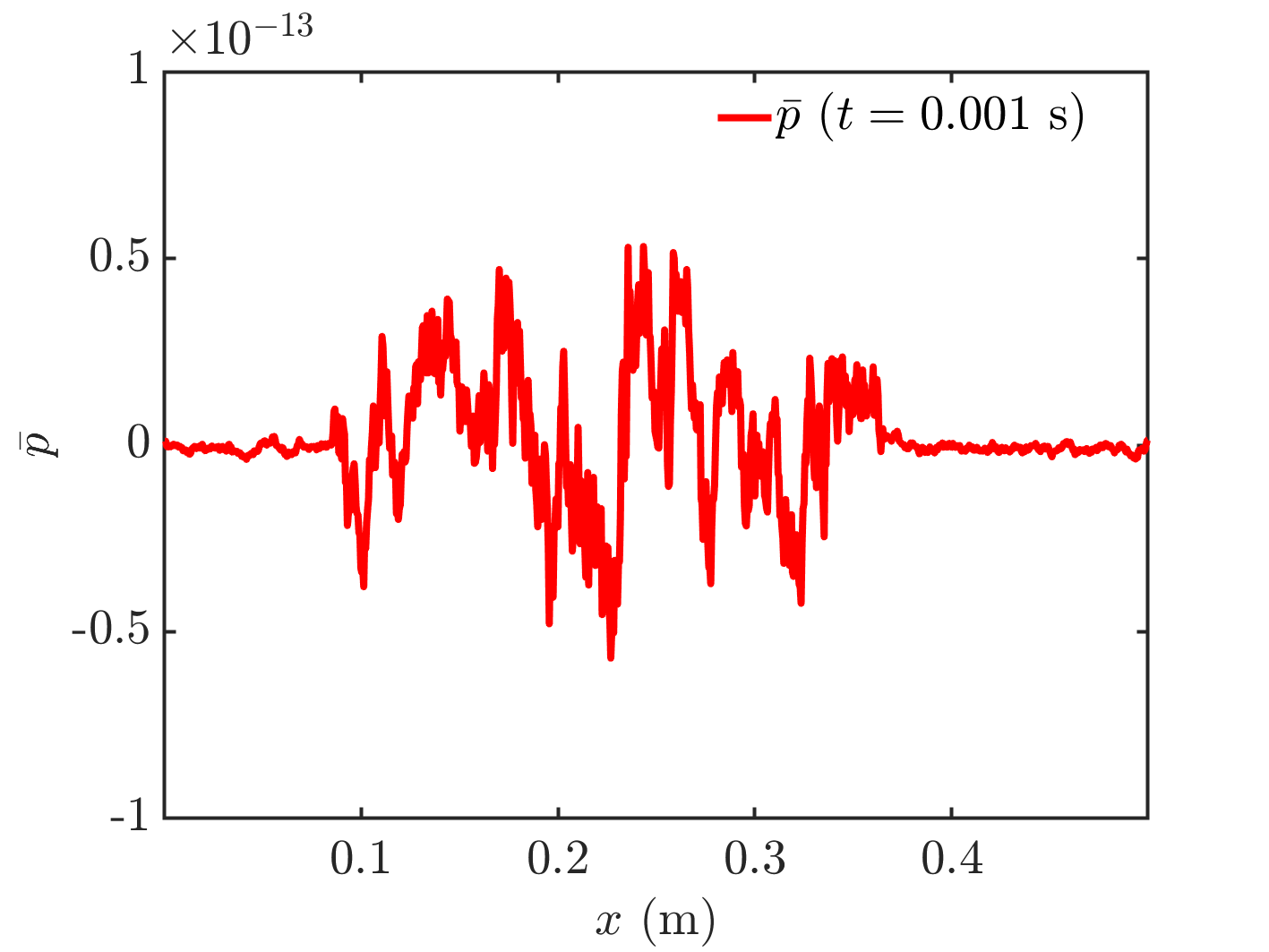}
    \caption{Pressure ($N=1024$)}
    \label{fig3b}
\end{subfigure}

\begin{subfigure}[b]{0.4\textwidth}
    \includegraphics[width=\textwidth]{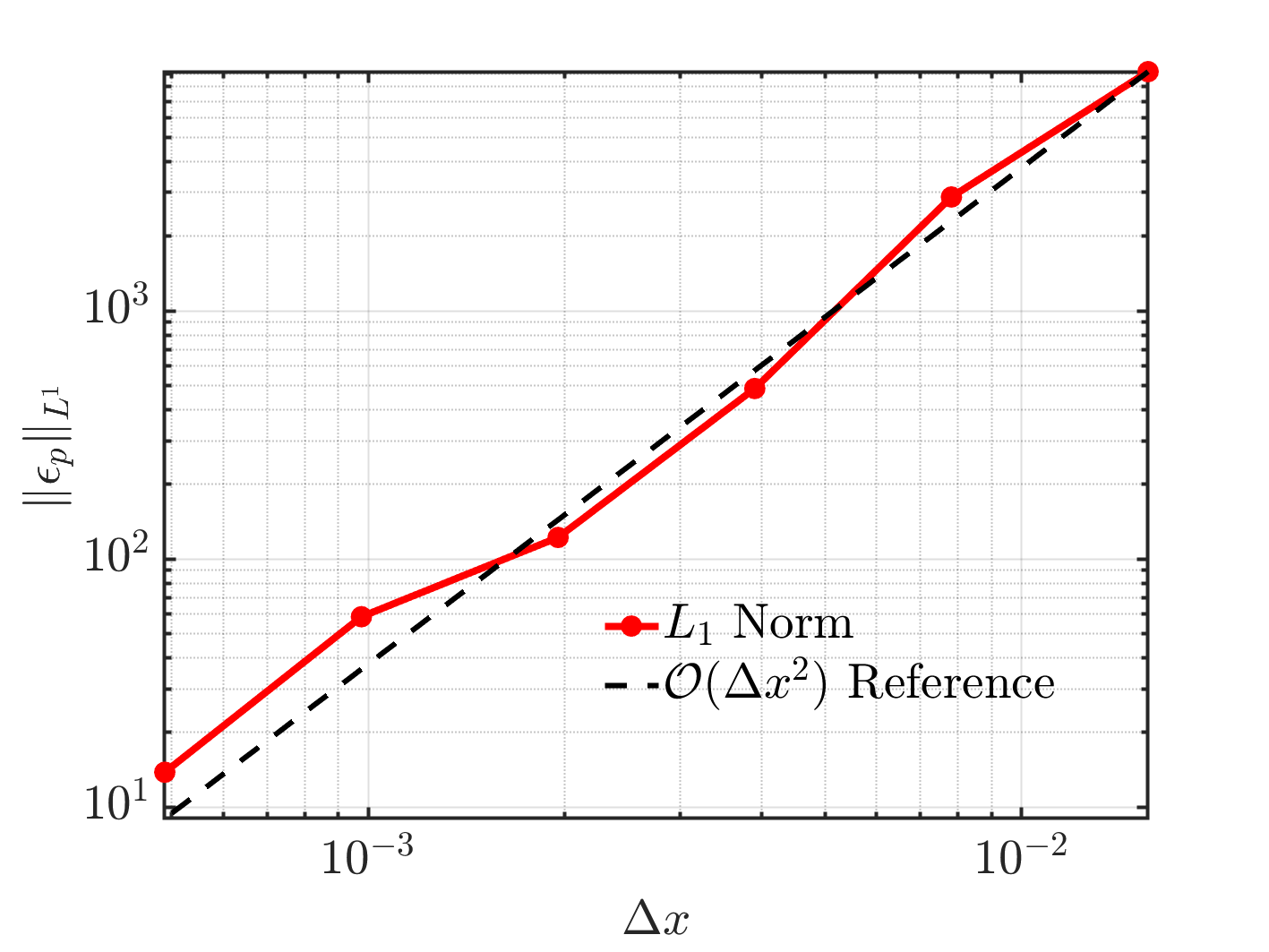}
    \caption{The L1 norm of the energy conservation error}
    \label{fig3c}
\end{subfigure}
\hspace{0.05\textwidth} 
\begin{subfigure}[b]{0.4\textwidth}
    \includegraphics[width=\textwidth]{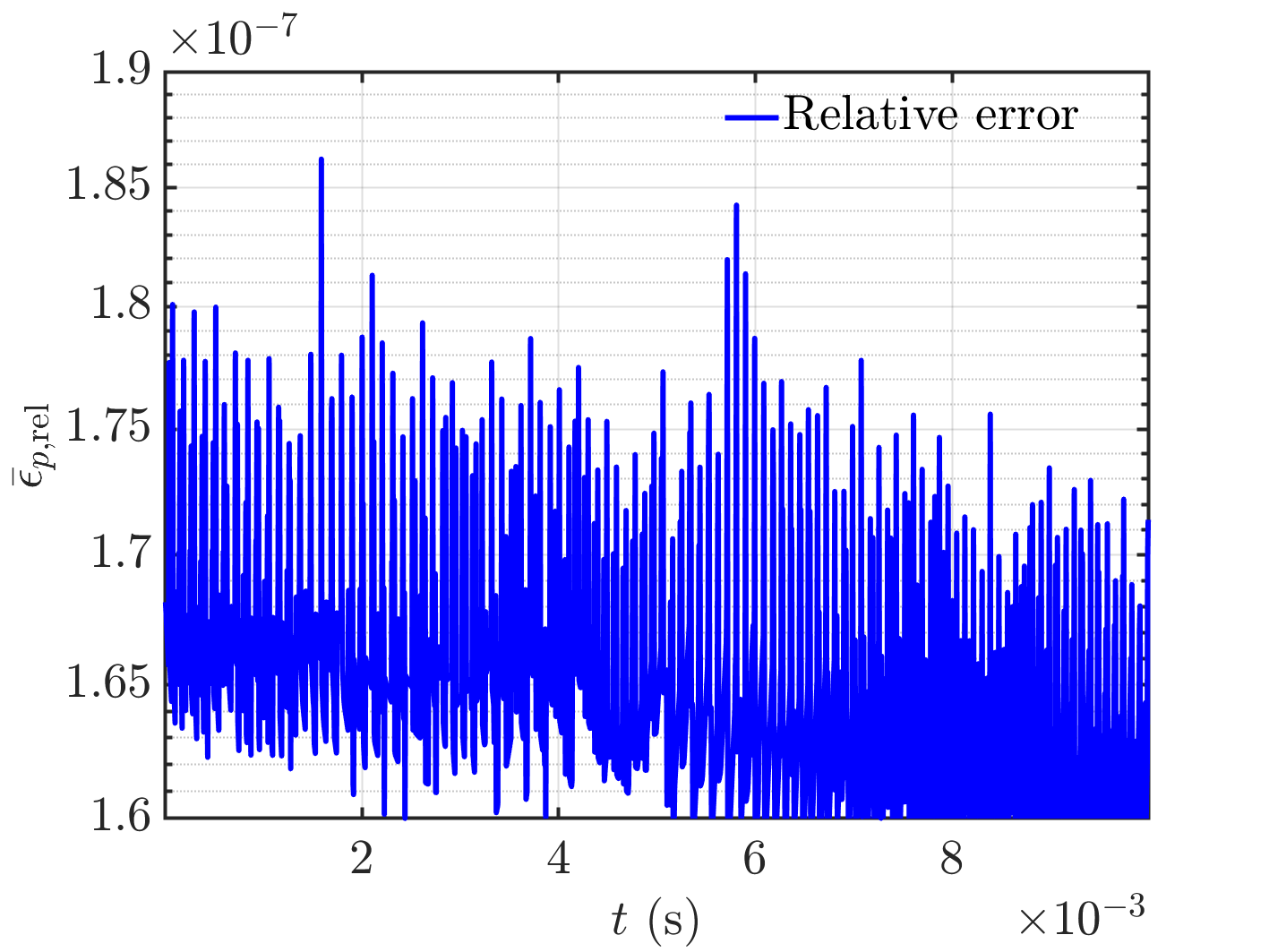}
    \caption{Relative energy conservation error ($N=1024$)}
    \label{fig3d}
\end{subfigure}

\caption{Verification of Energy Conservation Error in Smooth Advection.}
\label{fig3}
\end{figure}

With a CFL number of 0.25, we evaluated the re-projection energy conservation error on grids with $N=32, 64, 128, 256, 512, 1024$, for which the time step scales linearly with the grid spacing. The results are shown in Fig. \ref{fig3}. The profiles of density and pressure for $N=1024$ are displayed in Figs. \ref{fig3a} and \ref{fig3b}. After $t=0.001 \text{s}$, the wave has propagated by 0.1 m. The normalized pressure $\bar p = (p-p_0)/p_0$ is consistently maintained near machine precision, demonstrating that the RFQC method effectively suppresses pressure oscillations across the saturation line where thermodynamic properties vary significantly. Fig.\ref{fig3c} shows the convergence of the $L_1$ norm of the re-projection energy conservation error $\epsilon_p$ with respect to the grid size. The $L_1$ error is defined as:
\begin{equation}
\|\epsilon_p\|_{L_1}
=\sum_i \left| \epsilon_{p,i} \right|\Delta x .
\end{equation}
The error exhibits a clear second-order convergence trend, consistent with the theoretical analysis. Furthermore, the relative re-projection error $\bar{\epsilon}_{p,\mathrm{rel}}$ is defined as:
\begin{equation}
\bar{\epsilon}_{p,\mathrm{rel}}
=\frac{1}{L}
\left[\sum_i\left|\frac{\epsilon_{p,i}^{n}}{(\rho e_t)_{i}^{n}} \right|\Delta x\right].
\end{equation}
As shown in Fig.~\ref{fig3d}, for $N=1024$, $\overline{\epsilon}_{p}$ oscillates slightly around $1.7\times10^{-7}$ during the $0.01\text{s}$ simulation time, indicating that this error is extremely small relative to the internal energy and does not increase over time.

The convergence test results are summarized in Table \ref{tab:advection}. The data indicate that the RFQC method effectively suppresses pressure oscillations under conditions of sharp property variations, achieving high accuracy in energy conservation. Here,  $\bar{\epsilon}_p^\text{ave}$ represents the time-averaged value of the sum of relative energy errors over the simulation period, which also exhibits second-order convergence with grid refinement.
\begin{equation}
\bar{\epsilon}_p^\mathrm{ave}
=\frac{1}{T}\sum_{n}
\bar{\epsilon}_{p,\mathrm{rel}}^\mathrm{n}\Delta t_n =\frac{1}{T L}\sum_{n}
\left[\sum_i\left|\frac{\epsilon_{p,i}^{n}}{(\rho e_t)_{i,\mathrm{proj}}^{n}} \right|\Delta x\right]\Delta t_n .
\end{equation}

\begin{table}[H]
    \centering
    \caption{Verification of Energy Conservation Error in Smooth Advection}
    \label{tab:advection}
    \begin{tabular}{cccccc}
        \toprule
       N & $\mathrm{d}x$ (m) & $L_1$ error($\epsilon_p$)  & Order & Relative error($\bar{\epsilon}_p^\text{ave}$)  & Order\\
        \midrule
        32 &  $1.5625\times 10^{-2}$ &  $9.2120 \times 10^3$ & N/A&  $1.3879 \times 10^{-4}$ & N/A  \\
        64 &  $7.8125 \times 10^{-3}$ & $2.8787 \times 10^3$ & 1.68& $3.5786 \times 10^{-5}$ & 1.96  \\
        128 &  $3.9063 \times 10^{-3}$ & $4.8683 \times 10^2$ & 2.56& $9.1819 \times 10^{-6}$ & 1.96  \\
        256 &  $1.9531 \times 10^{-3}$ & $1.2221 \times 10^2$ & 1.99& $2.3436 \times 10^{-6}$ & 1.97   \\
        512 &  $9.7656 \times 10^{-4}$ & $5.8541 \times 10^1$ & 1.06& $6.1378 \times 10^{-7}$ &  1.93  \\
        1024 &  $4.8828 \times 10^{-4}$ & $1.3829 \times 10^1$ &2.08& $1.6710 \times 10^{-7}$ &  1.88  \\
        \bottomrule
    \end{tabular}
\end{table}

\subsection{Transcritical Riemann Problem Verification}
Transcritical flow is characterized by a continuous transition where the fluid state evolves from a liquid-like to a gas-like state without traversing the two-phase region.  During this process, the fluid crosses the pseudo-boiling line, leading to sharp variations in thermodynamic properties. Throughout the transition, the fluid remains in a stable single phase, avoiding any first-order phase transition \cite{Bai2025}.

In scramjet engines operating at high Mach numbers, fuel absorbs heat in cooling channels to reach a supercritical state, subsequently undergoing a transcritical transition upon injection into the lower-pressure combustor. Based on this physical background, the initial conditions for the transcritical Riemann problem are selected as shown in Table \ref{tab:case1}. The computational domain length $L=1\text{m}$, with a simulation time of $t=0.8 \text{ms}$, a grid size of $N=500$, and a CFL number of 0.1.

\begin{table}[H]
    \centering
    \caption{Initial Values for the Transcritical RP Calculation}
    \label{tab:case1}
    \begin{tabular}{cccccc}
        \toprule
       State & $P$ (Pa) & $\rho$ ($\mathrm{kg \,m^{-3}}$) & $T$ (K) & $u$ ($\mathrm{m \, s^{-1}}$) \\
        \midrule
        Left&  $2 \times 10^6$ & 200 & 665.1& 80  \\
        Right&  $1 \times 10^5$ & 2 &1026.8 & 0  \\
        \bottomrule
    \end{tabular}
\end{table}

\begin{figure}[h]
\centering
\begin{subfigure}[b]{0.4\textwidth}
    \includegraphics[width=\textwidth]{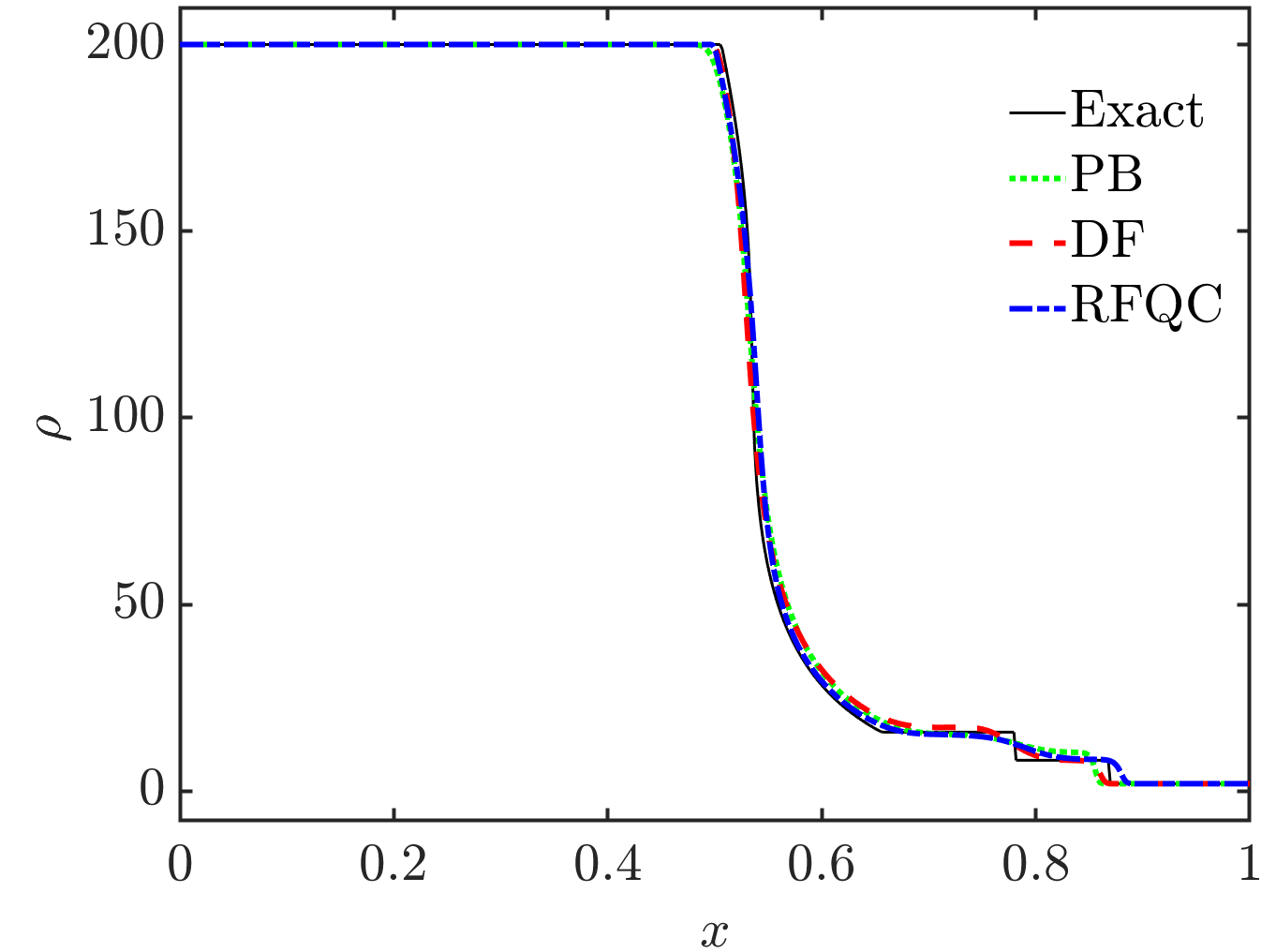}
    \caption{Density}
    \label{fig4a}
\end{subfigure}
%\hfill 
\hspace{0.05\textwidth} % 
\begin{subfigure}[b]{0.4\textwidth}
    \includegraphics[width=\textwidth]{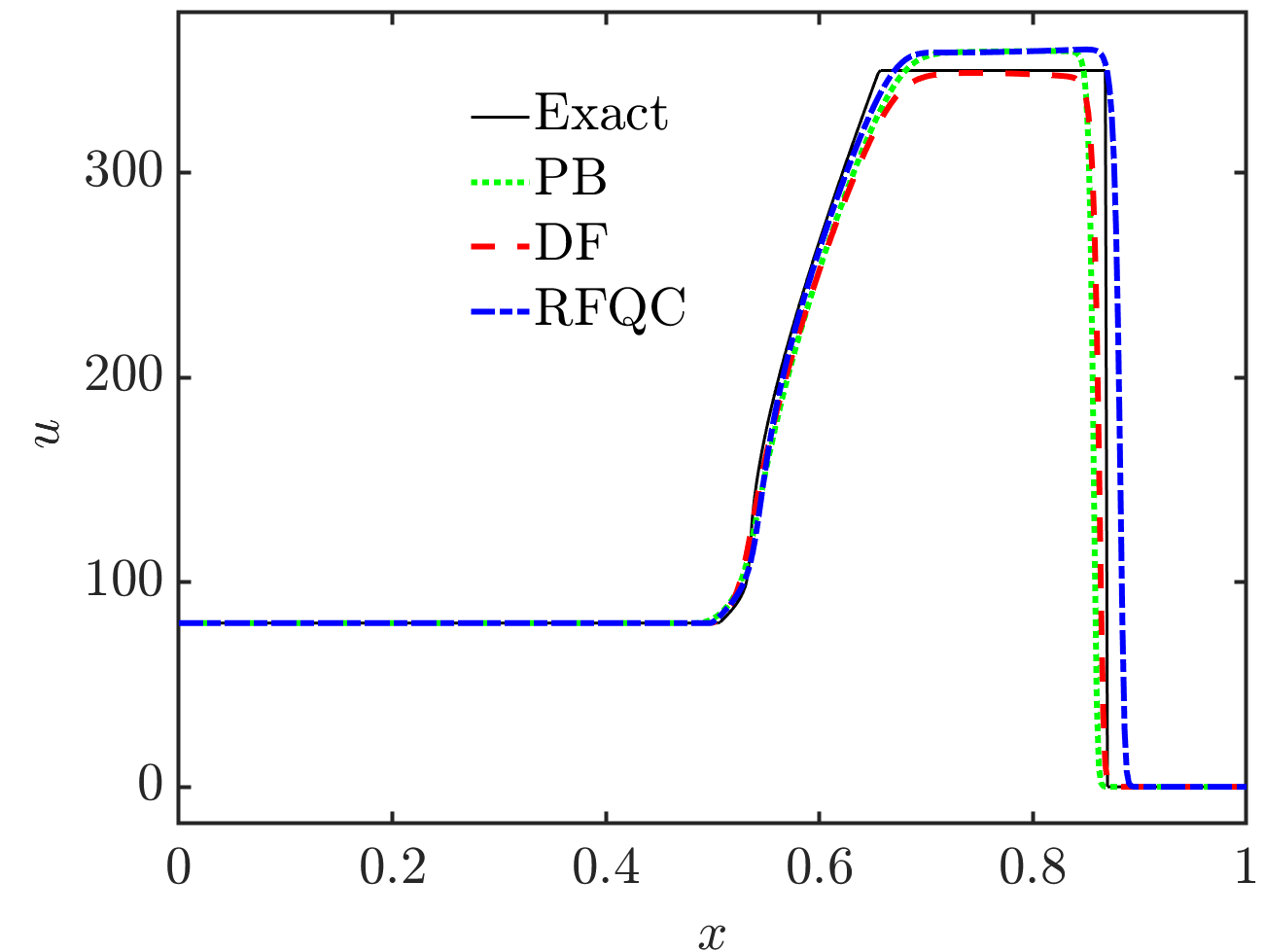}
    \caption{Velocity}
    \label{fig4b}
\end{subfigure}

\begin{subfigure}[b]{0.4\textwidth}
    \includegraphics[width=\textwidth]{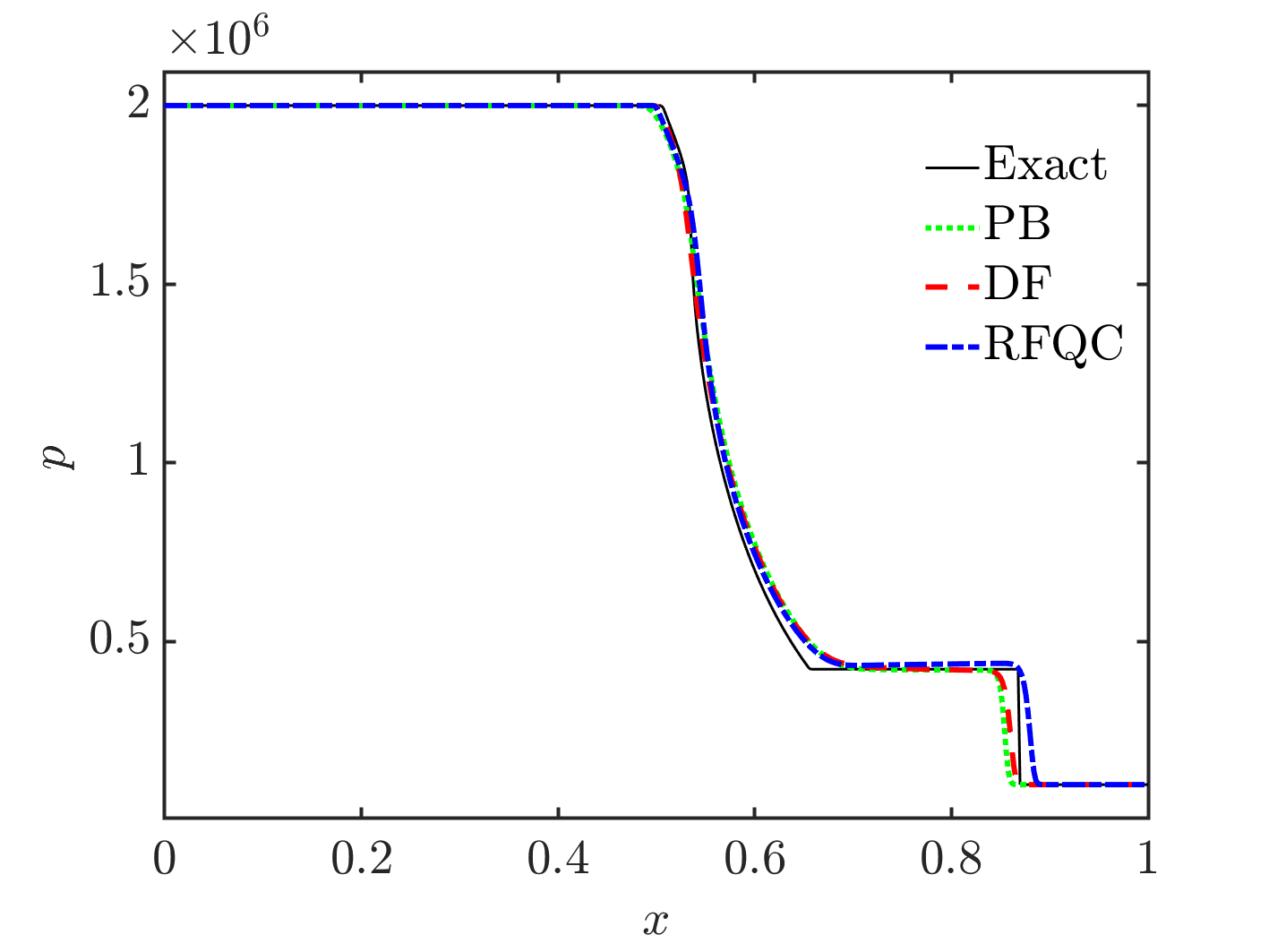}
    \caption{Pressure}
    \label{fig4c}
\end{subfigure}
\hspace{0.05\textwidth} 
\begin{subfigure}[b]{0.4\textwidth}
    \includegraphics[width=\textwidth]{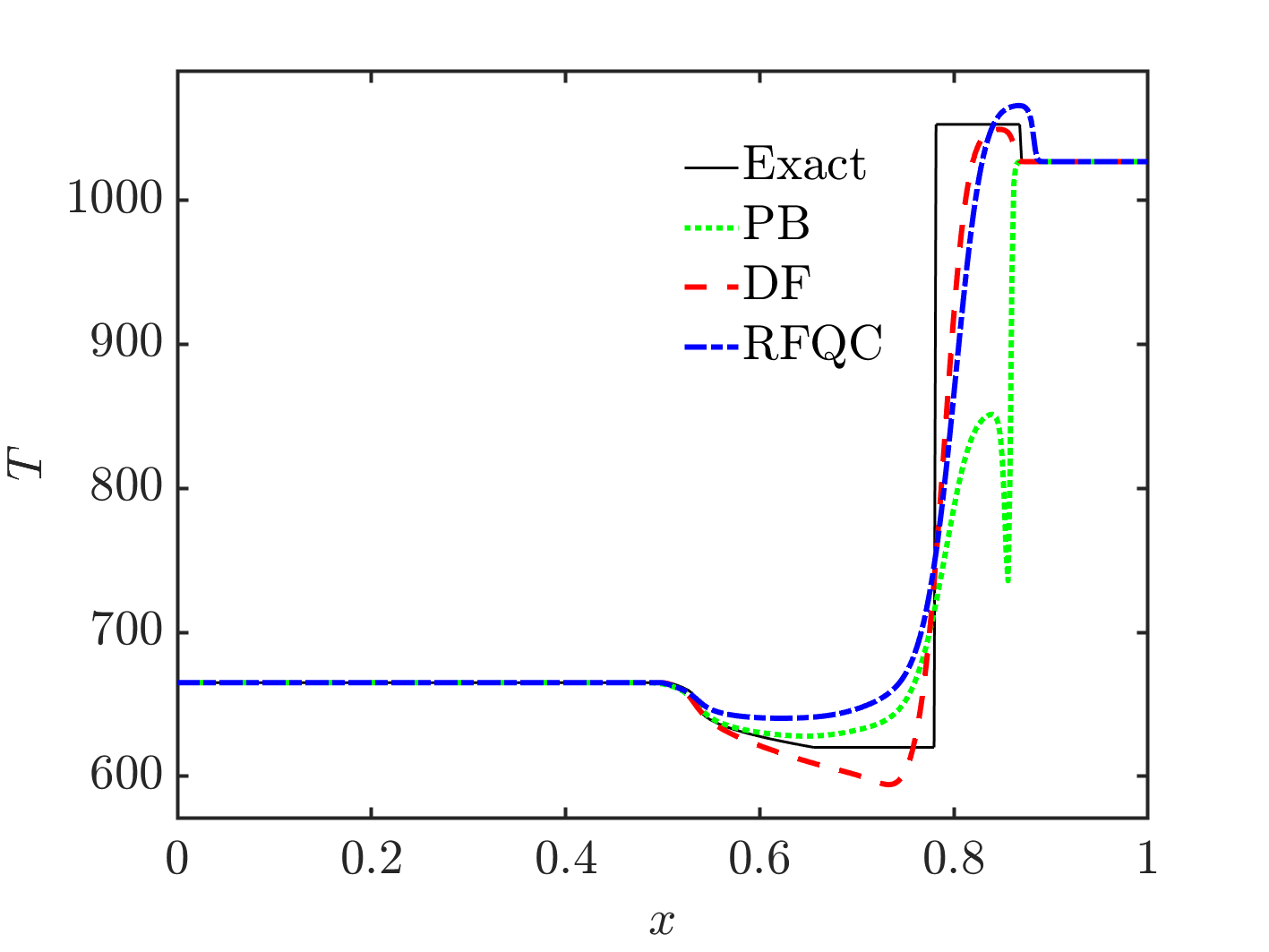}
    \caption{Temperature}
    \label{fig4d}
\end{subfigure}
\caption{Results of the Transcritical RP Calculation.}
\label{fig4}
\end{figure}

Fig. \ref{fig4} compares the results of the three methods against the exact solution for the transcritical Riemann problem.  Except for the PB method, which exhibits significant temperature deviations near the shock wave (Fig. \ref{fig4d}), both the DF and RFQC methods show good agreement with the exact solution. Specifically, the RFQC method slightly overestimates the temperature within the rarefaction wave, while the DF method underestimates it. This discrepancy is attributed to conservation errors induced by the transcritical property variations across the rarefaction wave.

\subsection{Flash Evaporation Riemann Problem Verification}
\label{sec:FeRP}
"Flash evaporation" is a rapid phase transition driven by expansion, commonly encountered in fuel injection systems of high-pressure internal combustion engines and scramjets. In our previous work \cite{Bai2026}, we formalized the flash evaporation process as a Riemann problem where the expansion branch of the solution curve crosses the saturation line, within the framework of  the Homogeneous Equilibrium Model (HEM) and Vapor-Liquid Equilibrium (VLE) assumptions. We also established an exact solution framework based on Newton iteration. In the flash evaporation Riemann problem (FeRP), the fluid crossing the saturation line induces a discontinuity in the speed of sound, which in turn leads to rarefaction wave splitting and the formation of non-classical expansion shocks \cite{MenikoffPlohr1989}. The thermodynamic properties of the fluid change sharply during this process, posing a significant challenge to the stability and accuracy of numerical methods. Based on this physical background, the computational domain is set to $L=1\text{m}$, with a simulation time of $t=0.8 \text{ms}$, a grid size of $N=500$, and a CFL number of 0.1.

In scramjet engines, fuel absorbs heat in the cooling channels and is injected into the relatively low-pressure combustor, where flash evaporation often occurs due to expansion. When the initial temperature is high, the fluid will be completely evaporated. In this scenario, the expansion branch of the Riemann solution crosses the saturation line twice, generating split rarefaction waves and non-classical expansion shocks. The initial conditions corresponding to this physical process are listed in Table \ref{tab:case2}.

\begin{table}[H]
    \centering
    \caption{Initial Values for the Completely Evaporated FeRP Calculation}
    \label{tab:case2}
    \begin{tabular}{cccccc}
        \toprule
       State & $P$ (Pa) & $\rho$ ($\mathrm{kg \,m^{-3}}$) & $T$ (K) & $u$ ($\mathrm{m \, s^{-1}}$) \\
        \midrule
        Left&  $2 \times 10^6$ & 300 & 652.4& 60  \\
        Right&  $1 \times 10^5$ & 2 & 1026.8 & 0  \\
        \bottomrule
    \end{tabular}
\end{table}

\begin{figure}[h]
\centering
\begin{subfigure}[b]{0.4\textwidth}
    \includegraphics[width=\textwidth]{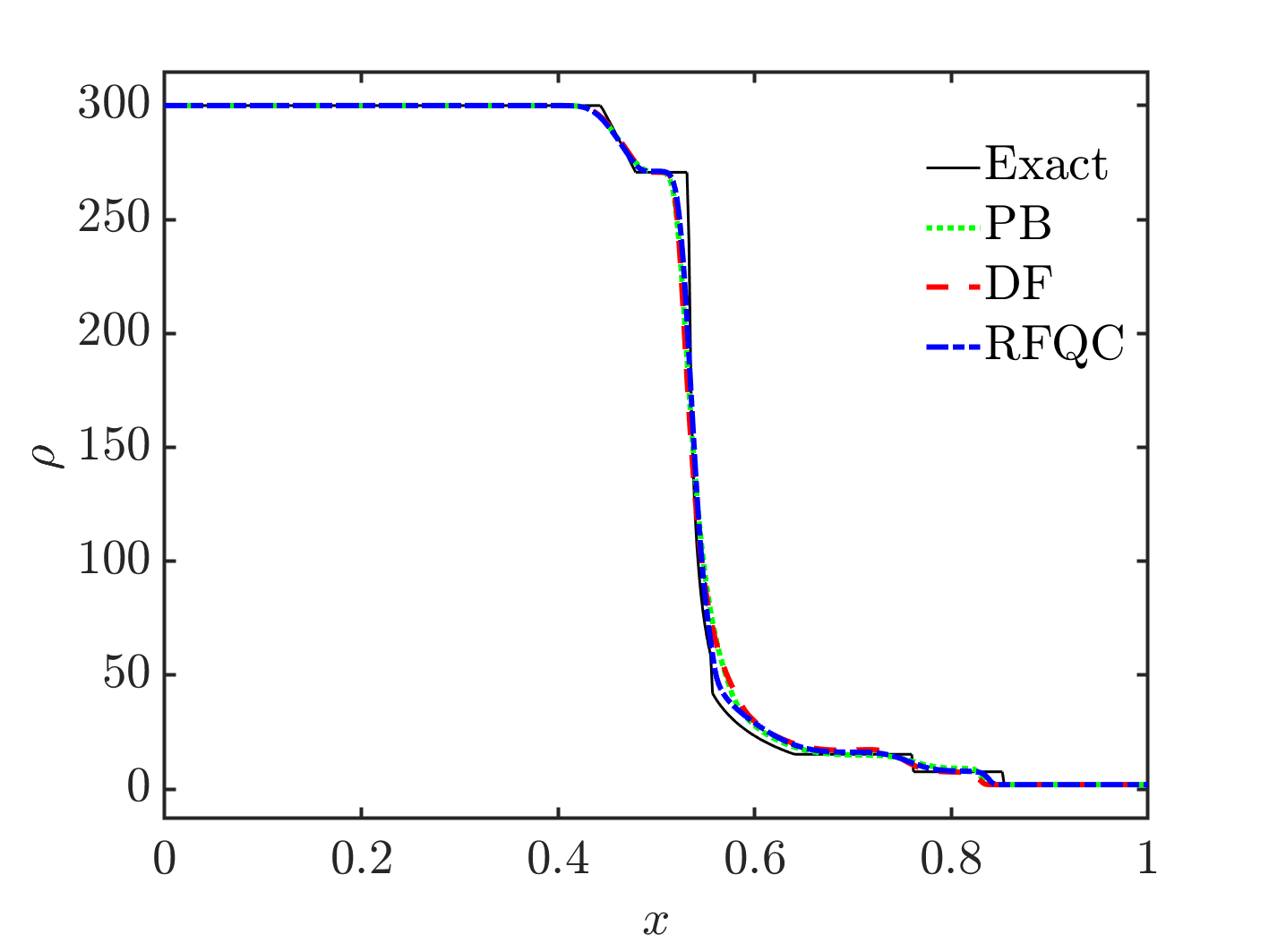}
    \caption{Density}
    \label{fig5a}
\end{subfigure}
%\hfill 
\hspace{0.05\textwidth} % 
\begin{subfigure}[b]{0.4\textwidth}
    \includegraphics[width=\textwidth]{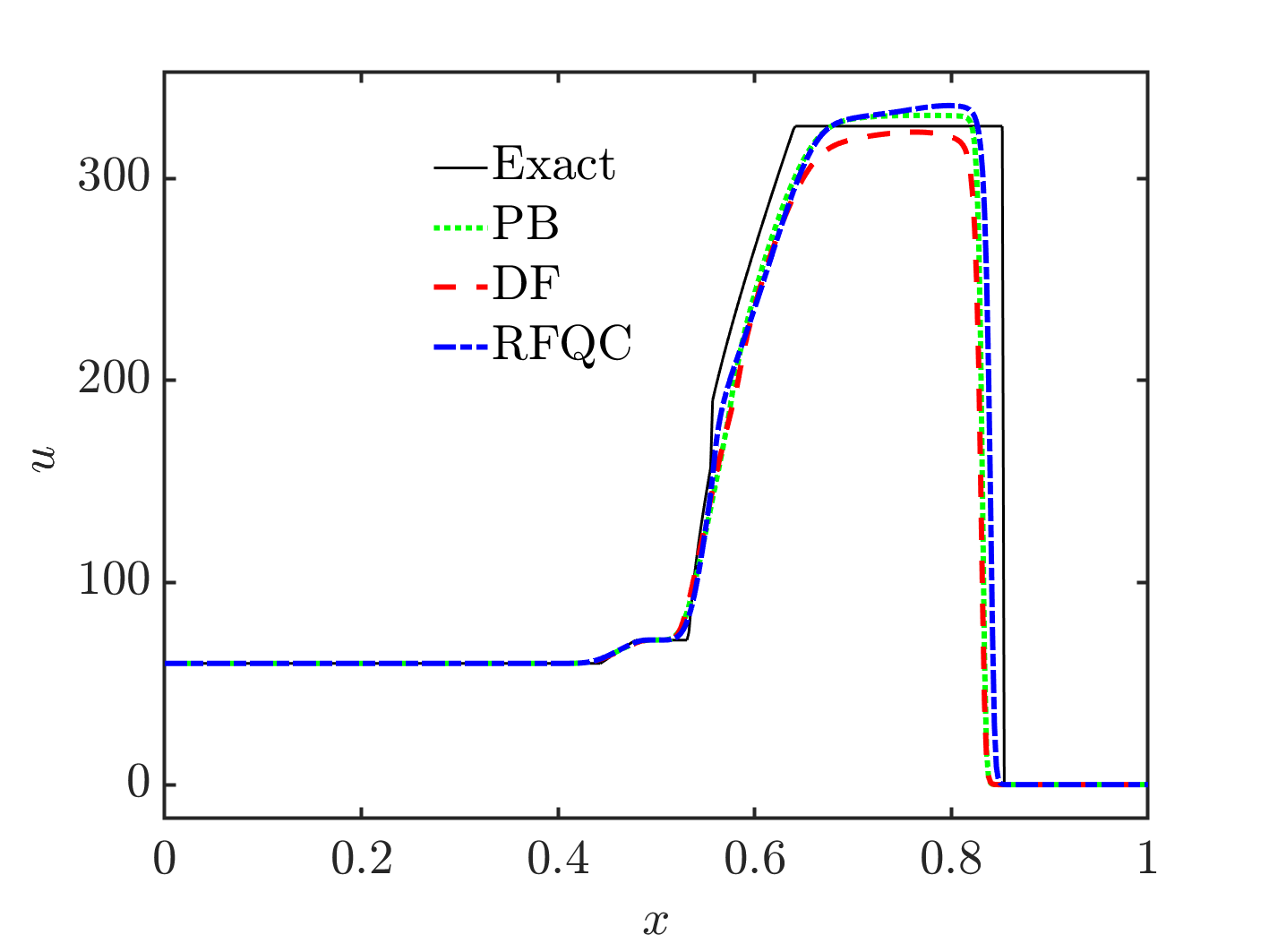}
    \caption{Velocity}
    \label{fig5b}
\end{subfigure}

\begin{subfigure}[b]{0.4\textwidth}
    \includegraphics[width=\textwidth]{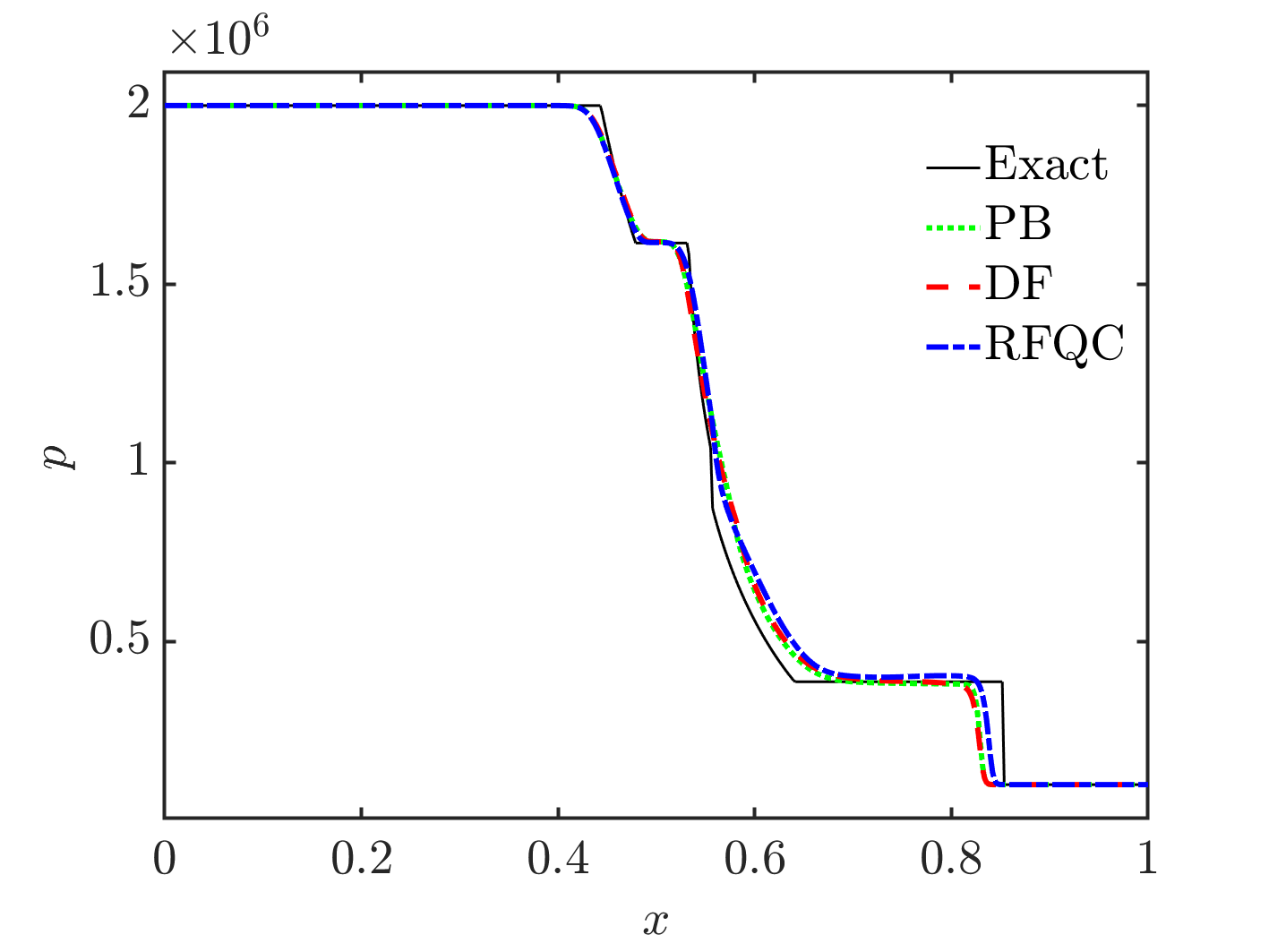}
    \caption{Pressure}
    \label{fig5c}
\end{subfigure}
\hspace{0.05\textwidth} 
\begin{subfigure}[b]{0.4\textwidth}
    \includegraphics[width=\textwidth]{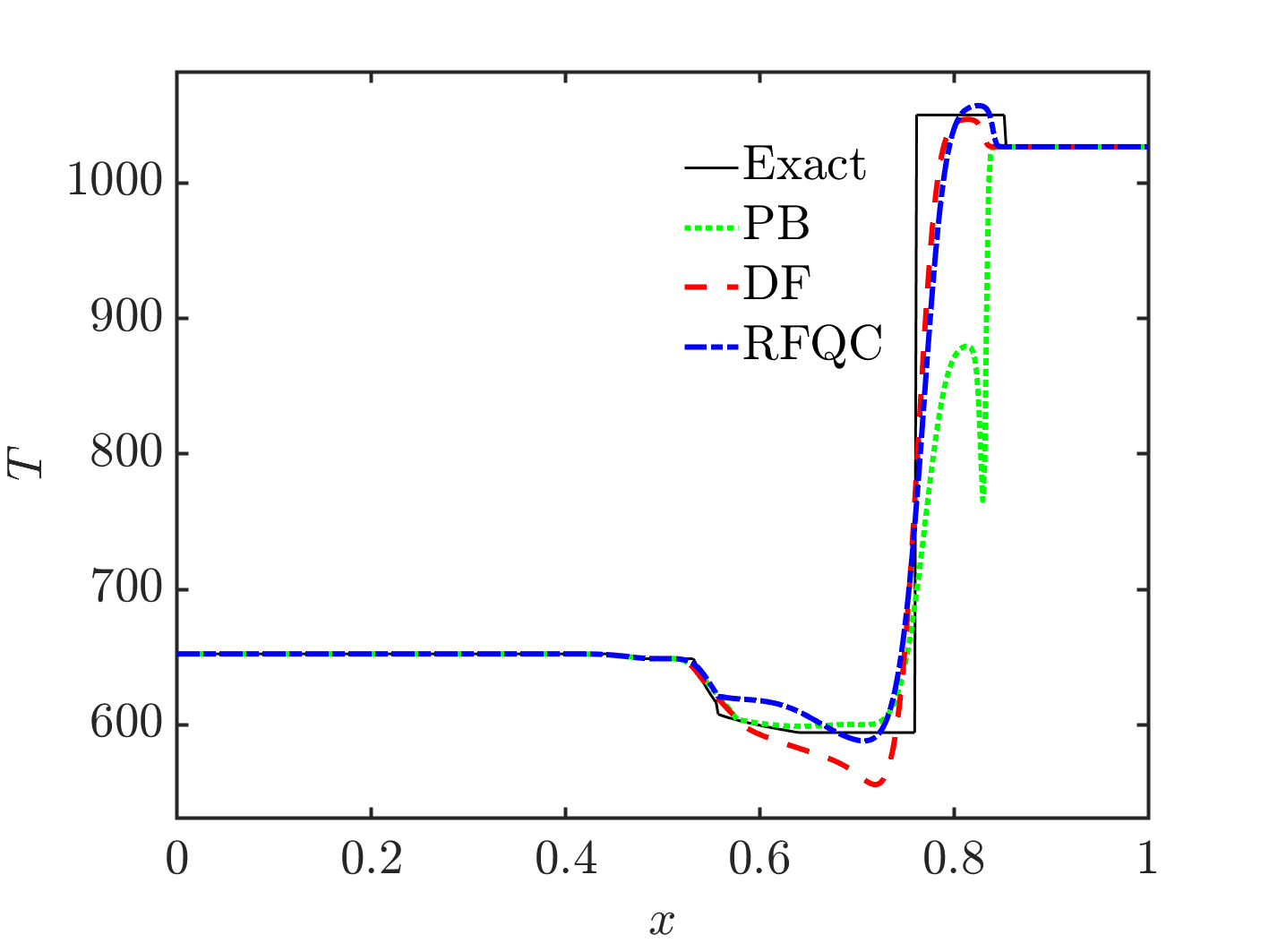}
    \caption{Temperature}
    \label{fig5d}
\end{subfigure}
\caption{Results of the Completely Evaporated FeRP Calculation.}
\label{fig5}
\end{figure}

Fig. \ref{fig5} presents the comparison of the three methods against the exact solution for the FeRP. The rarefaction wave splitting and expansion shock phenomena within the expansion branch are clearly visible in the exact solutions shown in Figs. \ref{fig5a} and \ref{fig5b}. Except for the PB method, which exhibits noticeable temperature deviations near the shock (Fig. \ref{fig5d}), both the DF and RFQC methods agree well with the exact solution. Notably, RFQC captures the shock velocity more accurately than the other two methods.

In high-pressure internal combustion engines, lower initial fuel temperatures often lead to incomplete flash evaporation. Consequently, the expansion branch of the Riemann problem crosses the saturation line only once, resulting in split rarefaction waves. Nevertheless, the speed of sound undergoes a significant discontinuity, causing sharp variations in thermodynamic properties. The initial conditions for this incomplete flash evaporation case are listed in Table \ref{tab:case3}.

\begin{table}[H]
    \centering
    \caption{Initial Values for the Incompletely Evaporated FeRP Calculation}
    \label{tab:case3}
    \begin{tabular}{cccccc}
        \toprule
       State & $P$ (Pa) & $\rho$ ($\mathrm{kg \,m^{-3}}$) & $T$ (K) & $u$ ($\mathrm{m \, s^{-1}}$) \\
        \midrule
        Left&  $5 \times 10^6$ & 500 & 572.3& 20  \\
        Right&  $1 \times 10^5$ & 2 & 1026.8 & 0  \\
        \bottomrule
    \end{tabular}
\end{table}

\begin{figure}[h]
\centering
\begin{subfigure}[b]{0.4\textwidth}
    \includegraphics[width=\textwidth]{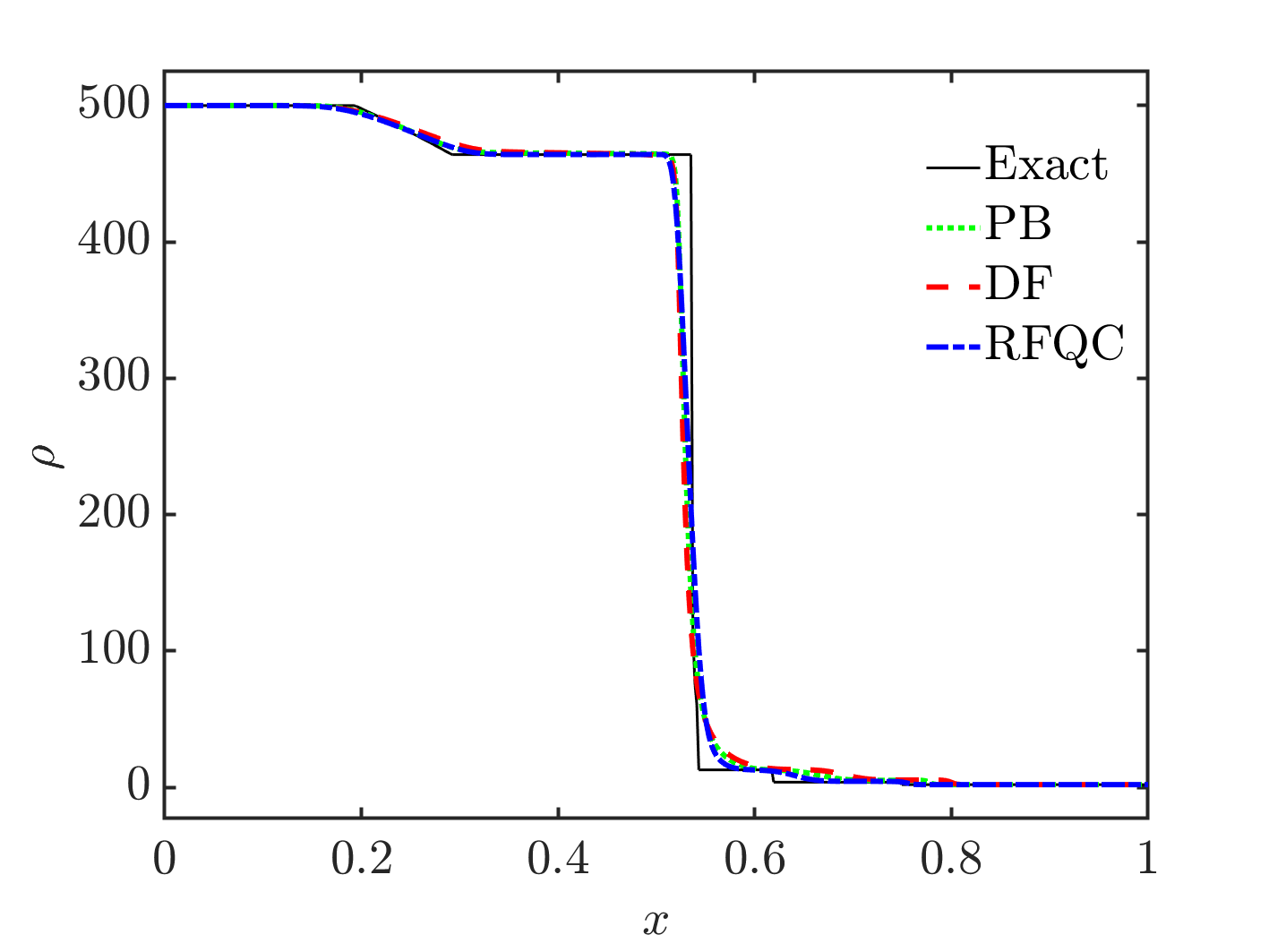}
    \caption{Density}
    \label{fig6a}
\end{subfigure}
%\hfill 
\hspace{0.05\textwidth} % 
\begin{subfigure}[b]{0.4\textwidth}
    \includegraphics[width=\textwidth]{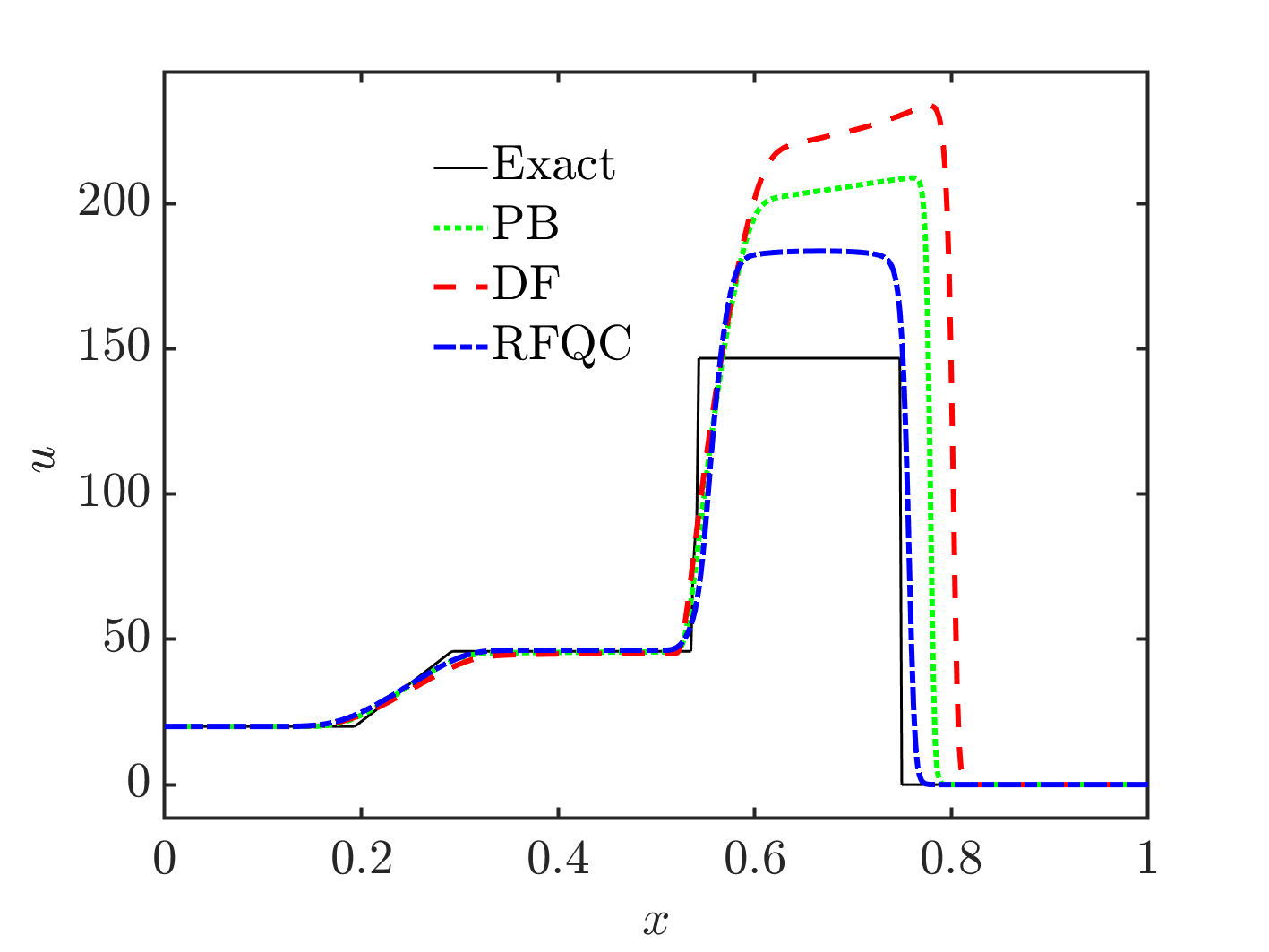}
    \caption{Velocity}
    \label{fig6b}
\end{subfigure}

\begin{subfigure}[b]{0.4\textwidth}
    \includegraphics[width=\textwidth]{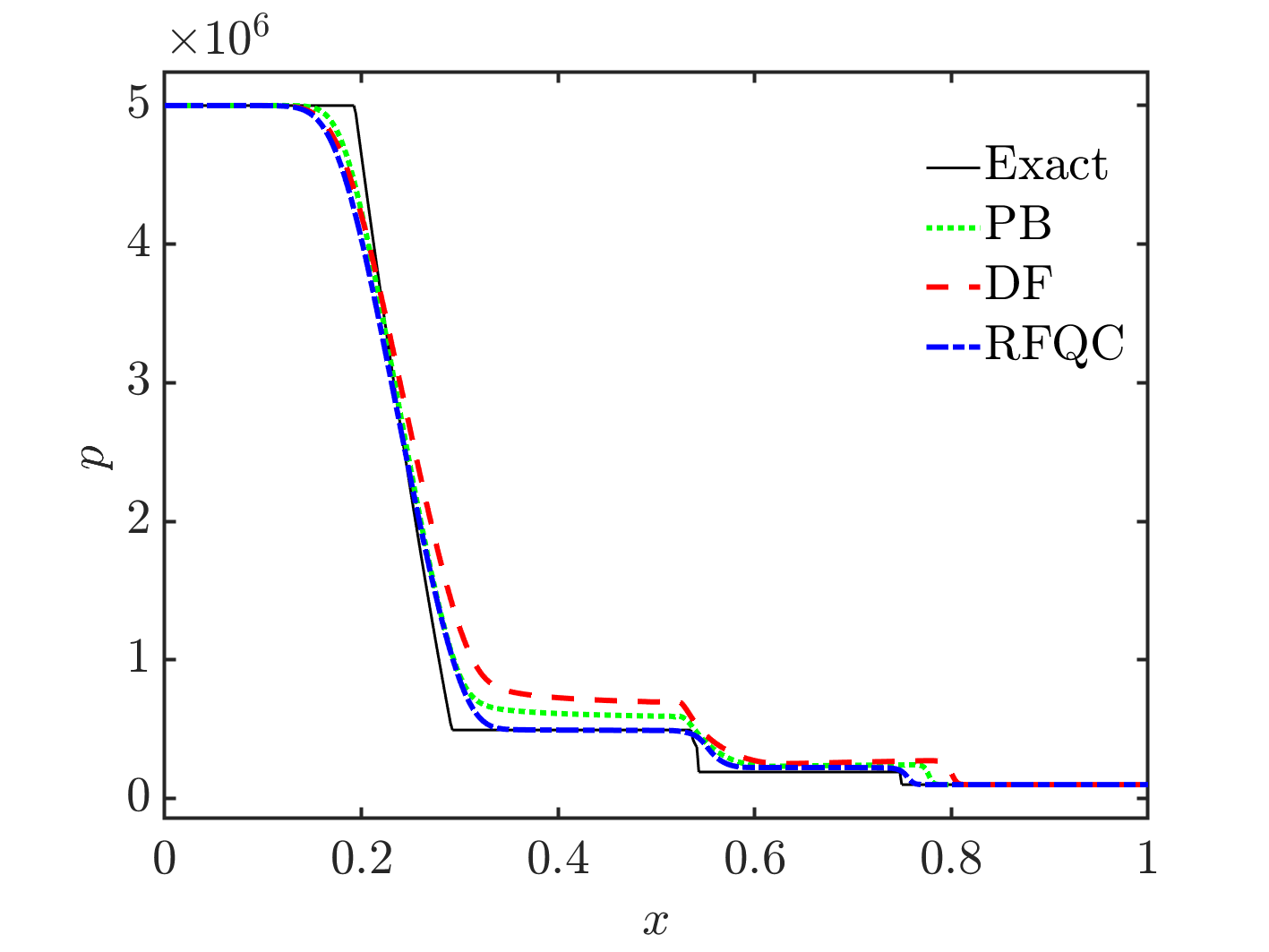}
    \caption{Pressure}
    \label{fig6c}
\end{subfigure}
\hspace{0.05\textwidth} 
\begin{subfigure}[b]{0.4\textwidth}
    \includegraphics[width=\textwidth]{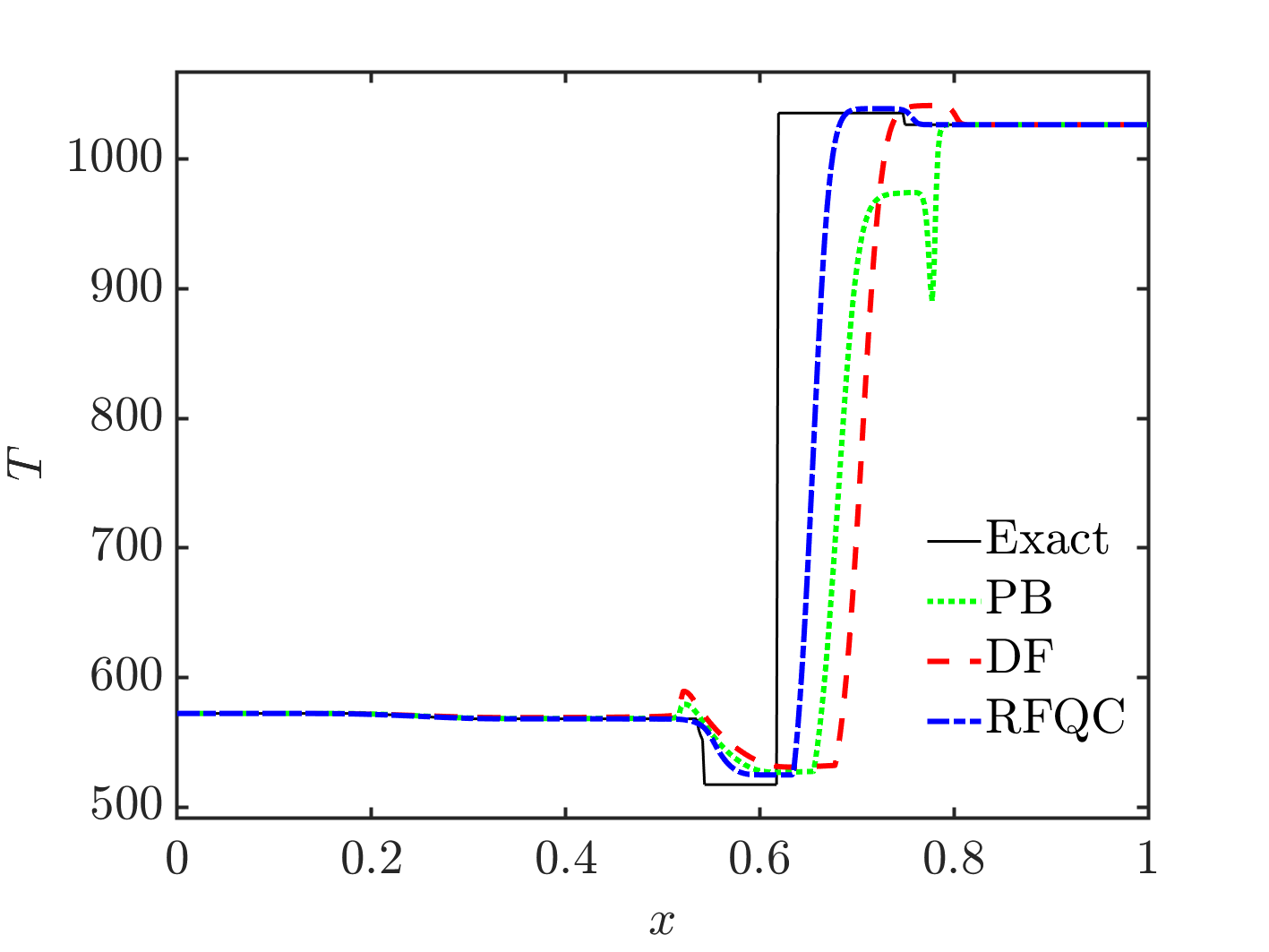}
    \caption{Temperature}
    \label{fig6d}
\end{subfigure}
\caption{Results of the Incompletely Evaporated FeRP Calculation.}
\label{fig6}
\end{figure}

Fig. \ref{fig6} presents the comparison of the three methods against the exact solution for the incomplete evaporation FeRP. In this case, the RFQC method demonstrates the best agreement with the exact solution. Apart from a deviation in the intermediate velocity, its predictions for pressure, temperature, and shock location are highly consistent with the exact solution. The DF method significantly overestimates the shock speed, intermediate velocity, and pressure. The PB method performs slightly better but still exhibits noticeable errors. Notably, both the DF and PB methods generate a non-physical temperature rise near the head of the rarefaction wave.

\subsection{Subcritical Liquid Impingement Verification}
Subcritical liquid impingement refers to the collision of liquid streams at subcritical pressures, a phenomenon frequently encountered in impinging injectors and fuel pipelines. Subcritical liquids are characterized by high stiffness; consequently, their collision generates intense shock waves. Based on this physical background, the initial conditions for the Riemann problem are selected as listed in Table \ref{tab:case4}. The computational domain length $L=1\text{m}$, with a simulation time of $t=0.5 \text{ms}$, a grid size of $N=500$, and a CFL number of 0.1.

\begin{table}[H]
    \centering
    \caption{Initial Values for the Riemann Problem Calculation}
    \label{tab:case4}
    \begin{tabular}{cccccc}
        \toprule
       State & $P$ (Pa) & $\rho$ ($\mathrm{kg \,m^{-3}}$) & $T$ (K) & $u$ ($\mathrm{m \, s^{-1}}$) \\
        \midrule
        Left&  $1 \times 10^6$ & 550 & 497.3 & 50  \\
        Right&  $1 \times 10^6$ & 550 & 497.3 & 0  \\
        \bottomrule
    \end{tabular}
\end{table}

\begin{figure}[ht]
\centering
\begin{subfigure}[b]{0.4\textwidth}
    \includegraphics[width=\textwidth]{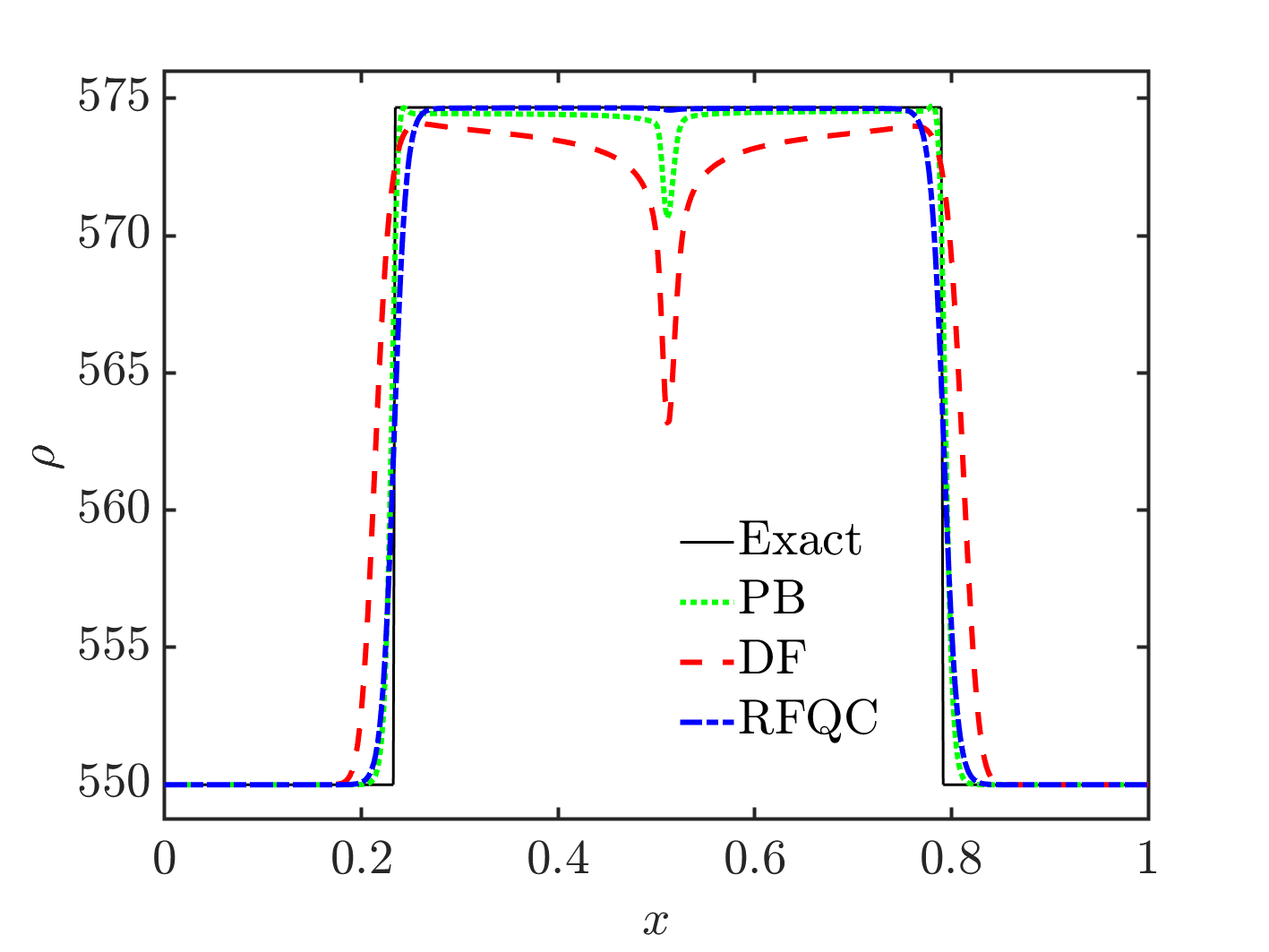}
    \caption{Density}
    \label{fig7a}
\end{subfigure}
%\hfill 
\hspace{0.05\textwidth} % 
\begin{subfigure}[b]{0.4\textwidth}
    \includegraphics[width=\textwidth]{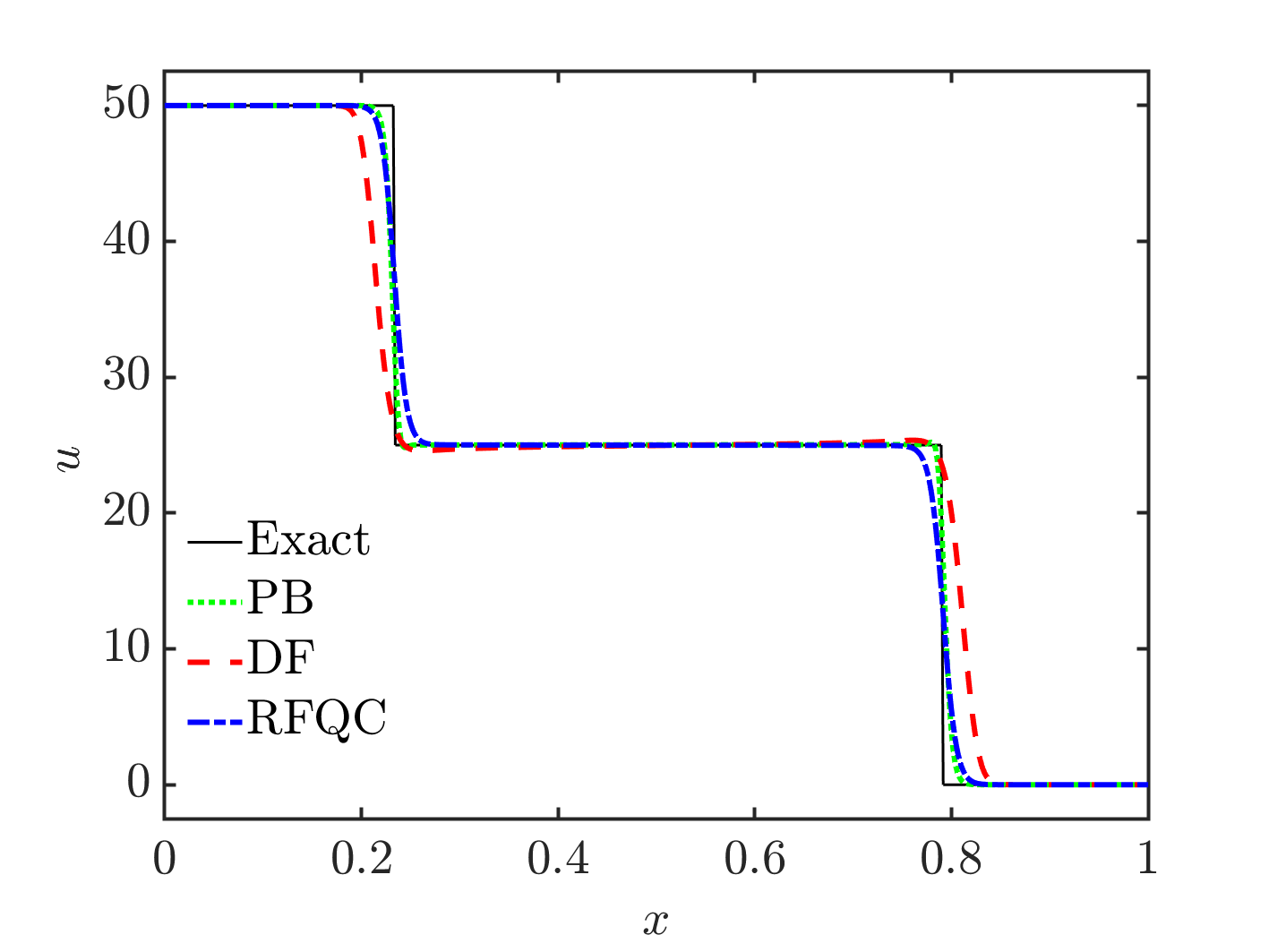}
    \caption{Velocity}
    \label{fig7b}
\end{subfigure}

\begin{subfigure}[b]{0.4\textwidth}
    \includegraphics[width=\textwidth]{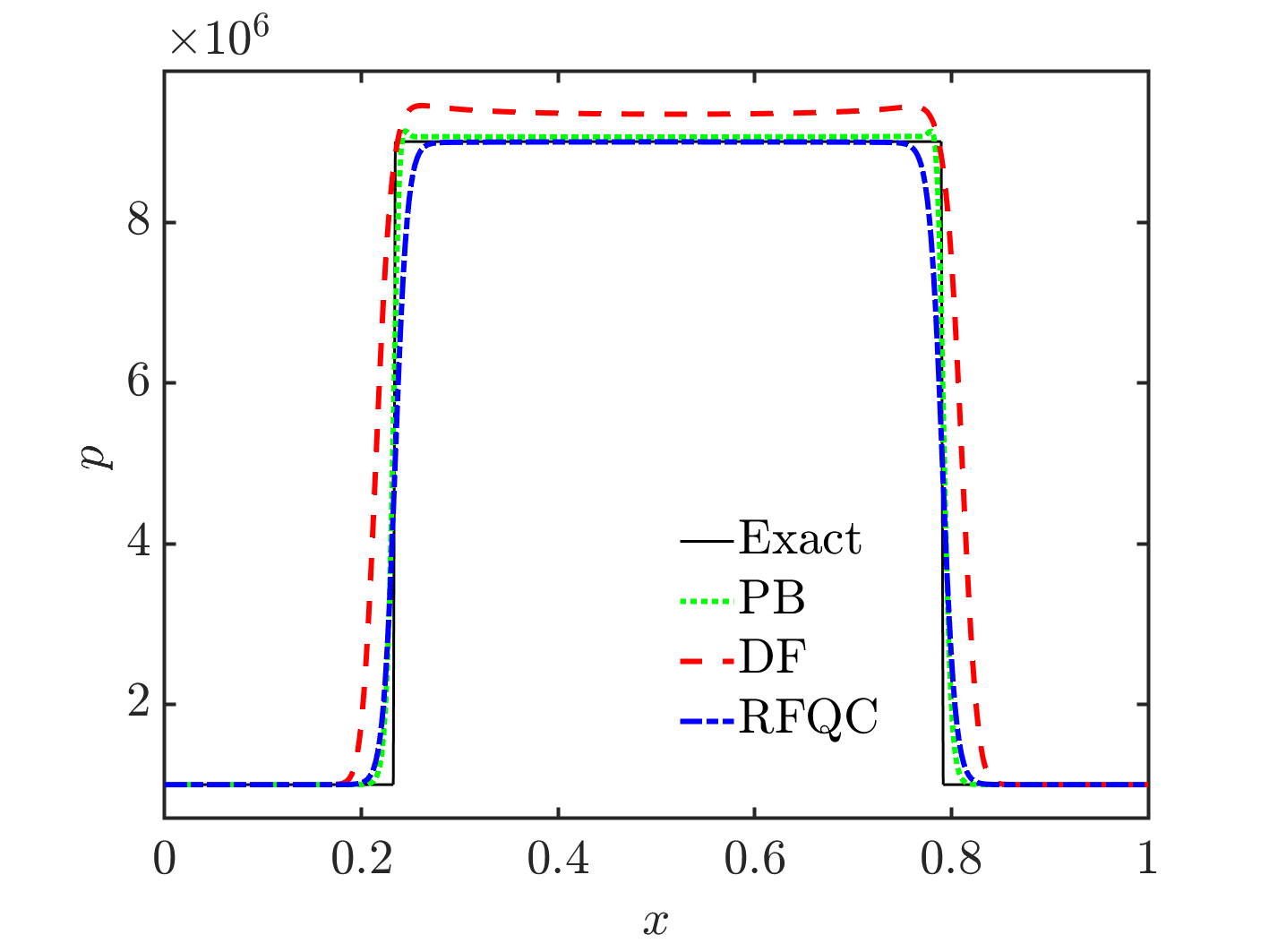}
    \caption{Pressure}
    \label{fig7c}
\end{subfigure}
\hspace{0.05\textwidth} 
\begin{subfigure}[b]{0.4\textwidth}
    \includegraphics[width=\textwidth]{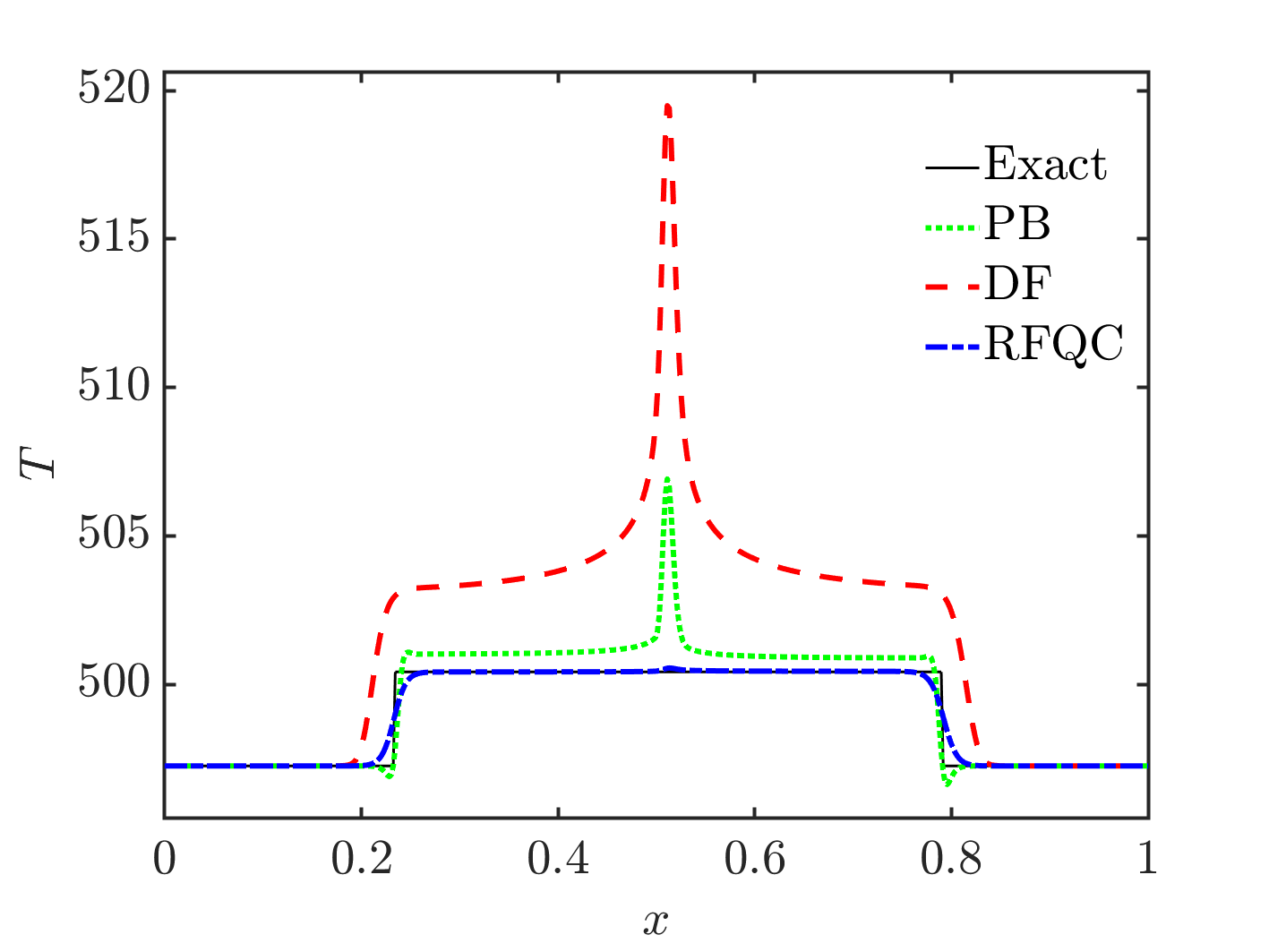}
    \caption{Temperature}
    \label{fig7d}
\end{subfigure}
\caption{Results of the Subcritical Fluid Impinging RP Calculation.}
\label{fig7}
\end{figure}

Fig. \ref{fig7} presents the comparison of the three methods against the exact solution for the subcritical impinging Riemann problem. In this case, the RFQC results show excellent agreement with the exact solution. However, both the DF and PB methods exhibit local non-physical artifacts near the initial discontinuity. These errors manifest primarily as a density dip and a temperature spike. This type of error corresponds to the well-known overheating error in CFD \cite{Menikoff1994}, which is commonly encountered in shock-shock collision and shock-wall reflection problems, while the RFQC method demonstrates the best performance among them.

In summary, the RFQC method exhibits excellent accuracy for various real fluid Riemann problems involving strong thermodynamic nonlinearities. Under an identical first-order discretization framework, the RFQC method outperforms both the PB and DF methods. Furthermore, since traditional conservative schemes suffer from severe numerical instabilities under these extreme conditions—leading to spurious pressure oscillations and even computational divergence—their results are omitted from this section.

\section{High-Order and Two-Dimensional Simulations}
To verify the applicability of the proposed algorithm in complex flows, we extend the RFQC scheme to high-order space-time formulations and two dimensions. We simulate a 1D shock-interface interaction, a 2D advection test, a 2D shock-droplet interaction, as well as 2D transcritical and phase-change jet flows of n-dodecane. For all these cases, the time advancement employs the third-order SSP Runge-Kutta scheme \cite{Shu1988}, and the spatial reconstruction uses the third-order WENO scheme \cite{Jiang1996} with Gaussian quadrature \cite{Titarev2004} (the phase-change jet simulation employs the second-order MUSCL reconstruction with the minmod limiter \cite{Sweby1984}). The two-dimensional discrete formulation of the RFQC method is detailed in \ref{app:Discrete}.

It is worth noting that to ensure the oscillation-free property, the spatial reconstruction must be applied simultaneously to the three primitive variables $(\rho, u, p)$ and the two thermodynamic coefficients $(\xi, E_0)$, rather than to the conservative variables. This is consistent with the properties of the original Shyue's scheme; a comprehensive analysis can be found in Ref.\cite{Johnsen2006}, and detailed formulations are in \ref{appsub:high}. Furthermore, we emphasize that in the SSP Runge-Kutta time advancement, the thermodynamic re-projection of the RFQC method is performed only at the end of each full time step, and not within each Runge-Kutta sub-step.

\subsection{Shock-Interface Interaction Problem}
In supersonic flow environments, the interaction between shock waves and fuel-air interfaces is an important mechanism driving the fuel droplet breakup and atomization. The robust calculation of such interactions is a prerequisite for numerical simulations of high-speed propulsion systems. Based on this, we establish a test case involving the interaction between a shock wave and an n-dodecane interface. The computational domain length $L=1\text{m}$, with a grid size of $N=500$ and a CFL number of 0.3. The specific initial conditions are listed in Table \ref{tab:case5}.

\begin{table}[H]
    \centering
    \caption{Initial Values for the Shock-Interface Interaction Problem}
    \label{tab:case5}
    \begin{tabular}{ccccccc}
        \toprule
       State & Domain (m) & $P$ (Pa) & $\rho$ ($\mathrm{kg \,m^{-3}}$) & $T$ (K) & $u$ ($\mathrm{m \, s^{-1}}$) \\
        \midrule
        Left& $(0,0.5)$&  $2 \times 10^5$ & 550 & 493.4 & 0  \\
        Middle& $(0.5,0.8)$&  $2 \times 10^5$ & 5 & 832.8 & 0  \\
        Right& $(0.8,1.0)$&  $2 \times 10^5$ & 5 &  832.8 & -800  \\
        \bottomrule
    \end{tabular}
\end{table}

\begin{figure}[H]
\centering
\hspace*{0.01\textwidth}
\begin{subfigure}[b]{0.41\textwidth}
    \includegraphics[width=\textwidth]{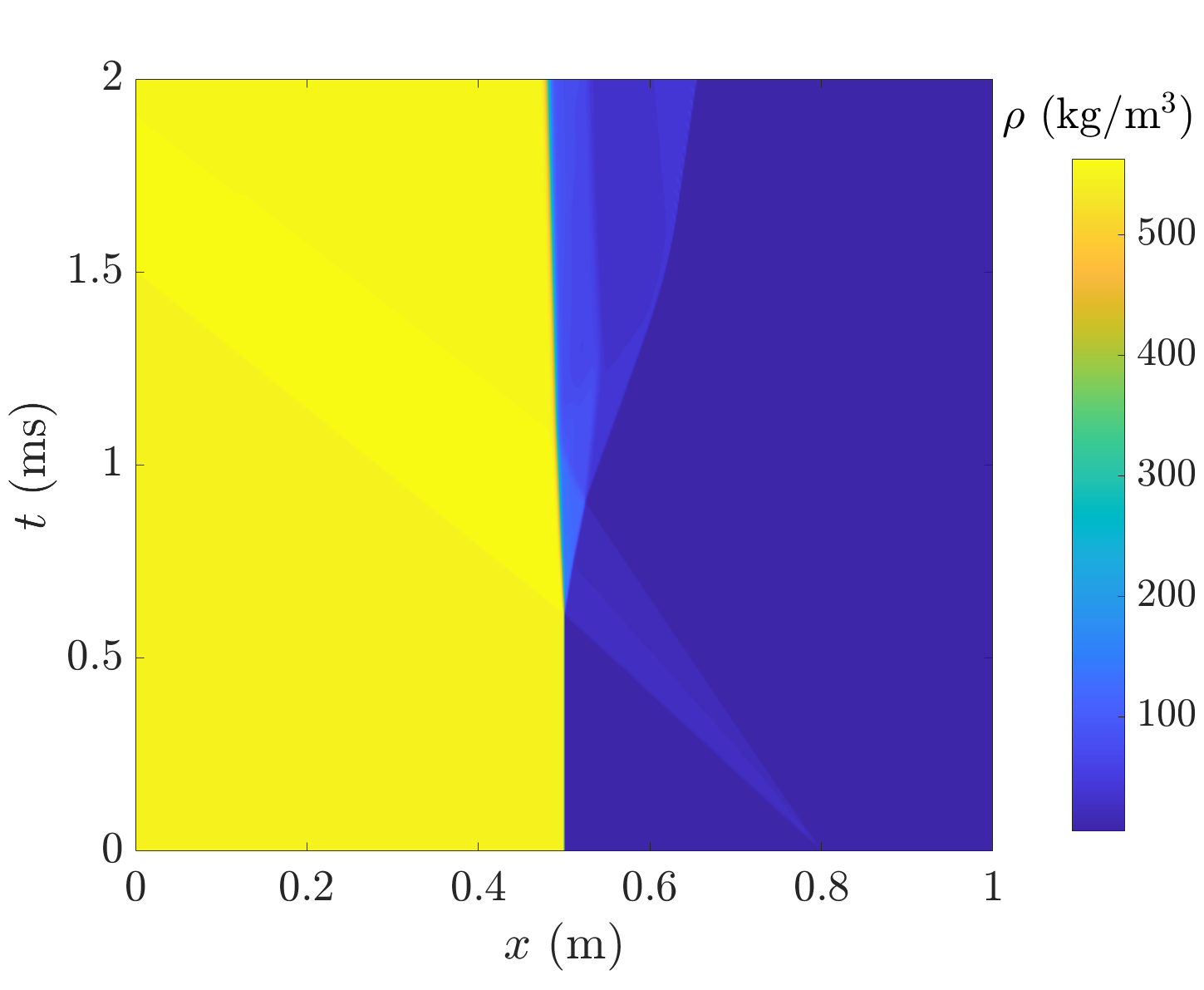}
    \caption{Density}
    \label{fig8a}
\end{subfigure}
%\hfill 
\hspace{0.007\textwidth} % 
\begin{subfigure}[b]{0.4\textwidth}
    \includegraphics[width=\textwidth]{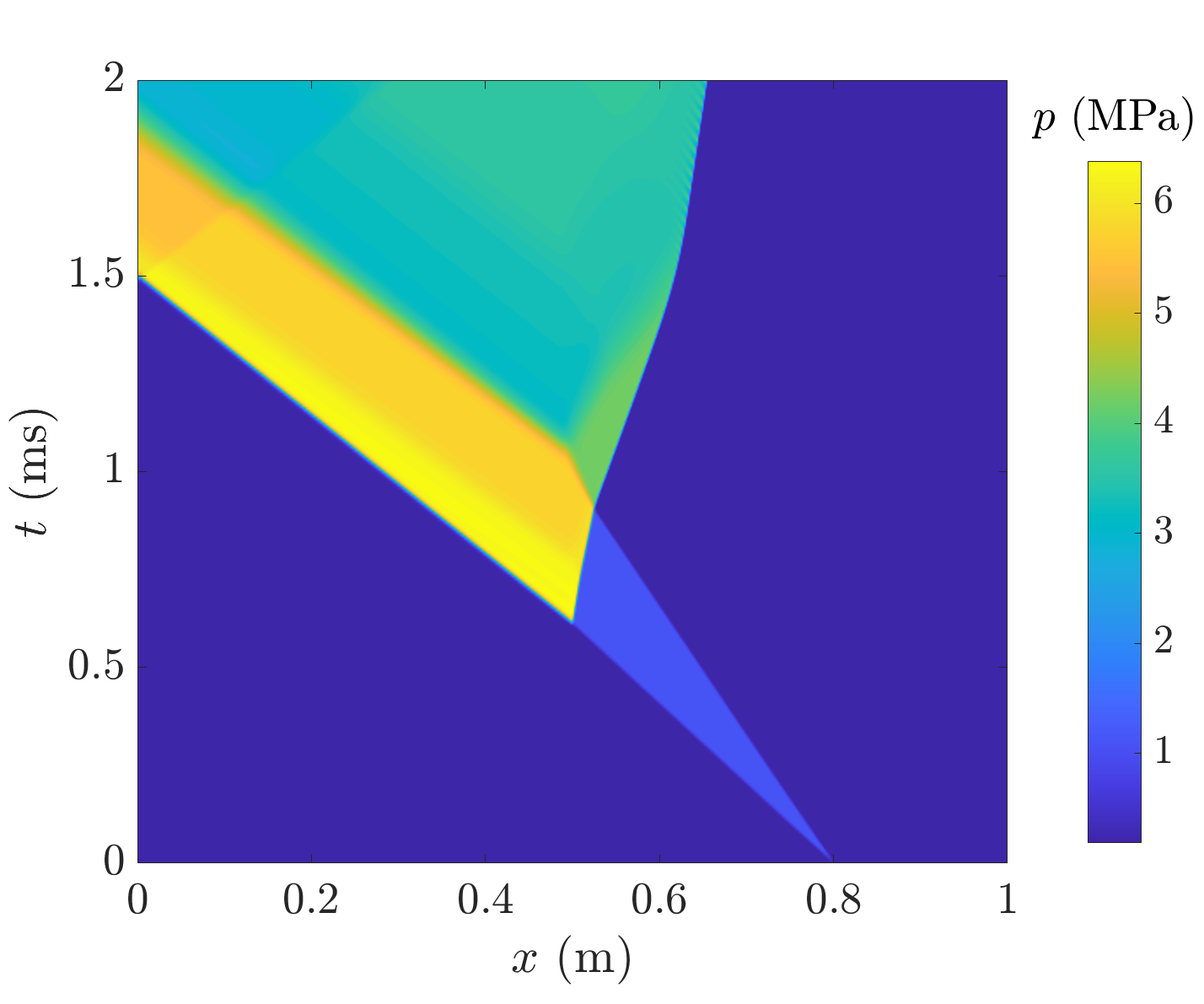}
    \caption{Pressure}
    \label{fig8b}
\end{subfigure}

\begin{subfigure}[b]{0.4\textwidth}
    \includegraphics[width=\textwidth]{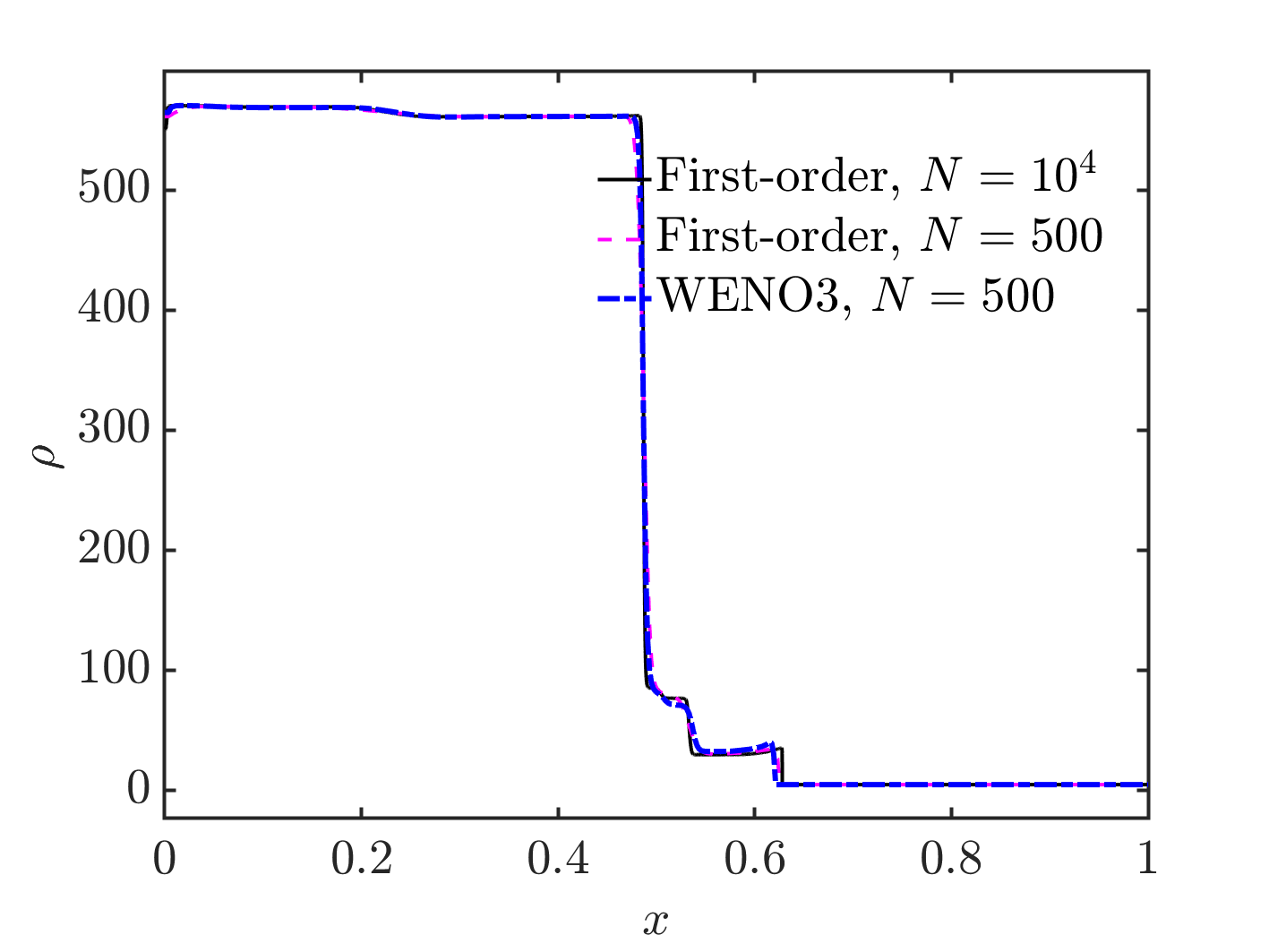}
    \caption{Density comparison, $t=15 \mathrm{ms}$}
    \label{fig8c}
\end{subfigure}
\hspace{0.04\textwidth} 
\begin{subfigure}[b]{0.4\textwidth}
    \includegraphics[width=\textwidth]{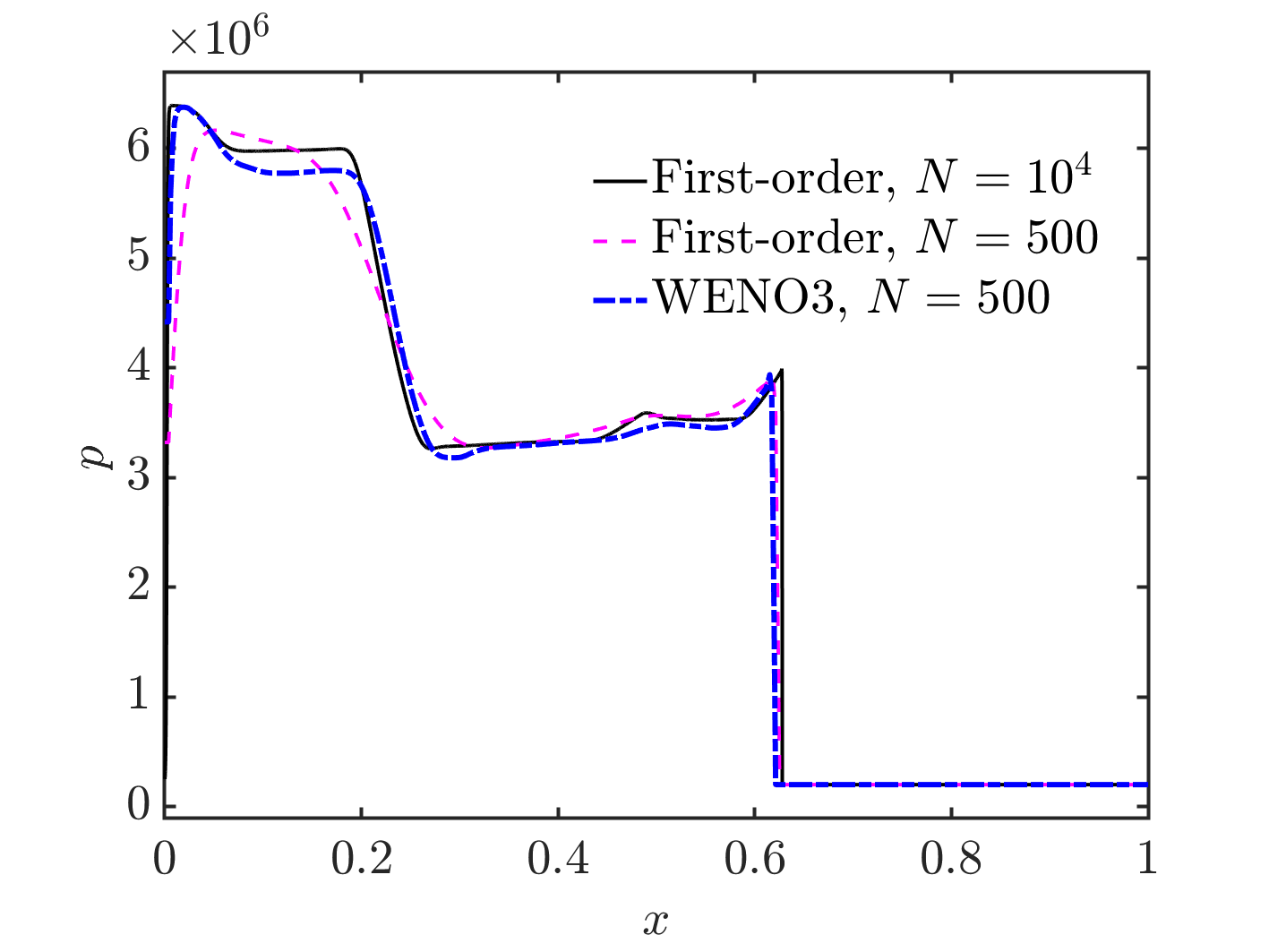}
    \caption{Pressure comparison, $t=15 \mathrm{ms}$}
    \label{fig8d}
\end{subfigure}

\caption{Results of the Shock-Interface Interaction Calculation.}
\label{fig8}
\end{figure}

Fig. \ref{fig8} displays the simulation results for the shock-interface interaction problem. The initial velocity discontinuity at $x=0.8\text{m}$ generates an incident shock wave propagating to the left. At $t=0.6\text{ms}$, this shock impacts the vapor-liquid interface near $x=0.5\text{m}$. Subsequently, stronger transmitted and reflected shocks are generated at the interface. The post-shock pressure rises to $6 \times 10^6\text{Pa}$ and propagates into both the vapor and liquid phases, while a rarefaction region gradually develops in the center. The RFQC method accurately captures these fundamental physical features. In contrast, both the Double Flux and Pressure-Based methods suffer from numerical divergence under the same grid and discretization scheme. This divergence manifests primarily as severe pressure oscillations and negative temperatures immediately upon the shock-interface interaction. Consequently, the results for these two methods are not presented.

Additionally, since this problem lacks an exact solution, the computational results at $t=15\mathrm{ms}$ are compared against a highly resolved first-order simulation ($N=10^4$). Figs. \ref{fig8c} and \ref{fig8d} show that over a long period, the current method accurately captures the shock position and yields post-shock pressure and density that agree well with the reference.

\subsection{Two-Dimensional Advection Test}
\label{sub:2D_advection}
To verify the pressure oscillation-free property of the RFQC method under multi-dimensional high-order conditions, the 1D density wave advection test from Section \ref{sec:convergencetest} is extended to 2D. The computational domain is a $1\,\text{m} \times 1\,\text{m}$ square with periodic boundary conditions on all sides. The flow field is initially uniform with a velocity of $u = v = 100\,\text{m/s}$ and a constant pressure of $p_0 = 10^6\,\text{Pa}$. Centered at the domain middle, the initial density field spans both the liquid phase and the two-phase region, governed by the following distribution function:
\begin{equation}
\rho(x,y) = 400 + 100 \sin\left(\frac{2\pi r}{L}\right),
\label{eq:2D_advection_rho}
\end{equation}
where $L=1\,\text{m}$ and $r = \sqrt{(x-0.5)^2+(y-0.5)^2}$.

A uniform square grid with $\Delta =1 \times 10^{-2}\,\text{m}$ is utilized, and the CFL number is set to 0.5. The final simulation time is $t=0.01\,\text{s}$, which corresponds to one full advection period.

\begin{figure}[H]
\centering
\begin{subfigure}[b]{0.47\textwidth}
    \includegraphics[width=\textwidth]{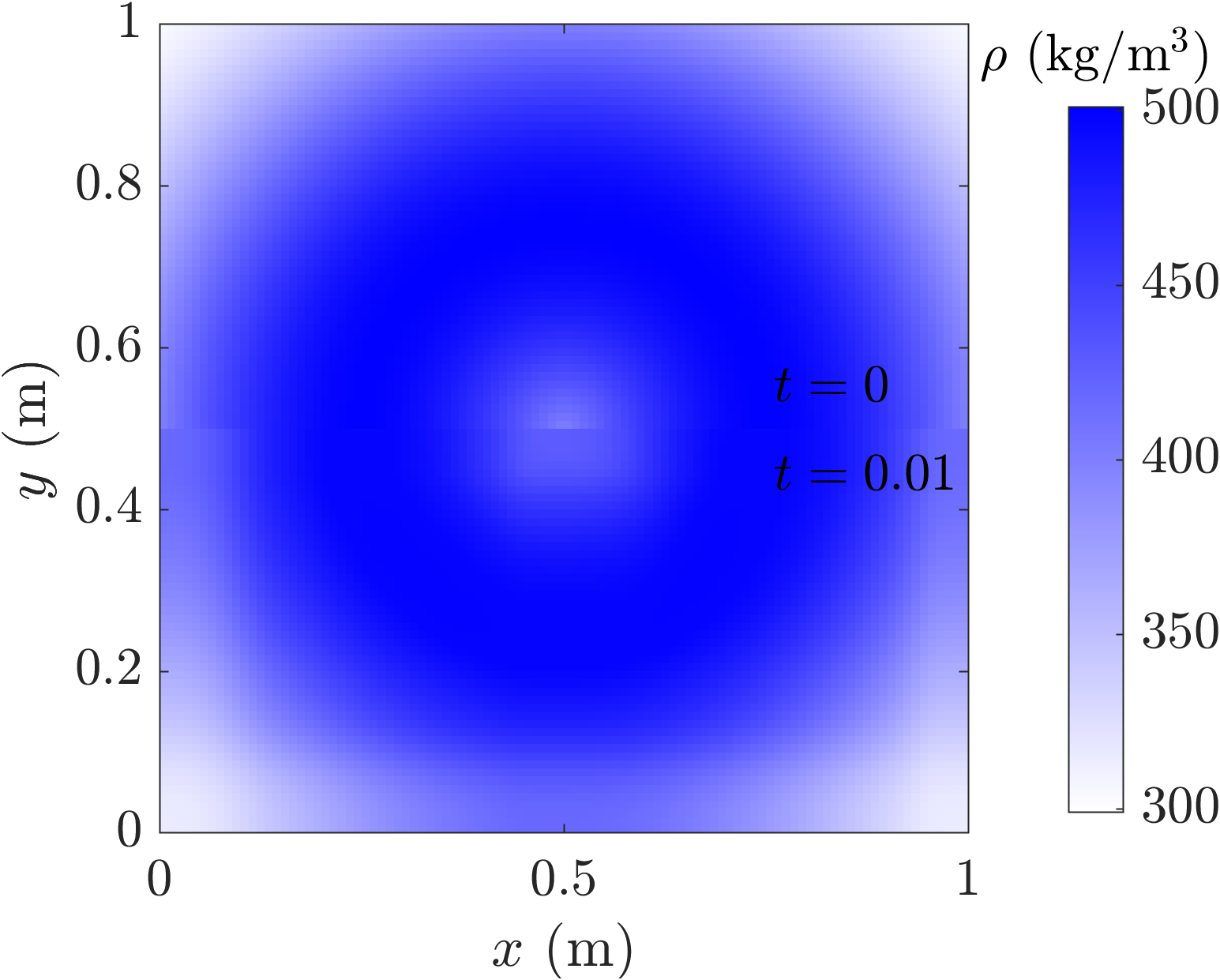}
    \caption{Density}
    \label{fig9a}
\end{subfigure}
\hfill
\begin{subfigure}[b]{0.47\textwidth}
    \includegraphics[width=\textwidth]{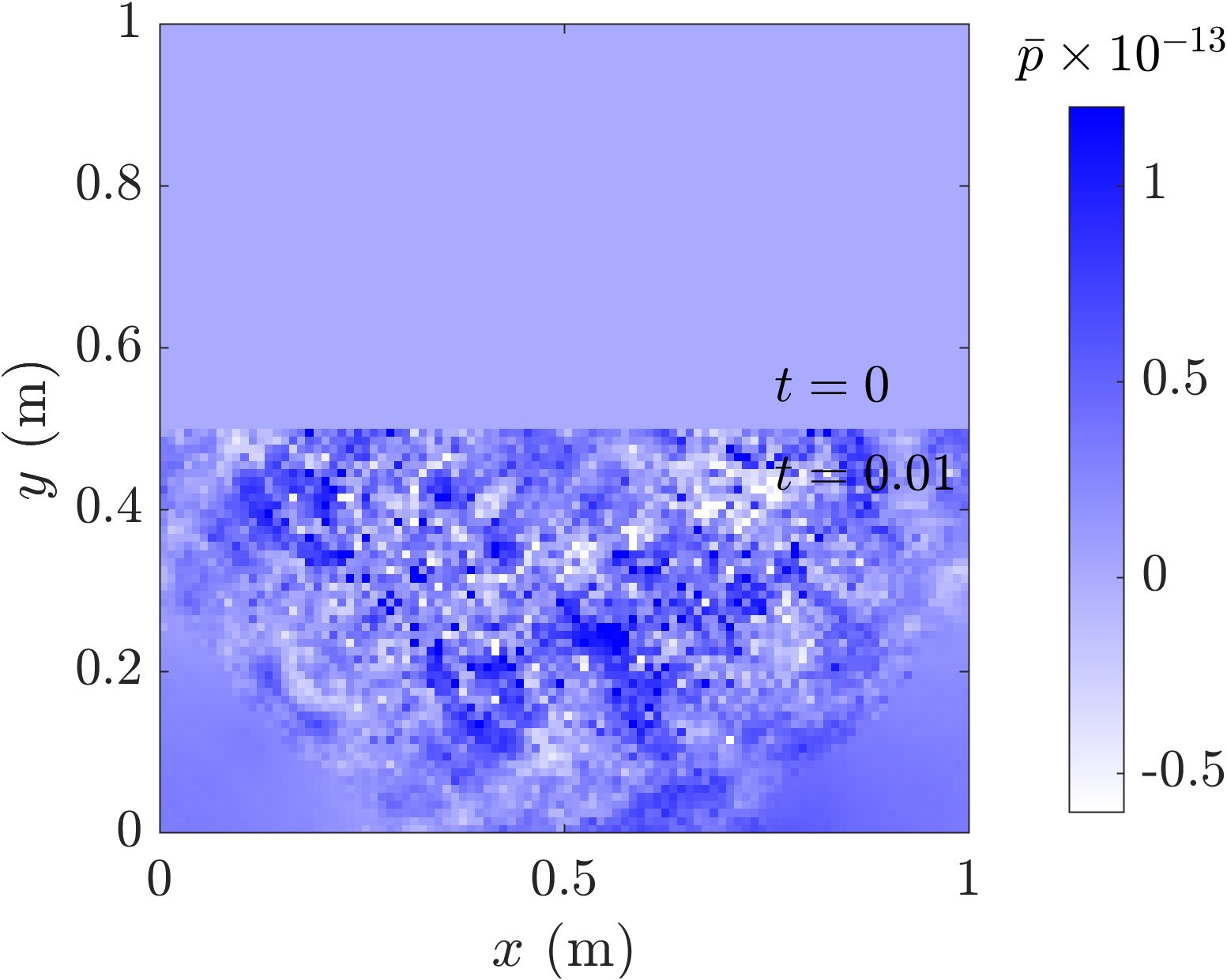}
    \caption{Pressure}
    \label{fig9b}
\end{subfigure}
\caption{Results of the Advection Test.}
\label{fig9}
\end{figure}

Figure \ref{fig9} shows the computational results at the initial state and after one period. As shown in Fig. \ref{fig9a}, the density wave returns to its initial position after one period, though its peak value decreases slightly due to numerical dissipation. Figure \ref{fig9b} shows the normalized pressure $\bar p = (p-p_0)/p_0$ field, indicating that the pressure equilibrium is strictly maintained with oscillations remaining near machine precision (on the order of $10^{-13}$). This result confirms that the proposed method preserves its oscillation-free property within a multi-dimensional high-order finite volume framework.

\subsection{Two-Dimensional Shock-Droplet Interaction}

\begin{figure}[h]
\centering 
\includegraphics[width=0.45\textwidth]{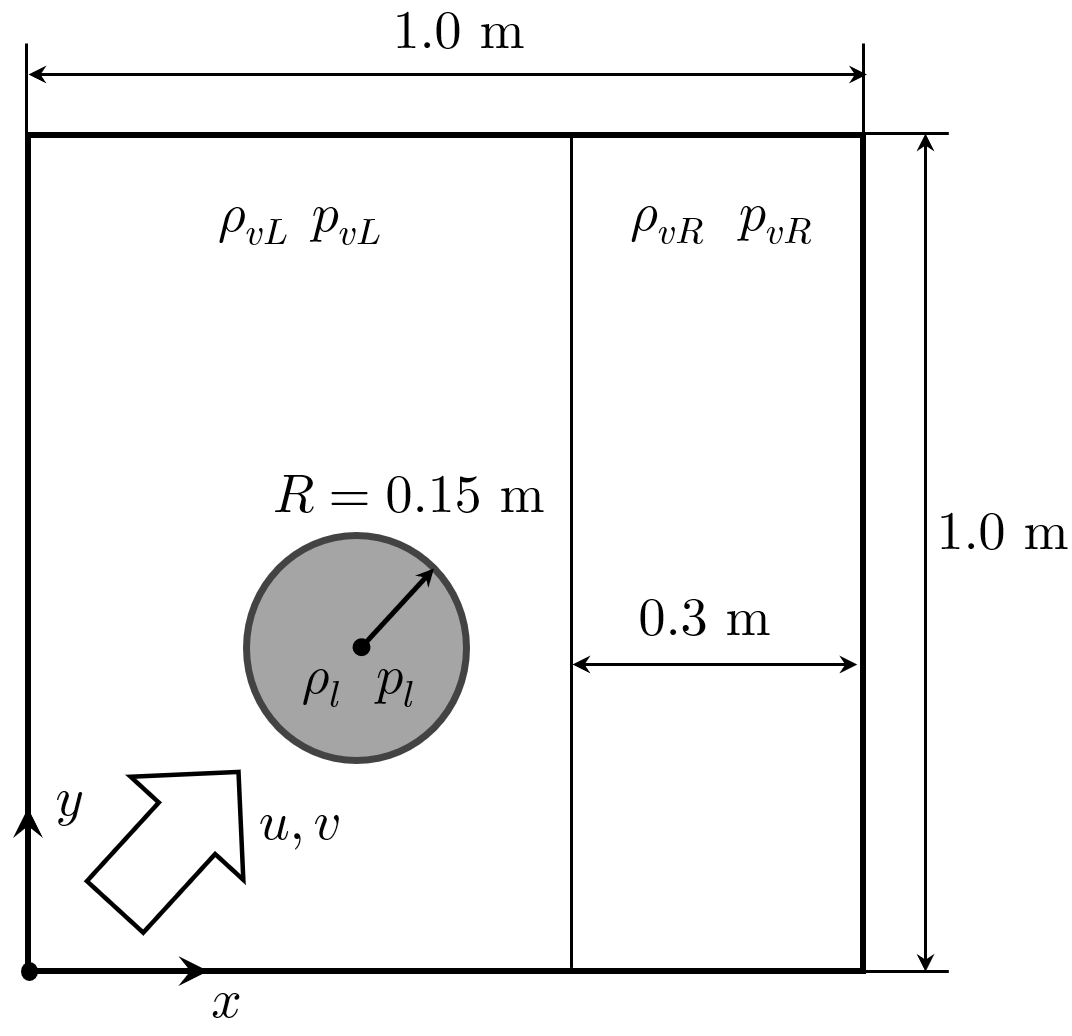} 
\caption{Computational Conditions for Shock Wave-Droplet Interaction.}
\label{fig10}
\end{figure}

This case simulates the interaction between a shock wave and a moving droplet under supercritical pressure. It aims to demonstrate that the RFQC method can maintain pressure equilibrium in two-dimensional flows. The computational domain is a square region with a side length of $L=1\text{m}$. The schematic and boundary conditions are shown in Fig. \ref{fig10}. Initially, the flow field contains a liquid n-dodecane droplet with a radius $R=0.15\text{m}$ and a density $\rho_l = 550 \text{kg/m}^3$, centered at $(0.4, 0.4)$. The droplet is surrounded by n-dodecane vapor with a density of $\rho_{vL} = 20 \text{kg/m}^3$. A pressure and density discontinuity is introduced at $x=0.7\text{m}$ to induce a shock wave: in the region $x<0.7\text{m}$, the fluid (including both the droplet and vapor) is at $p_l = p_{vL} = 2 \times 10^6 \text{Pa}$; in the region $x>0.7\text{m}$, the vapor density $\rho_{vR} = 40 \text{kg/m}^3$ and the pressure $p_{vR} = 6 \times 10^6 \text{Pa}$. The initial velocity is uniform throughout the entire field, with $u=v=100\text{m/s}$.

\begin{figure}[H]
\centering
\begin{subfigure}[b]{0.45\textwidth}
    \includegraphics[width=\textwidth]{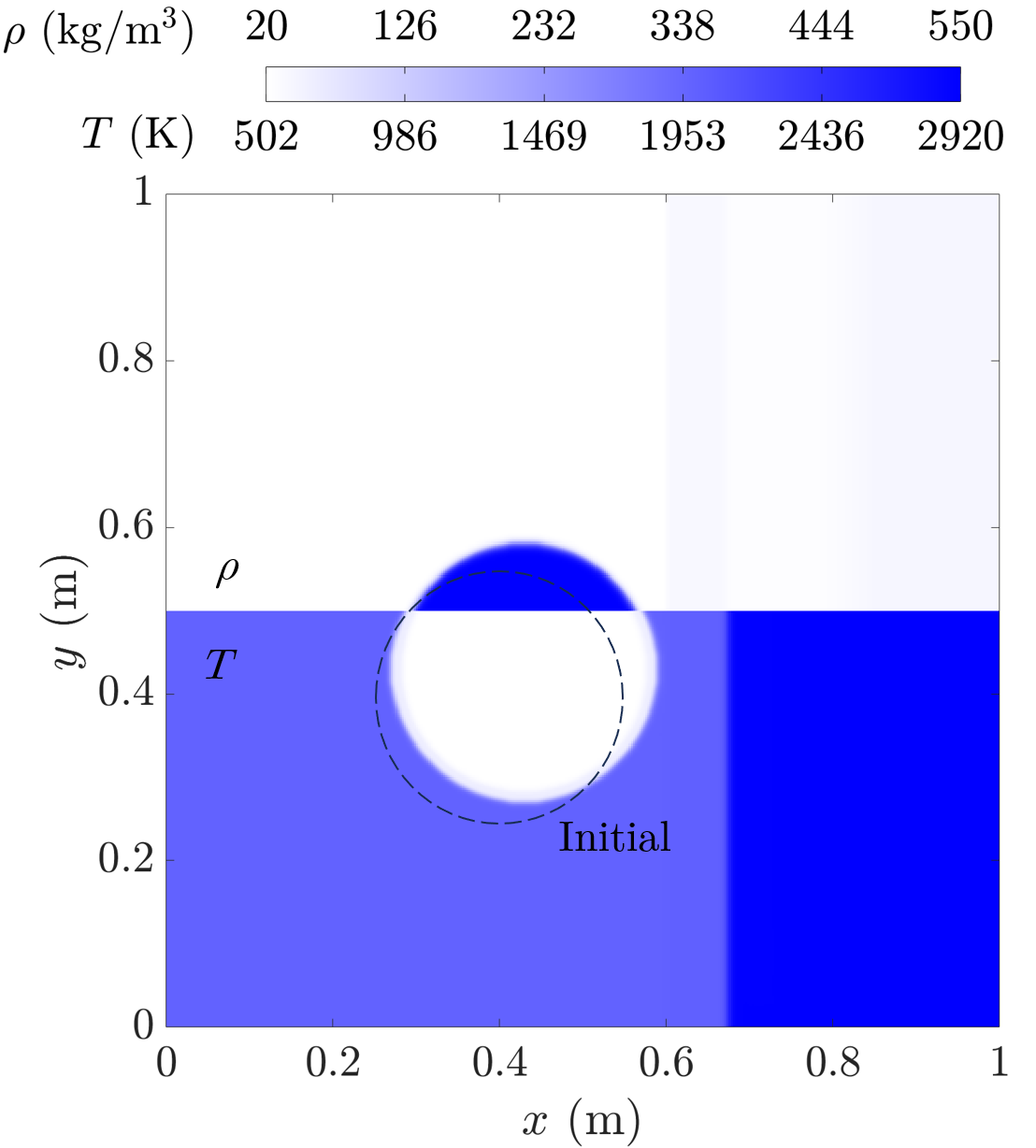}
    \caption{Density and temperature, $t=0.3\text{ms}$}
    \label{fig11a}
\end{subfigure}
\hspace{0.01\textwidth} % 
\begin{subfigure}[b]{0.45\textwidth}
    \includegraphics[width=\textwidth]{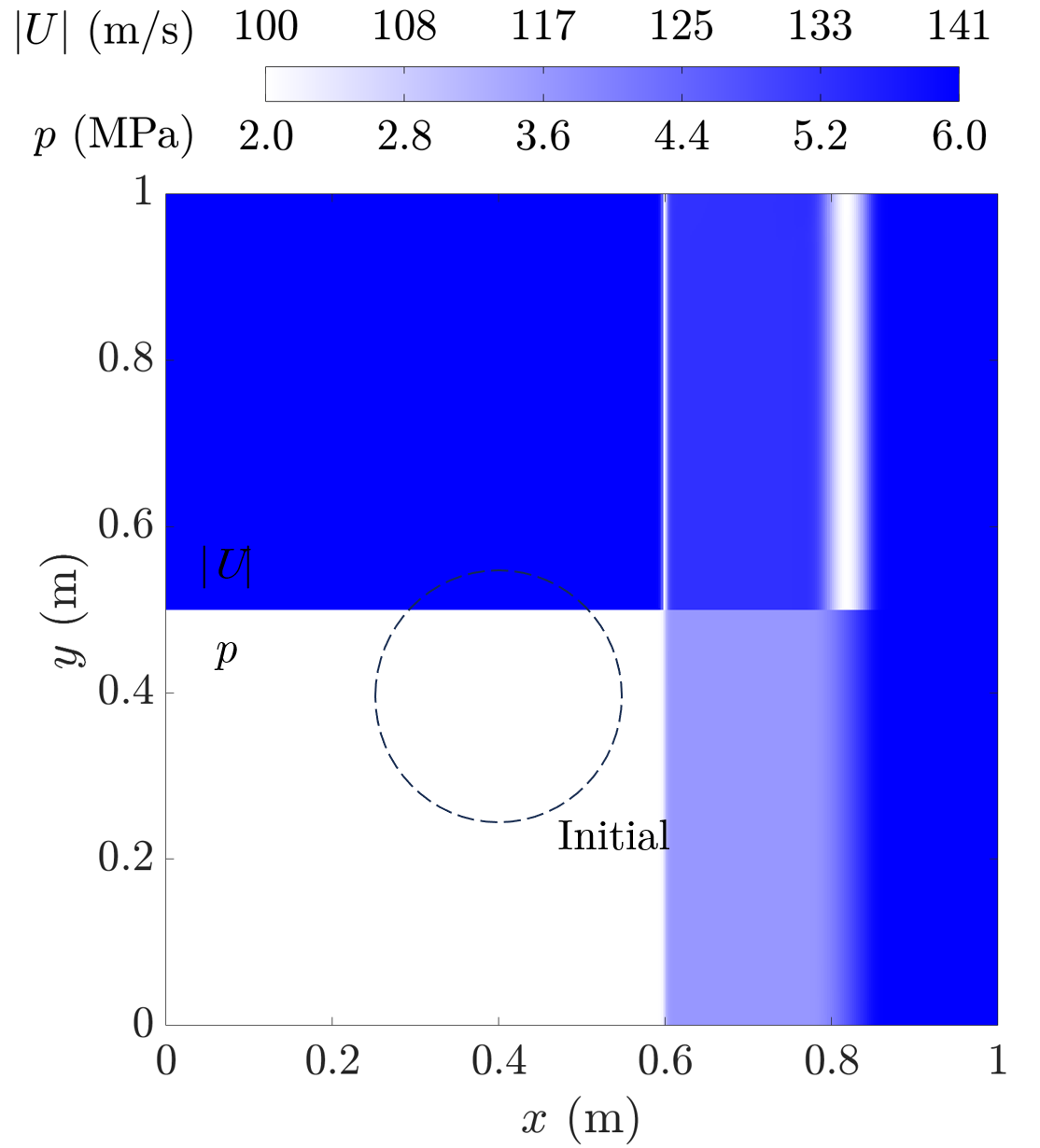}
    \caption{Pressure and velocity, $t=0.3\text{ms}$}
    \label{fig11b}
\end{subfigure}
\begin{subfigure}[b]{0.45\textwidth}
    \includegraphics[width=\textwidth]{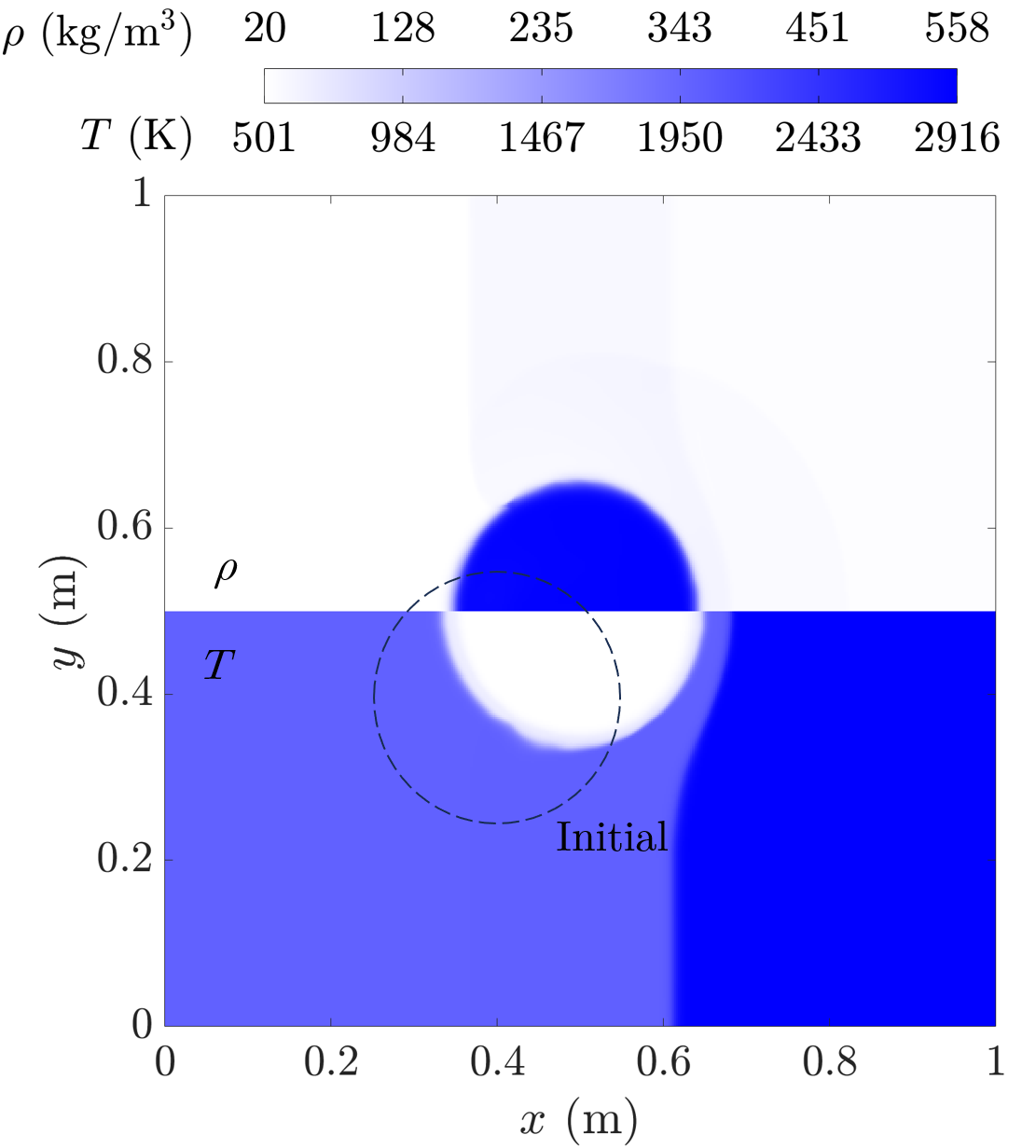}
    \caption{Density and temperature, $t=1.0\text{ms}$}
    \label{fig11c}
\end{subfigure}
\hspace{0.01\textwidth} % 
\begin{subfigure}[b]{0.45\textwidth}
    \includegraphics[width=\textwidth]{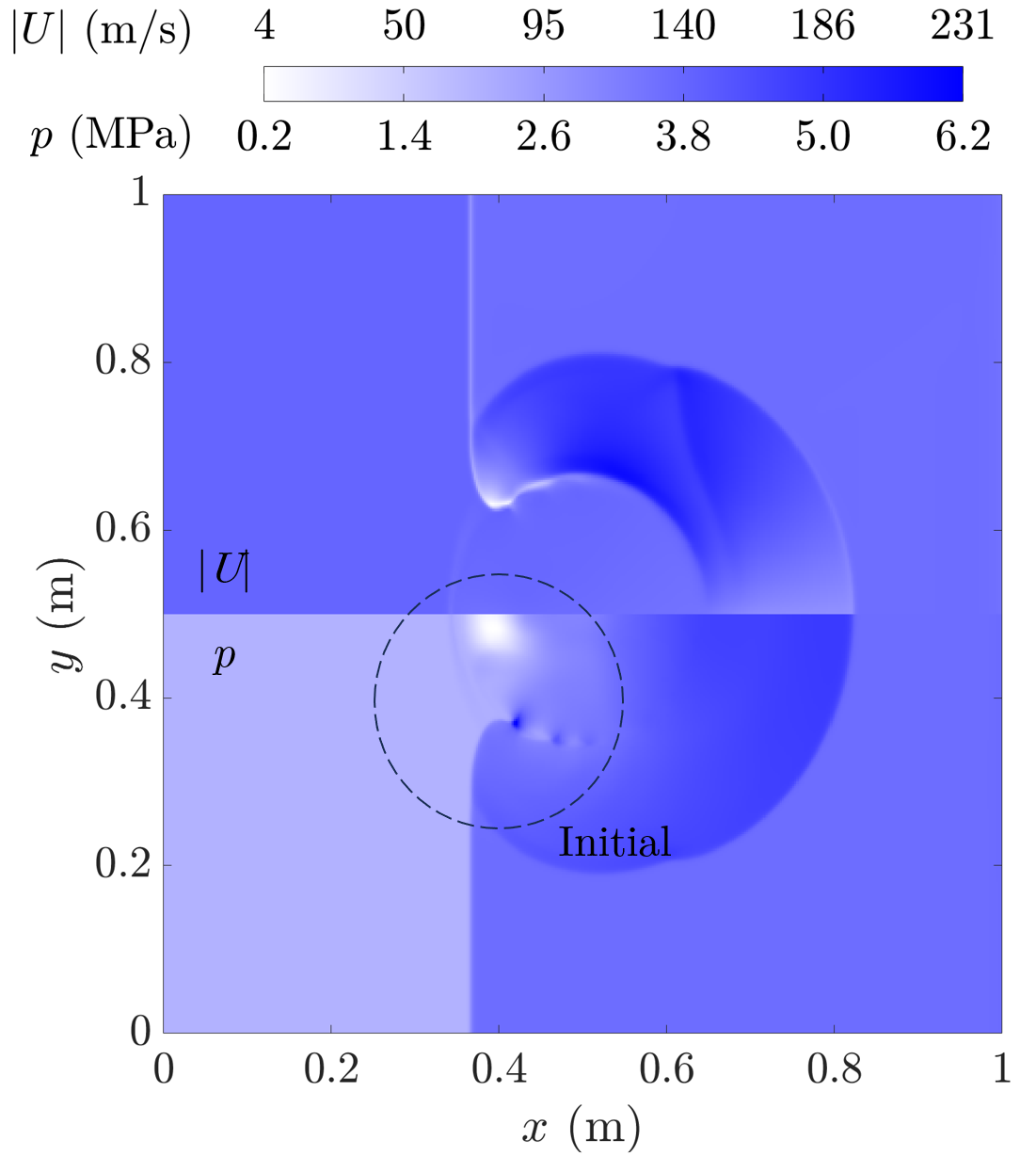}
    \caption{Pressure and velocity, $t=1.0\text{ms}$}
    \label{fig11d}
\end{subfigure}
\caption{Results of the Shock Wave-Droplet Interaction.}
\label{fig11}
\end{figure}

The computational domain is discretized using a uniform quadrilateral mesh with a grid size of $\Delta = 2.5 \times 10^{-3}\text{m}$, and the CFL number is set to 0.3. Fig. \ref{fig11} shows the flow field contours at different times. At $t=0.3\text{ms}$, the shock wave has formed and propagated to approximately $x=0.6\text{m}$, and the droplet centroid has moved to approximately $(0.43, 0.43)$. As shown in Fig. \ref{fig11b}, the pressure and velocity around the droplet remain constant, indicating that pressure oscillations are eliminated. At $t=1\text{ms}$, the droplet centroid reaches approximately $(0.5, 0.5)$, as shown in Fig. \ref{fig11c}. The shock wave has reached the trailing edge of the droplet, inducing complex pressure wave propagation within the droplet, while the pressure distribution across the vapor-liquid interface remains smooth. The computational results demonstrate that the RFQC method effectively suppresses spurious oscillations at the interface, ensuring computational stability.

\subsection{Two-Dimensional Transcritical and Phase-Change Jets}

\begin{figure}[H]
\centering 
\includegraphics[width=0.65\textwidth]{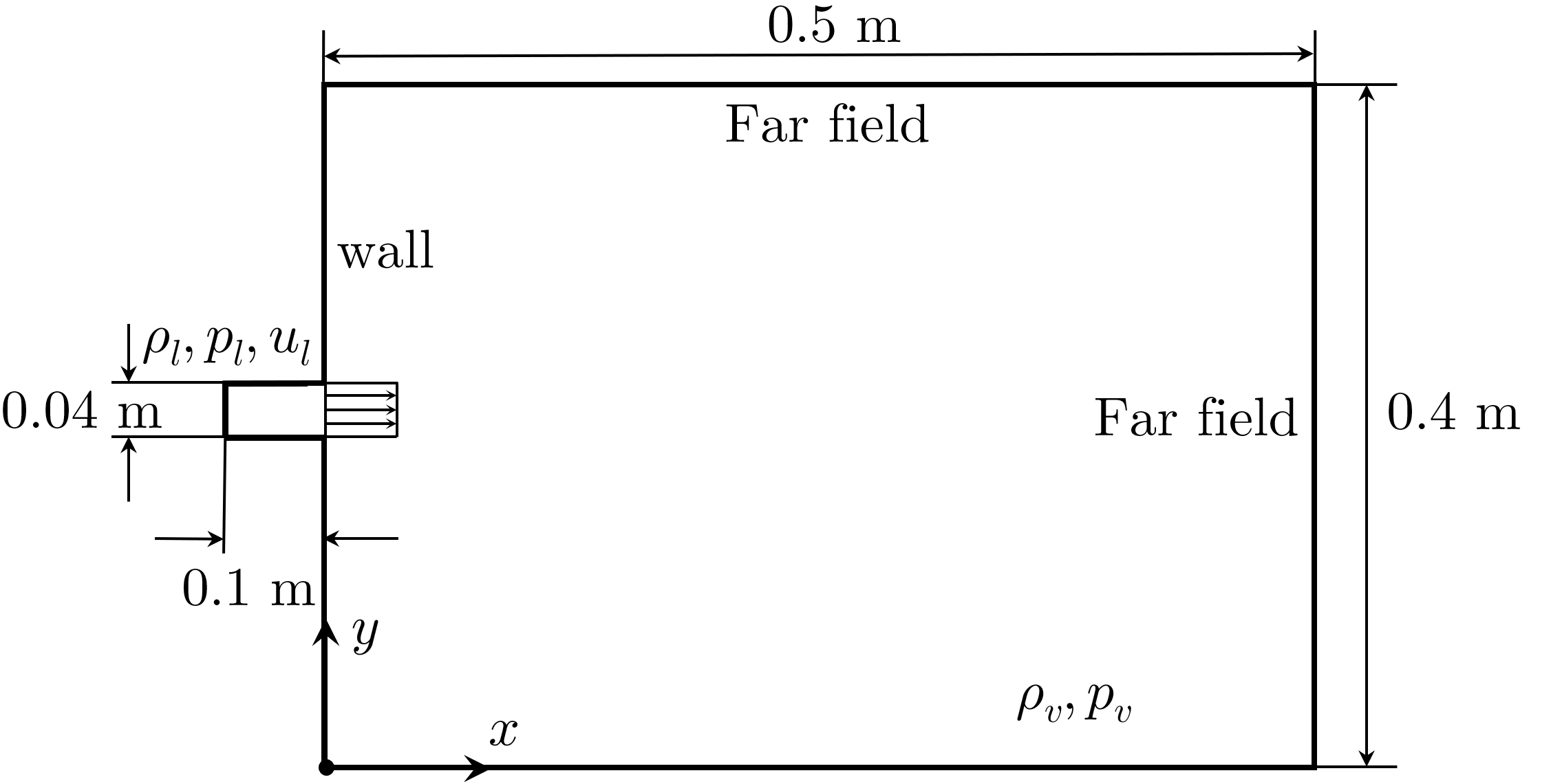} 
\caption{Computational Conditions for Transcritical and Phase-change Jet.}
\label{fig12}
\end{figure}

This case simulates the hot fuel injection process in high-pressure engines. The computational domain consists of a $0.5\text{m} \times 0.4\text{m}$ vapor region and a $0.1\text{m} \times 0.04\text{m}$ liquid region. The schematic and boundary conditions are shown in Fig. \ref{fig12}. Initially, the vapor region has a density of $\rho_{v} = 30 \text{kg/m}^3$, a pressure of $p_v = 2\times 10^6 \text{Pa}$, and is stationary. The liquid region has a pressure of $p_l = 2\times 10^6 \text{Pa}$ and a velocity of $u_l=250\text{m/s}$. The initial density of the jet determines its thermodynamic path after expansion; therefore, we simulate both transcritical and phase-change scenarios by setting different initial densities. For the transcritical jet, the liquid density is $\rho_{l} = 450 \text{kg/m}^3$, and for the phase-change jet, it is $\rho_{l} = 500 \text{kg/m}^3$. The simulation time is $t=1 \text{ms}$.

\begin{figure}[H]
\centering
\begin{subfigure}[b]{0.49\textwidth}
    \includegraphics[width=\textwidth]{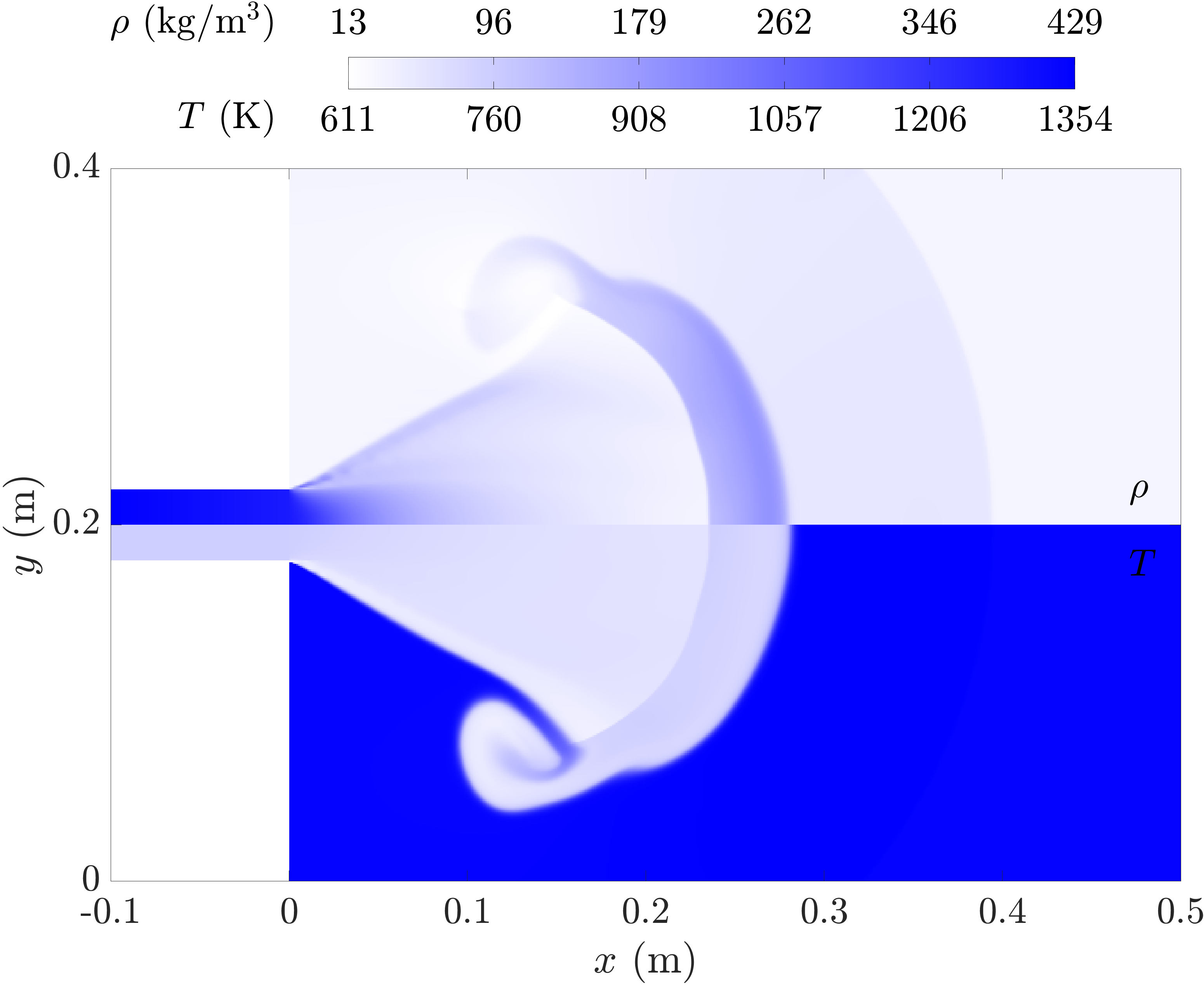}
    \caption{Density and temperature}
    \label{fig13a}
\end{subfigure}
\begin{subfigure}[b]{0.49\textwidth}
    \includegraphics[width=\textwidth]{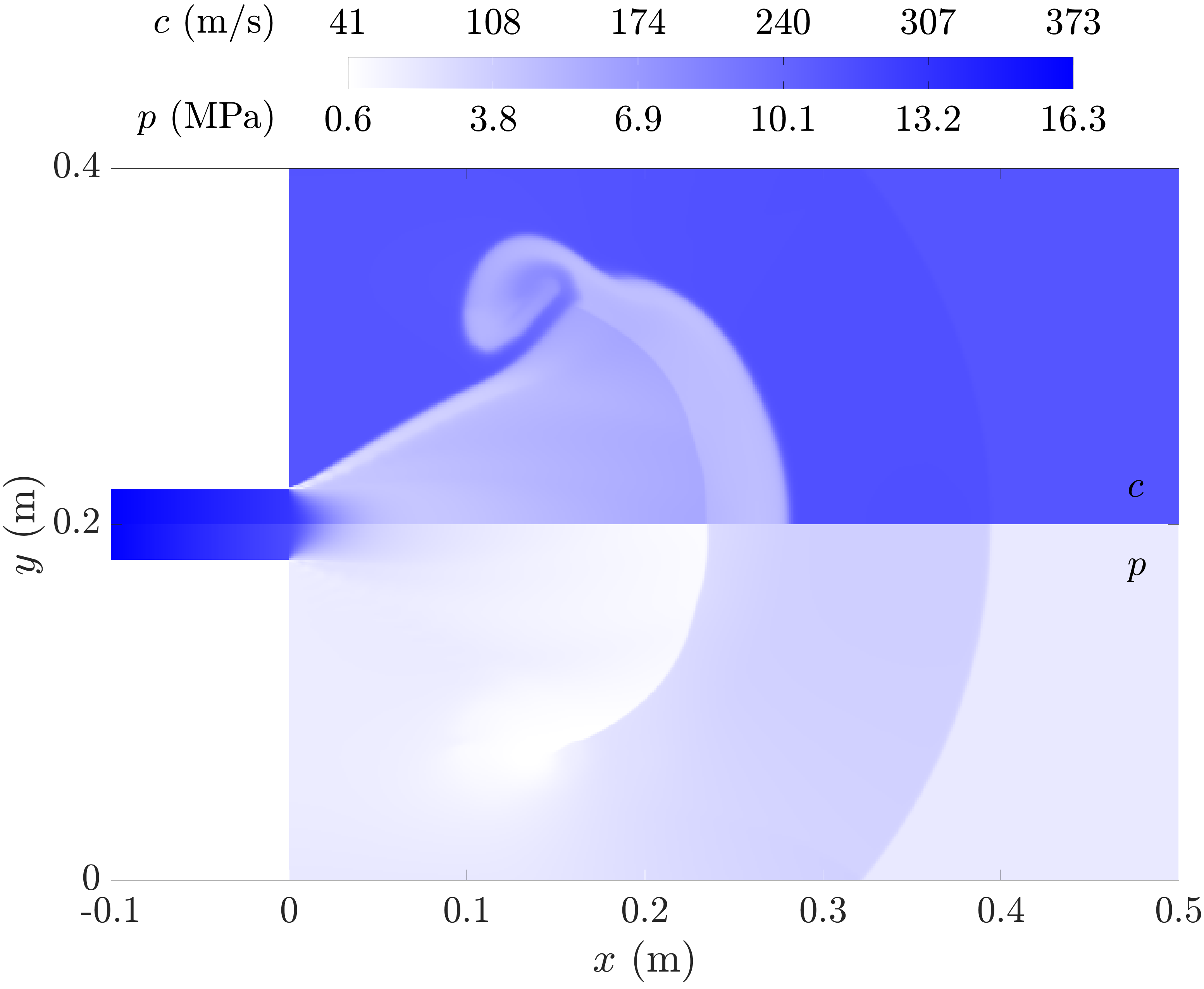}
    \caption{Sound speed and pressure}
    \label{fig13b}
\end{subfigure}
\caption{Results of the Transcritical Jet, $t=1 \text{ms}$.}
\label{fig13}
\end{figure}

The computational domain is discretized using a uniform quadrilateral mesh with a grid size of $\Delta = 1 \times 10^{-3}\text{m}$, and the CFL number is set to 0.3. Fig. \ref{fig13} displays the contours of the transcritical jet. Combined with the scatter distribution of the simulation points on the $\rho-p$ phase diagram (Fig. \ref{fig15a}), it is observed that the liquid phase expands rapidly along the isentrope after the simulation begins. The expansion path directly transitions into the vapor phase without entering the two-phase region, thus exhibiting thermodynamic transcritical behavior. A low sound speed region also appears at the boundary of the jet shear layer, which is caused by the direct homogeneous mixing of vapor and liquid within the shear layer. Furthermore, the over-expansion of the jet forms an intercepting shock (near $x=0.24\text{m}$), which recompresses the over-expanded vapor and increases its density.

\begin{figure}[H]
\centering
\begin{subfigure}[b]{0.49\textwidth}
    \includegraphics[width=\textwidth]{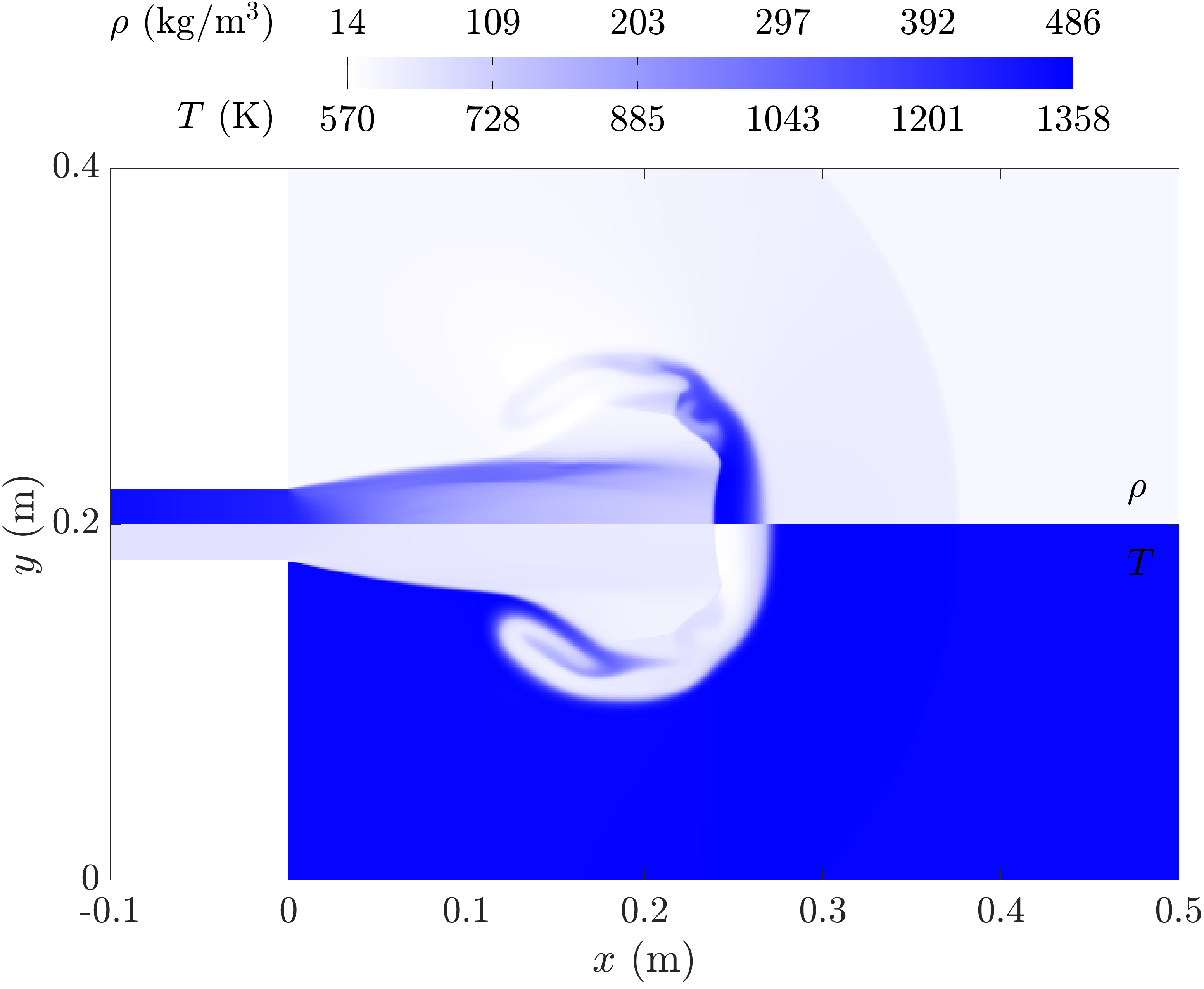}
    \caption{Density and temperature}
    \label{fig14a}
\end{subfigure}
\begin{subfigure}[b]{0.49\textwidth}
    \includegraphics[width=\textwidth]{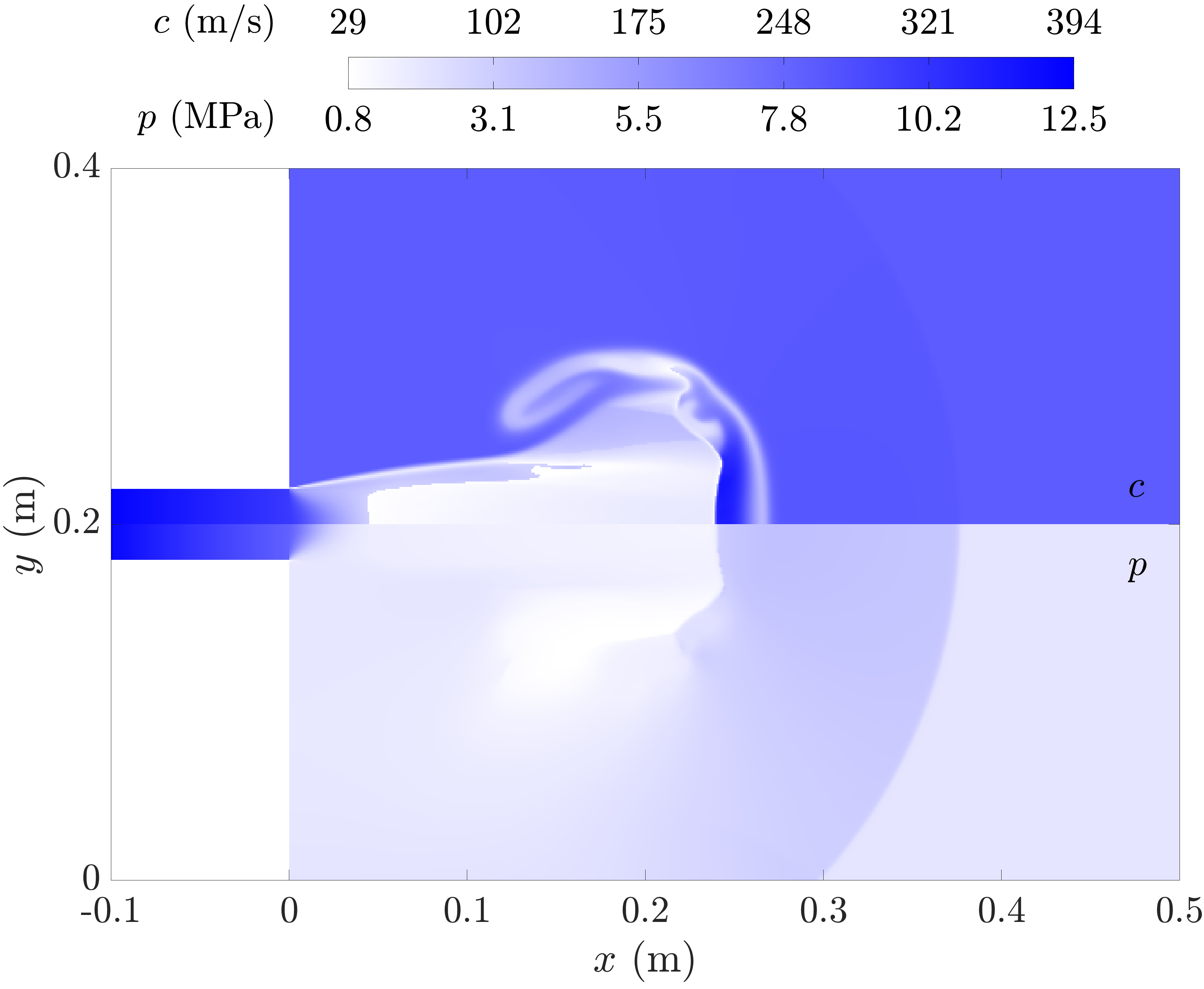}
    \caption{Sound speed and pressure}
    \label{fig14b}
\end{subfigure}
\caption{Results of the Phase-change Jet, $t=1 \text{ms}$.}
\label{fig14}
\end{figure}

\begin{figure}[H]
\centering
\begin{subfigure}[b]{0.49\textwidth}
    \includegraphics[width=\textwidth]{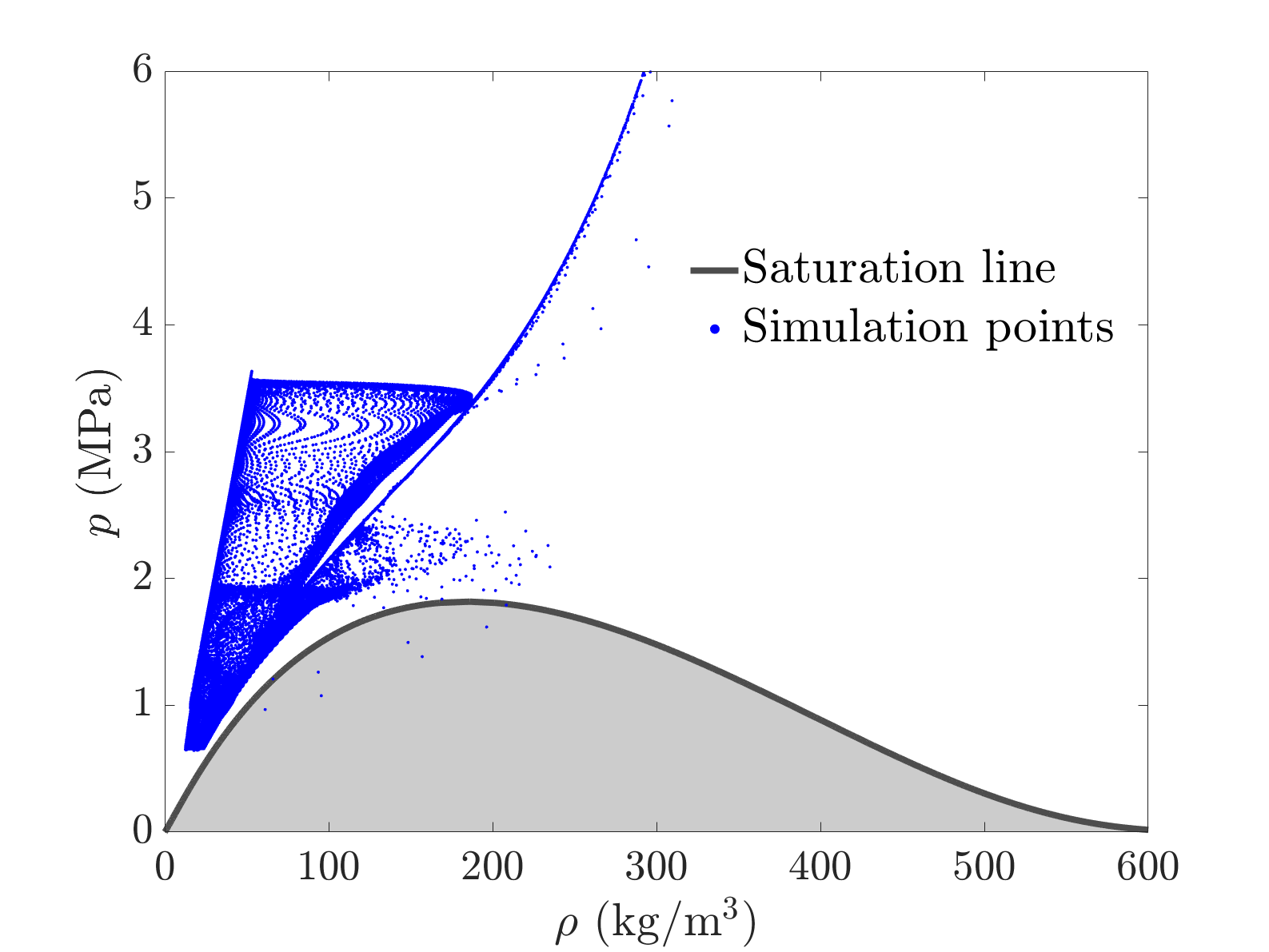}
    \caption{Transcritical Jet}
    \label{fig15a}
\end{subfigure}
\begin{subfigure}[b]{0.49\textwidth}
    \includegraphics[width=\textwidth]{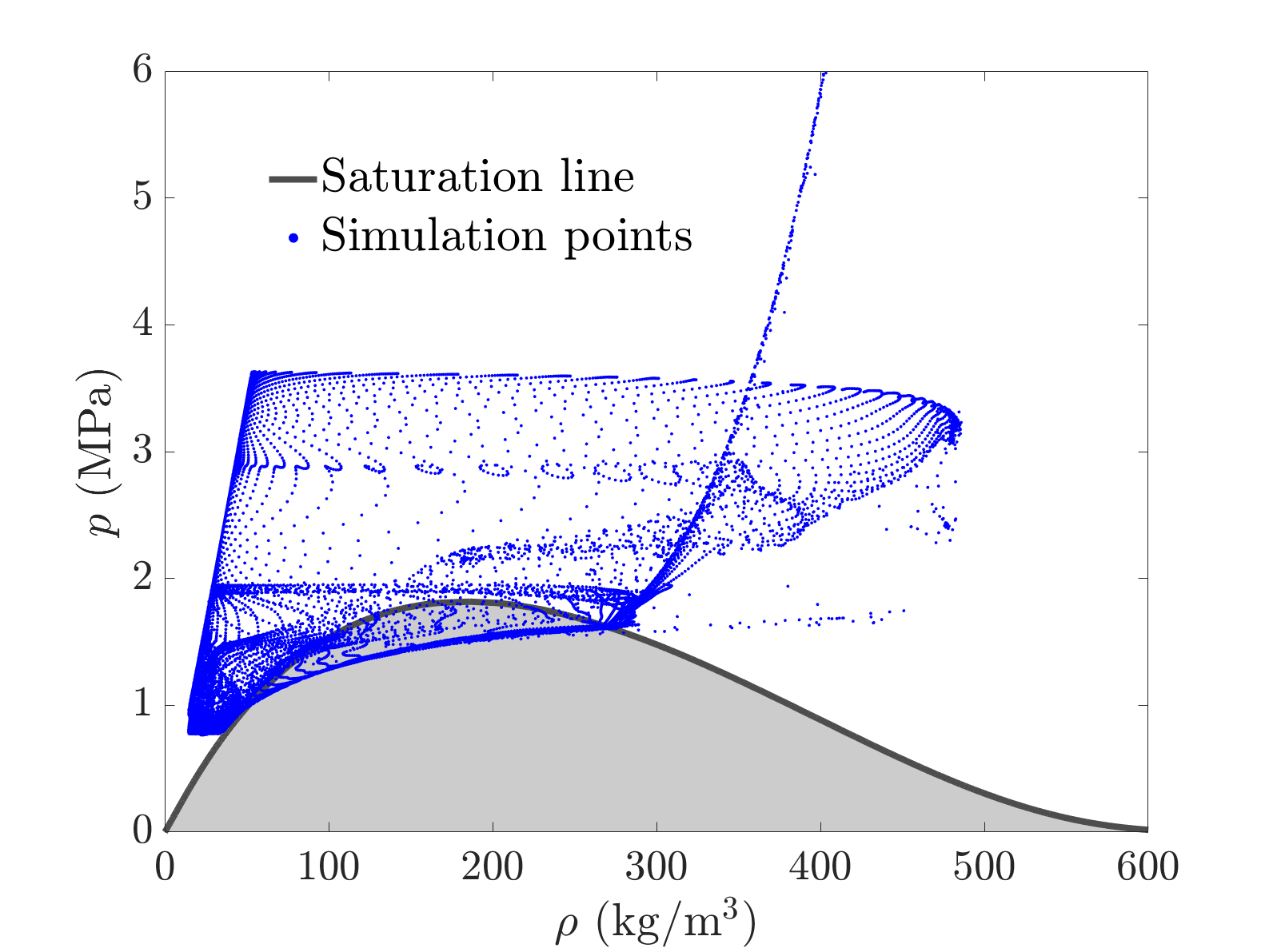}
    \caption{Phase-change Jet}
    \label{fig15b}
\end{subfigure}
\caption{Simulation points on $P-\rho$ phase diagram.}
\label{fig15}
\end{figure}

Fig. \ref{fig14} shows the contours of the phase-change jet. The scatter distribution of the simulation points on the $\rho-p$ phase diagram is shown in Fig. \ref{fig15b}. After the simulation begins, the liquid phase expands rapidly along the isentrope and enters the two-phase region, exhibiting phase change. This phase change causes a sharp drop in the speed of sound, which appears as a kink on the isentrope in Fig. \ref{fig15b}. As seen in Fig. \ref{fig14b}, a discontinuity in the sound speed contours occurs near $x=0.4\text{m}$, forming a large low-sound-speed region downstream. The over-expansion of the jet forms a very strong intercepting shock (near $x=0.24\text{m}$), which recompresses the low-density homogeneous mixture back into a high-density liquid.

Furthermore, we point out that the sound speed discontinuity induces rarefaction wave splitting, as demonstrated by the flash evaporation Riemann problem in Sec. \ref{sec:FeRP}. Theoretically, two different sound speeds exist at the location of the rarefaction wave splitting; therefore, small numerical deviations in density and pressure can easily cause the sound speed to jump between the two discontinuous values. This highly sensitive thermodynamic property makes the WENO scheme based on characteristic reconstruction prone to numerical instability, often leading to computational divergence. We recommend employing the second-order or first-order scheme for reconstruction in phase-change problems to ensure computational stability.

\section{Conclusion}
This paper extends the classic quasi-conservative five-equation scheme to general equations of state, developing a Real Fluid Quasi-Conservative (RFQC) scheme that achieves stable simulation of real fluids with strongly nonlinear thermodynamic properties. The accuracy of the method is verified against exact solutions through Riemann problems involving transcritical and phase-change fluids under Homogeneous Equilibrium (HEM) and Vapor-Liquid Equilibrium (VLE) assumptions. The main conclusions are as follows.

To address the issue of spurious pressure oscillations prevalent in traditional conservative schemes for real fluid calculations, this paper formalizes a generalized strategy involving energy-pressure freezing, flux evolution, pressure reconstruction, and thermodynamic re-projection. This formulation introduces a re-projection error, which acts as an effective mechanism to suppress spurious pressure oscillations. The success of such methods relies on the appropriate selection and evolution of the linearization coefficients. The Double Flux (DF) method utilizes the isentropic index to construct frozen coefficients and performs re-projection after a double flux evolution. In contrast, the RFQC method employs the Grüneisen coefficient and evolves it within Shyue’s quasi-conservative framework, thereby effectively eliminating the energy flux inconsistency error present in the DF method. Numerical experience suggests that the choice of the internal energy reference point for the Grüneisen parameter influences the algorithm's numerical characteristics. Therefore, constructing more optimal frozen coefficients will be a focus of future research.

Theoretical analysis demonstrates that the energy conservation error of the RFQC method is determined by the entropy production rate and the nonlinear curvature of the real fluid EoS. In smooth flow regions, the local error introduced by the thermodynamic projection is second-order with respect to the time step; in discontinuous regions, the error remains first-order, implying that the conservation error is dominated by time discretization and introduces no additional low-order spatial errors  beyond those inherent to shock capturing schemes. Numerical experiments further verify that with a first-order space-time discretization  scheme, the energy conservation error in smooth regions exhibits second-order accuracy. These results demonstrate the robustness and potential of the proposed method in handling extreme phase-change flows. However, maintaining high-order accuracy poses a challenge to the computational robustness of real-fluid simulations. To further apply this algorithm to practical engineering applications such as sprays in propulsion systems, simulations of multi-component and three-dimensional cases, as well as the development of high-order reconstruction schemes adapted to the RFQC framework, will be our primary research focus in the future.

\appendix
\section{Thermodynamic Relations}
\label{app:thermo}
Consider a single-component fluid under HEM and VLE assumptions. The specific thermodynamic relations are summarized as follows:

In the single-phase region, the equation of state takes $\rho$ and $T$ as independent variables, and $p$ has the form:
\begin{equation}
    p=p\left( \rho ,T \right)
    \label{equ:state1}
\end{equation}\par
In the two-phase equilibrium region, $p$ is a function of $T$. The $p-T$ curve is referred to as the saturation curve:
\begin{equation}
    p=p_{sat}\left(T \right)
    \label{equ:state2}
\end{equation}\par
Any point $(p,T)$ on the saturation curve corresponds to two density values: the saturated vapor density $\rho_v$ and the saturated liquid density $\rho_l$. The relationship between the two-phase mixture density $\rho$ and the saturation densities (also known as the lever rule) is:
\begin{equation}
    \rho=\alpha \rho_v +(1-\alpha)\rho_l
    \label{equ:rho2}
\end{equation}\par
where $\alpha$ is the vapor volume fraction. Eq. \ref{equ:rho2} is equivalent to:
\begin{equation}
    \alpha(\rho,T)=\frac{\rho-\rho_l(T)}{\rho_v(T) - \rho_l(T)}
    \label{equ:alpha}
\end{equation}\par
The expression for single-phase internal energy is:
\begin{equation}
    e(\rho,T)=\int_{{{T}_{0}}}^{T}{{{C}_{v}}\text{d}T}-\int_{{{\rho }_{0}}}^{\rho }{\frac{1}{{{\rho }^{2}}}\left( T\pderiv{p}{T}{\rho}-p \right)}\text{d}\rho +e_0
    \label{equ:e}
\end{equation}
where $C_v$ is the specific heat at constant volume. $C_v$ is also a function of $\rho$ and $T$, which often termed the caloric equation of state:
\begin{equation}
    {C_v}={C_v}(\rho ,T)
    \label{equ8}
\end{equation}\par
The internal energy expression for the two-phase mixture is:
\begin{equation}
    e(\rho,T)=\frac{\alpha \rho_v(T) e_v(T) +(1-\alpha)\rho_l(T) e_l(T)}{\rho}
    \label{equ:mix_e}
\end{equation}
where $e_v$ and $e_l$ are the internal energies at point $(p,T)$ on the vapor and liquid saturation lines, respectively.\par
The expression for single-phase entropy is:
\begin{equation}
    s(\rho,T)=\int_{{{T}_{0}}}^{T}{\frac{{{C}_{v}}}{T}\mathrm{d}T}-\int_{{{\rho }_{0}}}^{\rho }{\frac{1}{{{\rho }^{2}}}{\pderiv{p}{T}{\rho}}\mathrm{d}\rho }+s_0
    \label{equ:s}
\end{equation}\par
The two-phase mixture entropy expression is similar to that of internal energy. Defining $s_v$ and $s_l$ as the entropy at point $(p,T)$ on the saturation curve:
\begin{equation}
    s(\rho,T)=\frac{\alpha \rho_v(T) s_v(T) +(1-\alpha)\rho_l(T) s_l(T)}{\rho}
    \label{equ:mix_s}
\end{equation}

The speed of sound for the single-phase fluid is:
\begin{equation}
    {c^2}= \pderiv{p}{\rho}{s}=\pderiv{p}{\rho}{T}+\frac{T}{{{c}_{v}}{{\rho }^{2}}}\pderiv{p}{T}{\rho}^{2}
    \label{equ:c}
\end{equation}
The equilibrium speed of sound for the two-phase mixture can be calculated from the ratio of the derivatives of mixture entropy with respect to density and pressure. The simplified expression is:
\begin{equation}
    \frac{1}{\rho c_{eq}^2}=\frac{\alpha }{{\rho_v}c_v^2}+\frac{1-\alpha }{{\rho_l}c_l^2}+T\left( \frac{\alpha {\rho_v}}{{C_{p,v}}}{{\left( \frac{\text{d}s_v}{\text{d}p} \right)}^{2}}+\frac{(1-\alpha ){\rho_l}}{C_{p,l}}{{\left( \frac{\text{d}s_l}{\text{d}p} \right)}^{2}} \right)
    \label{equ:c_eq}
\end{equation}
where $c_v$, $c_l$ and $C_{p,v}$, $C_{p,l}$ are the speeds of sound and specific heats at constant pressure on the saturation line at $(p,T)$, respectively. They are also single-valued functions of $p$ or $T$. Additionally, $C_v$ and $C_p$ satisfy the thermodynamic relation \cite{Sedov1973}:
\begin{equation}
    {C_p}-{C_v}=\frac{T}{{\rho^2}}\frac{\left( {\partial p}/{\partial T}\;
\right)_\rho^2}{{\partial p}/{\partial \rho}}
    \label{equ:CpCv}
\end{equation}

\section{Appendix: P-R Equation of State and Thermodynamic Quantities}
\label{app:PR_EOS}
All thermodynamic calculations in this paper employ the single-component P-R equation of state. For multi-component cases, please refer to \cite{Ma2017}. The P-R EoS is defined as:
\begin{equation}
p = \frac{RT}{V - b} - \frac{a}{V(V + b) + b(V - b)}
\end{equation}
where $R=8.31446[ \text{J}/( \text{mol}\cdot \text{K})] $ is the universal gas constant, $V$ is the \underline{molar volume} (molar volume $V$ is related to specific volume $v$ by $v = V/M$, where $M$ is the molar mass), $a$ is the energy parameter, and $b$ is the covolume parameter.
The coefficients of the P-R EoS are defined as:
\begin{align}
    &a(T) = a_c \cdot \alpha(T) = a_c \left[ 1 + c\left( 1 - \sqrt{T/T_c} \right) \right]^2 \\
    &a_c = 0.457236\frac{R^2 T_c^2}{P_c}\\
    &b = 0.077796 \frac{R T_c}{P_c} \\
    &c = 0.37464 + 1.54226\omega - 0.26992\omega^2 
\end{align}
where $R$ is the gas constant and $\omega$ is the acentric factor. Calculating internal energy and specific heat capacity requires the following derivatives:
\begin{align}
    \pderiv{p}{T}{V}&=\frac{R}{V-b}-\frac{{\mathrm{d}a}/{\mathrm{d}T}}{V^2+2Vb-b^2}\\
    \pderiv{p}{V}{T}&= -\frac{RT}{(V-b)^2} +\frac{2a(V+b)}{(V^2+2Vb-b^2)^2}\\
    \frac{\mathrm{d}a}{\mathrm{d}T} &= - \frac{c 
\sqrt{a \cdot a_c}}{\sqrt{T_c T}}\\
    \frac{\mathrm{d}^2 a}{\mathrm{d} T^2} &= \frac{a_c c (1+c)}{2 T \sqrt{T T_c}}
\end{align}

For general real fluids, thermodynamic parameters are typically calculated from ideal gas values and departure functions. In the following, the superscript zero (0) denotes the ideal gas value of the thermodynamic quantity. For the P-R EoS, the analytical integration expression for energy is:

\begin{equation}
e = e^0(T) + \frac{K_0}{M} \left( a - T \frac{\mathrm{d}a}{\mathrm{d}T} \right).
\end{equation}
where $K_0$ is a function related to specific volume:
\begin{equation}
K_0 = \int_{+\infty}^{V} \frac{1}{V^2 + 2bV - b^2} dV = \frac{1}{2\sqrt{2}b} \ln \left( \frac{V + (1 - \sqrt{2})b}{V + (1 + \sqrt{2})b} \right).
\end{equation}

Similarly, specific entropy can be expressed as a combination of the ideal gas value and a departure function. For the P-R EoS, its integrated value is:

\begin{equation}
s = s^0(T) + R \ln \frac{V - b}{V} - \frac{K_0}{M} \left( \frac{\mathrm{d}a}{\mathrm{d}T} \right).
\end{equation}

The specific heat at constant volume is calculated as:
\begin{equation}
C_v = \left( \frac{\partial e}{\partial T} \right)_{v} = C_v^0 - \frac{K_0}{M} T \left( \frac{\mathrm{d}^2a}{\mathrm{d}T^2} \right),
\end{equation}
The specific heat at constant pressure is calculated as:
\begin{equation}
C_p = \left( \frac{\partial h}{\partial T} \right)_{p} = C_p^0 - R - \frac{K_0}{M} T \left( \frac{\mathrm{d}^2a}{\mathrm{d}T^2}  \right)- \frac{T}{M} \frac{(\partial p / \partial T)_{V}^2}{(\partial p / \partial V)_{T}},
\end{equation}
where $C_v^0$ is the specific heat at constant volume for the ideal gas, and $C_p^0$ is the specific heat at constant pressure for the ideal gas.

\section{Thermophysical Parameters of Dodecane}
\label{app:dodecane}
Using the P-R equation of state to calculate the saturation properties of Dodecane requires the following parameters:
\begin{table}[H]
    \centering
    \caption{Thermodynamic parameters of Dodecane}
    \label{tab:dodecane1}
    \begin{tabular}{ccccc}
        \toprule
          $M$ ($\mathrm{kg\,kmol^{-1}}$) & $\omega$ & $T_c$ (K) & $\rho_c$ ($\mathrm{kg \,m^{-3}}$) & $p_c$ (MPa) \\
        \midrule
        170.33& 0.574 & 658.1 & 186 & 1.82\\
        \bottomrule
    \end{tabular}
\end{table}
The precise integration of internal energy and speed of sound using the P-R EoS requires the following parameters:
\begin{table}[h]
    \centering
    \caption{The parameters of the energy and entropy model}
    \label{tab:dodecane2}
    \begin{tabular}{cccc}
        \toprule
          $C_{v,\infty}^c$ ($\mathrm{J \, kg^{-1} \,K^{-1}}$) & $n $ & $e_\text{ref}$ ($\mathrm{J \, kg^{-1}}$)  \\
        \midrule
       $2.970 \times 10^3$& 0.613 & $1 \times 10^6$ \\
        \bottomrule
    \end{tabular}
\end{table}\par
For the specific meaning of the above parameters, please refer to \cite{GuardoneArgrow2005}. Here, $\rho_c$ is the calculation result of the P-R EoS based on $T_C$ and $P_c$, which is smaller than the NIST data \cite{NIST2023} ($\rho_c = 226\mathrm{kg \,m^{-3}}$). The calculation results of key thermodynamic variables such as saturation line properties and saturation speed of sound are in good agreement with NIST data.

\section{Discrete Formulation of the RFQC Method}
\label{app:Discrete}
The discretization of the RFQC method combined with the HLLC solver is 
generally identical to Shyue's original scheme, mainly following the work of Johnsen and Colonius \cite{Johnsen2006}. 

\subsection{Discretization of Euler equations and advection equations}
For the Euler equations, denoting the conservative state vector as $\mathbf{U} = (\rho, \rho u, \rho e_t)^T$ and the corresponding flux vector as $\mathbf{F} = (\rho u, \rho u^2 + p, (\rho e_t + p)u)^T$, the discretization is identical to the standard Godunov finite volume method. At cell $i$, the semi-discrete update equation is given by:
\begin{equation}
\frac{\mathrm{d} \mathbf{U}_i}{\mathrm{d} t} = 
- \frac{1}{\Delta x} \left( \mathbf{F}^{\mathrm{HLLC}}_{i+\frac12} - \mathbf{F}^{\mathrm{HLLC}}_{i-\frac12} \right),
\label{eq:Euler_discrete}
\end{equation}
where $\mathbf{F}^{\mathrm{HLLC}}_{i+\frac12}$ is the numerical flux calculated by the HLLC Riemann solver \cite{Toro2009} at the cell interface.

However, special attention is required for the advection equations. The basic principle of the discretization is that the thermodynamic coefficients $\boldsymbol{\Phi}$ must be transported consistently with the mass flux. Specifically, following the finite volume formulation for conservative variables, the advection equation \ref{equ:xi_eq} can be equivalently rewritten into a "flux and source" form:
\begin{equation}
\frac{\partial \boldsymbol{\Phi}}{\partial t} + \frac{\partial (u\boldsymbol{\Phi})}{\partial x}
=
\boldsymbol{\Phi}\frac{\partial u}{\partial x}.
\label{eq:advFluxSource}
\end{equation}

At cell $i$, the semi-discrete update equation for the advection equation is:
\begin{equation}
\frac{\mathrm{d} \boldsymbol{\Phi}_i}{\mathrm{d} t} = 
- \frac{1}{\Delta x} \left( u^{\mathrm{HLLC}}_{i+\frac12} \boldsymbol{\Phi}^{\mathrm{up}}_{i+\frac12} - u^{\mathrm{HLLC}}_{i-\frac12} \boldsymbol{\Phi}^{\mathrm{up}}_{i-\frac12} \right)
+ \frac{\boldsymbol{\Phi}_i}{\Delta x} \left( u^{\mathrm{HLLC}}_{i+\frac12} - u^{\mathrm{HLLC}}_{i-\frac12} \right),
\label{eq:Phi_split}
\end{equation}
where $\boldsymbol{\Phi}^\mathrm{up}$ is the upwind value determined by the sign of the HLLC mass flux:
\begin{equation}
\boldsymbol{\Phi}^{\mathrm{up}}_{i+\frac12} = 
\begin{cases}
\boldsymbol{\Phi}_{i}, &
(\rho u)^{\mathrm{HLLC}}_{i+\frac12} \ge 0, \\
\boldsymbol{\Phi}_{i+1}, &
(\rho u)^{\mathrm{HLLC}}_{i+\frac12} < 0 ,
\end{cases}
\label{eq:Phi_upwind}
\end{equation}
and $u^{\mathrm{HLLC}}$ is the interface velocity given by the HLLC solver:
\begin{equation}
u^{\mathrm{HLLC}}_{i+\frac12} \equiv \frac{(\rho u)^{\mathrm{HLLC}}_{i+\frac12}}{\rho^{\mathrm{up}}_{i+\frac12}}.
\label{eq:u_cons_def}
\end{equation}
Furthermore, this method can be naturally extended to multi-dimensional computations without dimensional splitting. In the two-dimensional framework, denoting the corresponding $y$-direction flux as $\mathbf{G}$, the semi-discrete update equation for the conservative variables $\mathbf{U} = (\rho, \rho u, \rho v, \rho e_t)^T$ at cell $(i,j)$ is directly expressed as:
\begin{equation}
\frac{\mathrm{d} \mathbf{U}_{i,j}}{\mathrm{d} t} = 
- \frac{1}{\Delta x} \left( \mathbf{F}^{\mathrm{HLLC}}_{i+\frac12,j} - \mathbf{F}^{\mathrm{HLLC}}_{i-\frac12,j} \right) 
- \frac{1}{\Delta y} \left( \mathbf{G}^{\mathrm{HLLC}}_{i,j+\frac12} - \mathbf{G}^{\mathrm{HLLC}}_{i,j-\frac12} \right).
\label{eq:Euler_discrete_2d}
\end{equation}

Correspondingly, the two-dimensional unsplit update equation for the thermodynamic coefficients $\boldsymbol{\Phi}$ incorporates the advection and the velocity divergence terms from both spatial directions:
\begin{equation}
\begin{aligned}
\frac{\mathrm{d} \boldsymbol{\Phi}_{i,j}}{\mathrm{d} t} = 
&- \frac{1}{\Delta x} \left( u^{\mathrm{HLLC}}_{i+\frac12,j} \boldsymbol{\Phi}^{x,\mathrm{up}}_{i+\frac12,j} - u^{\mathrm{HLLC}}_{i-\frac12,j} \boldsymbol{\Phi}^{x,\mathrm{up}}_{i-\frac12,j} \right) \\
&- \frac{1}{\Delta y} \left( v^{\mathrm{HLLC}}_{i,j+\frac12} \boldsymbol{\Phi}^{y,\mathrm{up}}_{i,j+\frac12} - v^{\mathrm{HLLC}}_{i,j-\frac12} \boldsymbol{\Phi}^{y,\mathrm{up}}_{i,j-\frac12} \right) \\
&+ \frac{\boldsymbol{\Phi}_{i,j}}{\Delta x} \left( u^{\mathrm{HLLC}}_{i+\frac12,j} - u^{\mathrm{HLLC}}_{i-\frac12,j} \right) 
+ \frac{\boldsymbol{\Phi}_{i,j}}{\Delta y} \left( v^{\mathrm{HLLC}}_{i,j+\frac12} - v^{\mathrm{HLLC}}_{i,j-\frac12} \right),
\end{aligned}
\label{eq:Phi_discrete_2d}
\end{equation}

\subsection{High-Order Reconstruction}
\label{appsub:high}
To achieve high-order spatial accuracy without generating spurious oscillations, a characteristic reconstruction based on the primitive variables $\mathbf{W} = (\rho, u_n, u_t, p, \xi, E_0)^T$ is employed. At each interface, the local frozen acoustic speed squared $c^2$ and impedance $\hat{Z}$ are evaluated with the arithmetic average of the adjacent cell states:

\begin{equation}c^2 = \frac{p_{\mathrm{avg}}(1 + \xi_{\mathrm{avg}}) + E_{0,\mathrm{avg}}}{\rho_{\mathrm{avg}}\xi_{\mathrm{avg}}}, \quad \hat{Z} = \rho_{\mathrm{avg}} \sqrt{c^2},\label{eq:acoustic_scale}
\end{equation}

The local states are projected onto the characteristic space to obtain the characteristic variable vector:
\begin{equation}\mathcal{W} = \left( \rho - \frac{p}{c^2} , \xi , E_0, u_t, \frac{p + \hat{Z} u_n}{2} , \frac{p - \hat{Z} u_n}{2} \right)^T.\label{eq:char_project}
\end{equation}

In the present work, the standard third-order WENO interpolation \cite{Jiang1996} is applied component-wise to $\mathcal{W}$ to obtain the reconstructed left and right states, $\hat{\mathcal{W}}_L$ and $\hat{\mathcal{W}}_R$. Finally, the physical primitive variables $\mathbf{W}_{L/R}$ at the quadrature points are recovered by:
\begin{equation}
\begin{aligned}
&p = \hat{\gamma}^{(5)} + \hat{\gamma}^{(6)}, \quad u_n = \frac{\hat{\gamma}^{(5)} - \hat{\gamma}^{(6)}}{\hat{Z}}, \quad u_t = \hat{\gamma}^{(4)}, \\
&\rho = \hat{\gamma}^{(1)} + \frac{p}{c^2}, \quad \xi = \hat{\gamma}^{(2)}, \quad E_0 = \hat{\gamma}^{(3)},
\end{aligned}
\label{eq:inverse_char}
\end{equation}
where $\hat{\gamma}^{(m)}$ denotes the $m$-th component of the reconstructed characteristic vector $\mathcal{W}$.

In multi-dimensional finite volume frameworks beyond second-order accuracy, Gauss quadrature is required to evaluate the face-averaged numerical fluxes \cite{Titarev2004}. For the present third-order scheme, a two-point Gauss quadrature rule is employed. Taking the vertical cell interface $(i+1/2, j)$ as an example, the face-averaged flux vector $\mathbf{F}^{\mathrm{HLLC}}_{i+1/2,j}$ (Eq. \ref{eq:Euler_discrete_2d}) is computed via a weighted summation at the quadrature points:
\begin{equation}
\mathbf{F}^{\mathrm{HLLC}}_{i+\frac12,j} = \sum_{g=1}^{2} \omega_g \mathbf{F}^{\mathrm{HLLC}}\left(\mathbf{W}_{L,g}, \mathbf{W}_{R,g}\right),
\label{eq:Gauss_F}
\end{equation}
where $\omega_1 = \omega_2 = 1/2$ are the quadrature weights. To evaluate the interface fluxes, all components of the left and right primitive variable vectors $\mathbf{W}_{L/R,g}$ are reconstructed at these specific Gauss quadrature locations along the cell interface:
\begin{equation}
y_g = y_j \pm \frac{\sqrt{3}}{6}\Delta y, \quad (g=1,2).
\label{eq:Gauss_points_y}
\end{equation}

Analogously, the horizontal fluxes $\mathbf{G}^{\mathrm{HLLC}}_{i,j+\frac12}$ and all remaining interface variables in Eq. \eqref{eq:Phi_discrete_2d} are evaluated using the same quadrature rule.

It is worth noting that for schemes beyond second-order accuracy, the present method directly recovers the primitive variables $\mathbf{\bar{W}}$ from the cell-averaged conservative variables $\mathbf{\bar{U}}$, which is consistent with the method of Johnsen and Colonius \cite{Johnsen2006}. However, Coralic and Colonius \cite{Coralic2014} pointed out that directly using cell-averaged conservative variables degrades accuracy, and in principle, $\mathbf{U}$ should be reconstructed at multiple cell-volume quadrature points, converted to $\mathbf{W}$ pointwise, and then averaged to recover $\mathbf{\bar{W}}$ with the corresponding accuracy. Although this procedure is mathematically more rigorous, it can lead to numerical instabilities when simulating thermodynamically sensitive real fluids. Therefore, the method of Johnsen \cite{Johnsen2006} is retained here to guarantee computational robustness.

\end{document}